\documentclass[aps,twocolumn,superscriptaddress,groupedaddress]{revtex4}  
\usepackage{graphicx}  
\usepackage{dcolumn}   
\usepackage{bm}        
\usepackage{amssymb}   
\usepackage{amsmath}
\usepackage{cleveref}
\usepackage{mathtools}
\usepackage{appendix} 
\usepackage{comment}
\usepackage{enumitem}

\usepackage{color}
\usepackage{xcolor}
\newcommand{\si}{SASI}

\newcommand{\sn}{supernova}

\newcommand{\n}{neutrino}
\newcommand{\ns}{neutrinos}
\newcommand{\gw}{GW}

\newcommand{\hk}{Hyper-K}

\newcommand{\nsi}{no-SASI}
\newcommand{\kur}{KKHT}

\begin{document}



\title{Characterizing a supernova's Standing Accretion Shock Instability with neutrinos and gravitational waves} 
\author{Zidu Lin }
\affiliation{Department of Physics, University of Tennessee Knoxville, TN 37996, USA\\ }
\author{Abhinav Rijal}
\affiliation{Embry-Riddle Aeronautical University, Prescott campus, Prescott, AZ 86301, USA} 
\affiliation{ Institute Of Engineering (Pulchowk Campus), Tribhuvan University, Lalitpur 44600, Nepal}
\author{Cecilia Lunardini }
\affiliation{Department of Physics, Arizona State University, \\ 450 E. Tyler Mall, Tempe, AZ 85287-1504, USA}
\author{Manuel D. Morales}
\affiliation{Departamento de F\'isica, 
Universidad de Guadalajara, \\ Guadalajara, Jal., 44430, M\'exico}
\author{Michele Zanolin }
\affiliation{Embry-Riddle Aeronautical University, Prescott campus, AZ, USA}

\date{\today}

\begin{abstract}
 We perform a novel multi-messenger analysis for the identification and parameter estimation of the Standing Accretion Shock Instability (SASI) in a core collapse supernova with neutrino and gravitational wave (GW) signals. In the neutrino channel, this method performs a likelihood ratio test for the presence of SASI in the frequency domain. For gravitational wave signals we process an event with a modified constrained likelihood method. Using simulated supernova signals, the
 properties of the Hyper-Kamiokande neutrino detector, and O3 LIGO Interferometric data, we produce the two-dimensional probability density function (PDF) of the SASI activity indicator and calculate the probability of detection $P_\mathrm{D}$ as well as the false identification probability $P_\mathrm{FI}$. We discuss the probability to establish the presence of the SASI as a function of the source distance in each observational channel, as well as jointly. Compared to a single-messenger approach, the joint analysis results in $P_\mathrm{D}$ (at $P_\mathrm{FI}=0.1$) of SASI activities  that is larger by up to $\approx~40\%$ for a distance to the supernova of 5 kpc. We also discuss how accurately the frequency and duration of the SASI activity can be estimated in each channel separately. Our methodology is suitable for implementation in a realistic data analysis and a multi-messenger setting. 

\end{abstract}

\pacs{}

\maketitle

\section{Introduction}
With the recent detection of gravitational signals from  binary systems, we have entered the era of Multimessenger Astronomy  with gravitational waves (GW) \cite{LIGOScientific:2016aoc,LIGOScientific:2017vwq}. There is great hope that many more classes of sources will eventually be detected in GW in the future. One such class, with great scientific potential, is Core Collapse Supernovae (CCSNe) \cite{LIGOScientific:2019ryq,marek}. 

A CCSN is a prime multimessenger source, involving GWs, neutrinos, electromagnetic signals at several wavelengths, and possibly cosmic rays. 
In particular, neutrinos and gravitational waves play an important role, as they carry information about the early stages in the collapse as well as the causes for shock revival, while instead electromagnetic  observation mainly probe the later, post-shock breakout physics \cite{LIGOScientific:2019ryq}.

In the event of a supernova in our galaxy, a detection in the neutrino channel is guaranteed, as long as neutrino detectors at or above the kiloton scale are operational \cite{Lin:2019wwm}. The detection in GW is possible and might be achieved at ground based laser interferometers depending on the detailed GW morphology \cite{Szczepanczyk:2021bka}. The physics potential of a joint detection of \ns\ and GW from a galactic (or otherwise nearby) supernova has been explored by some pioneer works \cite{Halim:2021qll,Halim:2021yqa}, but has not been fully studied yet.

The simulated gravitational waves from CCSNe appear as stochastic processes in time domain but also  present
deterministic features in time-frequency domain. These features include the frequency evolution of the
 fundamental vibrational mode (f/g-mode) of the Proto-Neutron star (PNS)  \cite{Astone:2018uge,Kotake:2011yv,Torres-Forne:2019zwz}, as well as a deterministic imprint related to the hydrodynamic instability called Standing Accretion Shock Instability (SASI) \cite{Blondin:2005wz,Mezzacappa:2020lsn,Kuroda:2016bjd,Kotake:2006aq,Kotake_2009,Ohnishi:2005cv,Ohnishi_2008,Marek:2007gr,10.1111/j.1365-2966.2012.20333.x,Scheck:2007gw,Foglizzo:2006fu,Foglizzo:2015dma}.
The latter has a distinctive signature in neutrinos, in the form of quasi-periodic fluctuations of the neutrino luminosity, and therefore it is a natural candidate for multimessenger studies. 
 As its name indicates, SASI is a large scale, sloshing motion of the stalled shock front, which typically lasts for a fraction of a second post-collapse. 
 Depending on its amplitude and duration, it could have a critical role in promoting convection and therefore aiding the shock revival mechanism, especially in more massive progenitors that are prone to collapse into a black hole \cite{Blondin:2005wz}.

  The SASI descriptive parameters include frequency (in both GW and neutrino channels), duration, amplitude and GW polarization state.
 The average SASI frequency contains information about the average radius of the stalled shock front and coupling mechanism between the shock wave and the PNS \cite{Blondin:2005wz,Walk:2019miz}. Longer SASI duration could also appear in failed supernovae \cite{Walk:2019miz}. 
 
 It is of interest to analyze the statistical conditions needed to detect the presence of the SASI and to estimate its parameters for realistic detectors where noise is present.
 The noises, as well as the signal processing artifacts, of neutrino and GW detectors are different. However, in both the \n\ and \gw\ channels, noise can induce some energy in the SASI time-frequency regions, thus complicating the analyses. 
 Spectra properties of the SASI features in the neutrino luminosity were described in \cite{Mueller:2012sv, Lund:2010kh, Lund:2012vm} for a specific set of progenitors. The question of the detectability of SASI was discussed, not with respect to a specific algorithm, but in terms of  the 
 spectral amplitude relative to Cherenkov detector's shot noise. In \cite{Lund:2010kh,Lund:2012vm}, the shot noise was estimated by Fourier transforming a neutrino time series. The estimated shot noise became independent of frequency. Note however that when the statistical fluctuation of the neutrino signals itself dominate over the noise of the detector's background, the frequency independence  assumption may only serve as a rough approximation. The extension to a full SASI detection methodology in the \n\ channel
 was performed in \cite{Lin:2019wwm}, where some of the current authors proposed a procedure (which we named the ``SASI-meter") to detect the presence  of SASI with a desired statistical confidence, as well as obtain an estimate of the frequency for SASI candidates that pass a desired confidence threshold. In that work, we also pointed out that there is an intrinsic uncertainty in the frequency of SASI both in the GW and Neutrino channel related to the finite duration of the SASI.

 In \cite{Takeda:2021hmf}, an application of the Hilbert-Huang Transform (HHT) to a 3-D CCSN GW has been proposed for the SASI frequency and duration estimation with simulated Gaussian noise.

In \cite{PhysRevD.99.063018}
a Bayesian method that uses a training process on an existing database 
of GWs was proposed to identify the presence of SASI. Magneto rotational emission models were assumed not to contain SASI. Parameter estimation and false identification probability were not involved in that study. 
 In this paper, we further extend the theme of SASI detectability and parameter estimation in real interferometric noise. We analyse the probability that the presence of the SASI can be established, the intrinsic uncertainties of the SASI frequency and the SASI false identification probability.
 We use frequentist inference, which does not apply prior information of the SASI from any specific numerical simulation. 
 Though we use theoretical knowledge to identify conservative boundaries of the time frequency region of a GW signal where SASI contributions would be present.

 We consider a scenario where a CCSN detection has been established both in the (time-coincident) \n\ and \gw\ channels. In this framework, we focus on the estimation of physical parameters of the SASI hydrodynamic instability from the recorded neutrino luminosity and GW signatures. We extend our recent neutrino analysis \cite{Lin:2019wwm}, 
 with an estimate of the duration of the SASI and by including the GW channel both for the detection and for the estimation of the deterministic parameters. The wavelet decomposition of the GW data, recorded at a laser interferometer like LIGO \cite{Abbott_2018}, is performed using the  Coherent WaveBurst (cWB) algorithm \cite{Klimenko:2015ypf}, while the post-processing of the (simulated) data is a novel 
 element of this work. We introduce a new quantitative metric for the detection of the SASI in the GW channel, which computes the ratio of a collective Signal-to-Noise Ratio (SNR) of the wavelet components in the SASI time-frequency region versus the total SNR of the event.

In section \ref{sec:physics}, we review  the physical origins of SASI,  its deterministic parameters, and discuss an illustrative example. In section \ref{sec:method} we discuss the methodology of analyzing SASI-induced neutrino and GW modulations separately, and the results of the single messenger analyses. In section \ref{sec:multimessenger} we present a novel methodology of jointly analyzing the neutrino and GW signatures, and the results of this analysis. We also discuss the expected precision in estimating the oscillation frequencies, amplitude, starting time and duration of SASI-induced modulations on neutrino and GW signatures separately. For the frequency estimation, we present the results of the two channels separately\footnote{The relationship between the SASI frequency in the neutrino channel and in the GW channel is uncertain; this motivates our analyzing the two frequencies separately, see discussion in Sec. \ref{sec:physics}.}. 
The role of the detectors and of the distance to the CCSN is investigated. Finally, we discuss the application of this novel joint analysis to future CCSNe observations and conclude in section \ref{sec:conclusion}.

\section{Timing, frequency and duration of SASI}\label{sec:physics}

In this section, we review the origin of the signal processing features of the SASI like the timing with respect to the initial collapse, the duration, frequency content in the neutrino and GW channels, as well as some considerations on the amplitude and related GW polarization state (even if the current analysis does not make use of the GW polarization state).

The electron neutrino luminosity increases to its peak level over a $\sim 1-10$ ms time scale after the core bounce due to neutronization.  Over the same time frame, for a non-rotating progenitor, we do not expect a relatively strong GW emission, because the collapsed core and its immediate surroundings are nearly spherically symmetric
(see for example the GW signals in \cite{Szczepanczyk:2021bka}). A turbulent phase is expected to start after the shock stalls, at  $t \gtrsim 50$ ms post-bounce,  with parts of the shock collapsing under aspherical mass accretion. Such turbulence corresponds to a relatively stronger GW emission, which will be initially due to the fundamental g/f oscillatory mode of the PNS. When the accretion on the shock breaks spherical symmetry, parts of the shock are also susceptible to tangential forces that can amplify the SASI, see for example \cite{Blondin:2005wz} and references within.

As an illustrative example of our methodology,
we use the results of the 3D general relativistic (GR) simulation by Kuroda, Kotake, Hayama and Takiwaki (KKHT from here on \cite{Kuroda:2017trn}), in which SASI was found to leave an imprint both in \ns\ and \gw. 
Specifically, we use the numerically calculated neutrino event rate for Hyper-Kamiokande (\hk\ from here on)
 and the GW time series for the model S15.0 (SFHx) which is for a non-rotating progenitor having mass 15.0 $M_\odot$ and the equation of state SFHx. The KKHT model exhibits vigorous sloshing (as opposed to spiral) SASI motion. We obtained both the simulated neutrino and the GW signals for a representative observer direction, which is generic, and not special with respect to the sloshing SASI motion. The investigation of models with fast-rotating progenitors that show spiral SASI motion will be left for future works.

 \subsection{Physical origin of SASI}
 \label{subsub:physics}
 
Le us begin by reviewing basic analytical arguments on SASI.  
 There is agreement in stating that the SASI period depends on the mechanism that couples the shock wave and the surface of the PNS, as well as the total mass behind the shock wave. However more discussion is ongoing on the details as well as the best definitions for parameters like the PNS radius. Two equations about these issues are discussed in the following.
In \cite{Blondin:2005wz}, the coupling is stated to be acoustic because the advective effect is expected to operate on slower timescales.
According to \cite{Scheck:2007gw,Walk:2019miz}, SASI is due to an advective-acoustic cycle whose period
is given by the sum of the advective and acoustic time-scales
for perturbations traveling between the (angle-averaged) shock
radius $r_{sh}$ and the radius of \textit{maximum deceleration} $r_\nabla$ on the surface of the proto-neutron star:
\begin{equation}
T_{SASI}=\int_{r_\nabla}^{r_{sh}}\frac{dr}{|V_r|}+\int_{r_\nabla}^{r_{sh}}\frac{dr}{c_s-|V_r|}.
\label{eq:SASIT}
\end{equation}
Here $V_r$ is the average radial velocity of the outgoing material within the average shock radius and $c_s$ is the average speed of sound in the same region. 
Eq. (\ref{eq:SASIT}) shows that fluctuations in the shock radius will induce variations in period of the SASI. 
The size of these variations is about $20 \%$ in some numerical simulations of CCSNe with direct collapse into black holes \cite{Walk:2019miz}. The  dependence of the central frequency on the mass behind the shock has also been estimated, either by physical arguments \cite{10.1093/mnras/stz990,Szczepanczyk:2021bka,Murphy_2009,Mueller:2012sv} or by
mode analysis \cite{10.1093/mnras/stx3067,Morozova:2018glm,Torres-Forne:2018nzj,Torres-Forne:2019zwz,Sotani:2020dnh}. According to the fitting formulas derived in \cite{Torres-Forne:2019zwz}, the typical frequency of the \gw\  emission associated with
SASI activity \cite{Richardson:2021lib} is given by:
\begin{equation}
 \begin{split}
f^{SASI}_{GW}= 2\times 10^2~\mathrm{ Hz}\sqrt{\frac{m_{sh}}{r^3_{sh}}}-8.5~\mathrm{Hz} \bigg( \frac{m_{sh}}{r^3_{sh}}\bigg)~.
\label{eq:Sasi_freq2}
\end{split}
\end{equation} 
Here $m_{sh} = M_{sh}/M_\odot$, and $r_{sh} = R_{sh}/100~\mathrm{ km}$, where $M_{sh}$, and $R_{sh}$ are the total mass behind the shock, and the average shock radius, respectively. $M_{sh}$ can be approximated by the PNS mass, $M_{sh}\simeq M_{PNS}\simeq 1.4~M_\odot$, because most of the mass is confined within the central object. Before shock revival, typically  $R_{sh}\approx 150$ km. With this choice of parameters, we find $f^{SASI}_{GW}\approx 125$ Hz \cite{Richardson:2021lib}.
 The connection between $f^{SASI}_{GW}$ and the actual frequency of the hydrodynamical instability is still an open question. 
 Since each full cycle of the SASI amounts to two periods of an associated quadrupole deformation, the frequency of the GWs generated directly by the SASI, or by the resonant excitation of the PNS by the SASI, is expected to be roughly twice the
 frequency of the instability itself, which is also the expected frequency of the fluctuations in the neutrino luminosity, $f^{SASI}_\nu$. Therefore, one might expect $f^{SASI}_{GW} \sim 2 f^{SASI}_\nu$. Some authors also find that, for observing directions along the SASI axis, the SASI signature in the GW channel could have two main components, one at $f \sim f^{SASI}_\nu$ and the other at $f \sim 2 f^{SASI}_\nu$ \cite{Mueller:2012sv}. In other numerical works, a simpler connection, $f^{SASI}_{GW} \sim f^{SASI}_\nu$ emerges\cite{Kuroda:2017trn}. Motivated by the currently evolving discussion, here $f^{SASI}_{GW}$ and  $f^{SASI}_\nu$ will therefore be treated independently, without any assumption linking them.

 We note (see also \cite{Lin:2019wwm}) that the SASI frequency suffers of a  
 minimum uncertainty given by the inverse of the duration of the SASI 
 (e.g., a 100 ms SASI would have an intrinsic uncertainty of 10 Hz) because of the line broadening induced by temporal windowing. In this regard, given that the duration is the same for the neutrino and GW signatures, this fixed duration's relative impact would be smaller in the GW channel if $f^{SASI}_{GW} \sim 2 f^{SASI}_\nu$.
 Given that the GW features related to the SASI are expected to develop a specific preferential rotational axis, the polarization is expected to be mainly elliptical \cite{Jolien:2011}. Here however, the polarization signatures will not be considered in detail; their inclusion is postponed to future work.

 \subsection{Illustrative example} \label{sec:waveform}

To illustrate the methodology, we use the results of the  \kur\ model S15.0, for both the \n\ and the \gw\ channels. The simulated \gw\ waveform contains both the g-mode and the SASI. In 2D time-frequency maps, the g-mode has a slope of $\sim 3000~\mathrm{ s^{-2}}$, and it starts roughly before 100 ms post-bounce.  The SASI has duration of approximately 50 ms, and frequency $\sim$127 Hz.

\begin{figure*}[htp]
	\centering

	\includegraphics[width=0.45\textwidth]{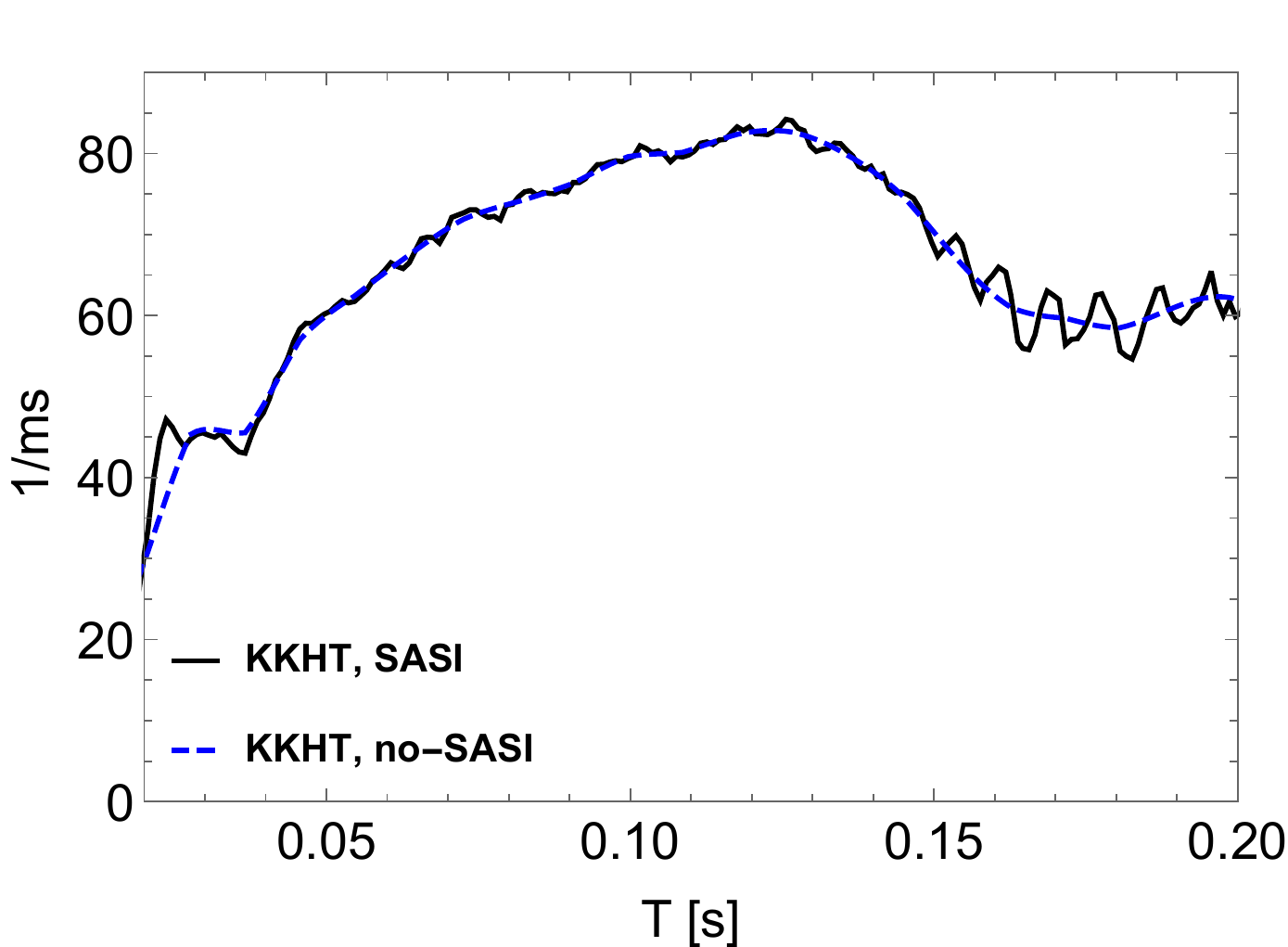}
	\includegraphics[width=0.45\textwidth]{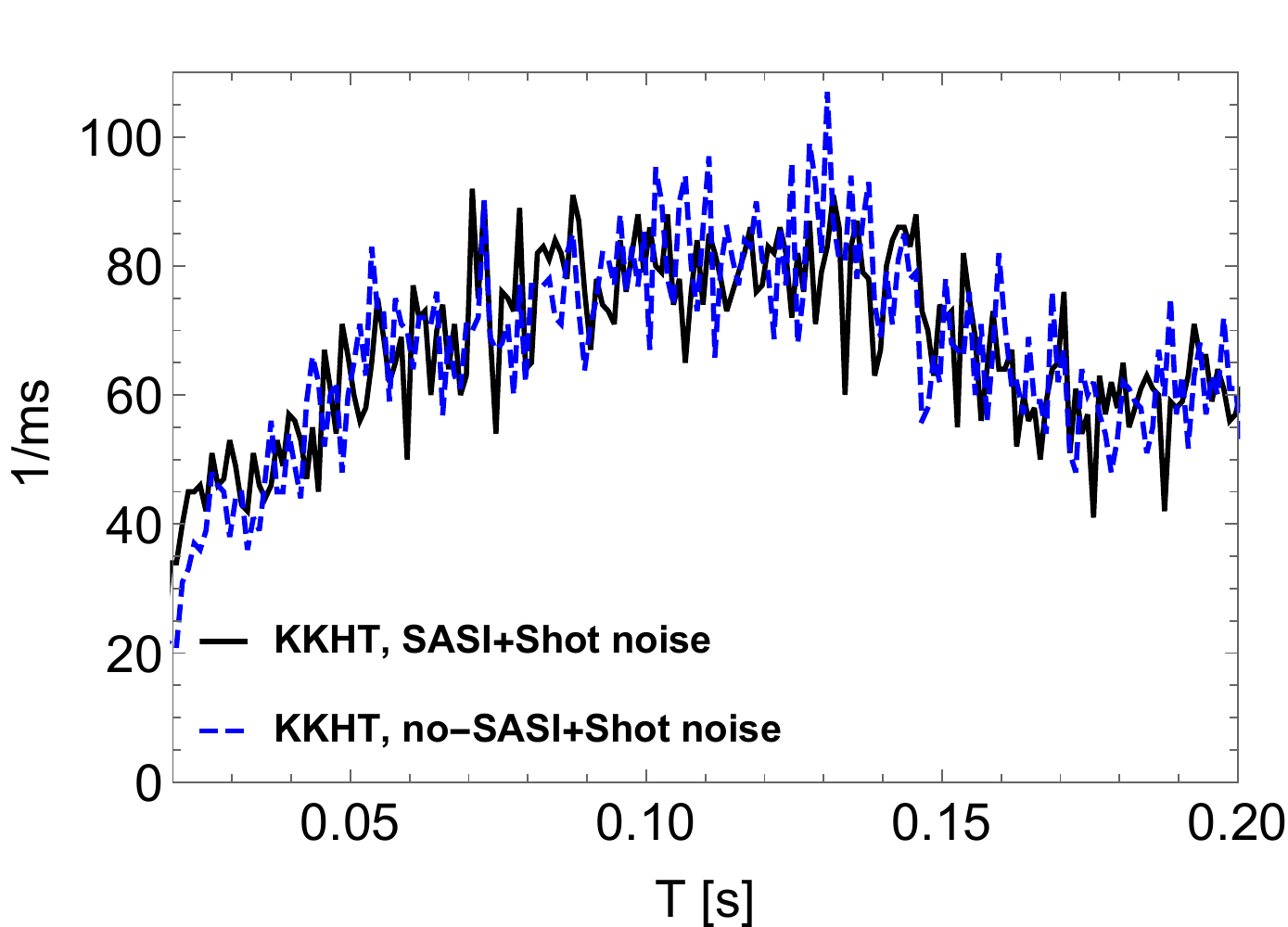}
	
	\caption{Neutrino event rates with and without SASI-induced imprints predicted by KKHT~\cite{Kuroda:2017trn}  at Hyper-K for a supernova at distance $D=10$ kpc. The curve without SASI is obtained by applying a low pass filter on the original signal. In the left (right) panel the rates are shown without (with) the realistic Poisson fluctuations  -- i.e., the neutrino shot noise -- which are driven by the signal itself.
}
	\label{fig:KurodaT}
\end{figure*}

For the \n\ event rates, the signal with the SASI removed (\nsi\ from here on) was generated in the way described in \cite{Lin:2019wwm}. The original time sequence was made smoother (thus eliminating the high frequency oscillations due to SASI) by taking the event rates averaged over eight time bins, each of width $\Delta = 1$ ms, and performing a polynomial interpolation of these averaged rates. See Fig. \ref{fig:KurodaT} for the predicted original and smoothed-out \n\ event rates. The figure also shows a realistic version of the same rates, 
where, for illustration, we include a realization of the statistical fluctuations of the number of events in each bin (neutrino shot noise). 

To illustrate the GW pipeline we prepared two sets of simulated \gw\ waveforms and corresponding \n\ event rates (numbers of events in time bins of 1 ms width) at the Hyper-K detector. One set includes the original \kur\ output containing the SASI for \ns\ and \gw. The other was produced from the original one by artificially removing the SASI imprints. We use the two waveforms to derive the Receiver Operating Curves (ROCs) to identify the SASI in GW data (see section \ref{sec:method}). The two sets are shown in Figs. \ref{fig:cross} and \ref{fig:SASI_spec}; a brief description of how they were prepared is given below. 

For GW, we resampled the original \kur\ waveform (run S15.0) to $16384~\mathrm{Hz}$ to match the standard sampling frequency of the LIGO noise and used waveform data after 57.37 ms as the earlier part has neither g-mode nor SASI components. We apply a two-fold filtering to the S15.0 waveform as follows:

\begin{enumerate}
    \item The S15.0 waveform was split into two segments based on time intervals: one ($t<156.25$ ms, S15.0-E for Early) where the SASI is not present and the other ($t>156.25$ ms, S15.0-L for Late) where it is present. This ``cut-off" was chosen particularly for this waveform (to produce the no SASI injection). The GW SASI-meter does not include any such cut-off. 
     Then, a band-stop Butterworth digital filter \cite{Oppenheim:1975} was applied to the S15.0-L segment to remove frequencies from $60$ Hz to $200$ Hz, thus removing the SASI (fS15.0-L segment, where f is for filtered).\\
    \item We re-joined the S15.0-E and fS15.0-L segments, and applied a low pass filter to the resulting waveform to address the residual discontinuity in the junction point by removing frequency components greater than 2000 Hz. The final result is the new, filtered waveform, fS15.0. 
\end{enumerate}

Fig. \ref{fig:cross} shows the two polarizations with the two models in the time domain, with their corresponding spectrograms in the time-frequency domain. 

\begin{center}
 \begin{figure}[htp]
	\includegraphics[width=0.4\textwidth]{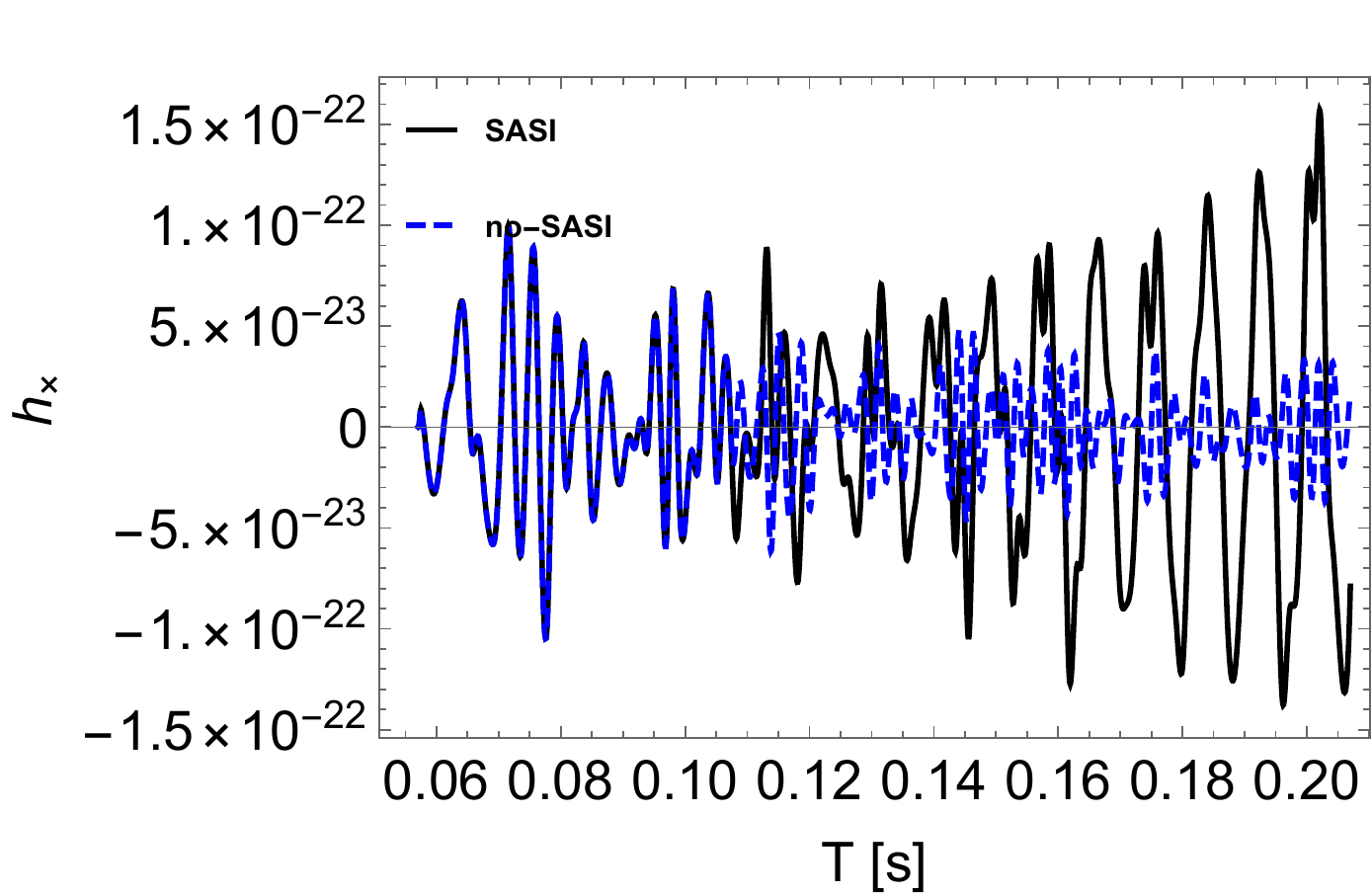}
	\includegraphics[width=0.4\textwidth]{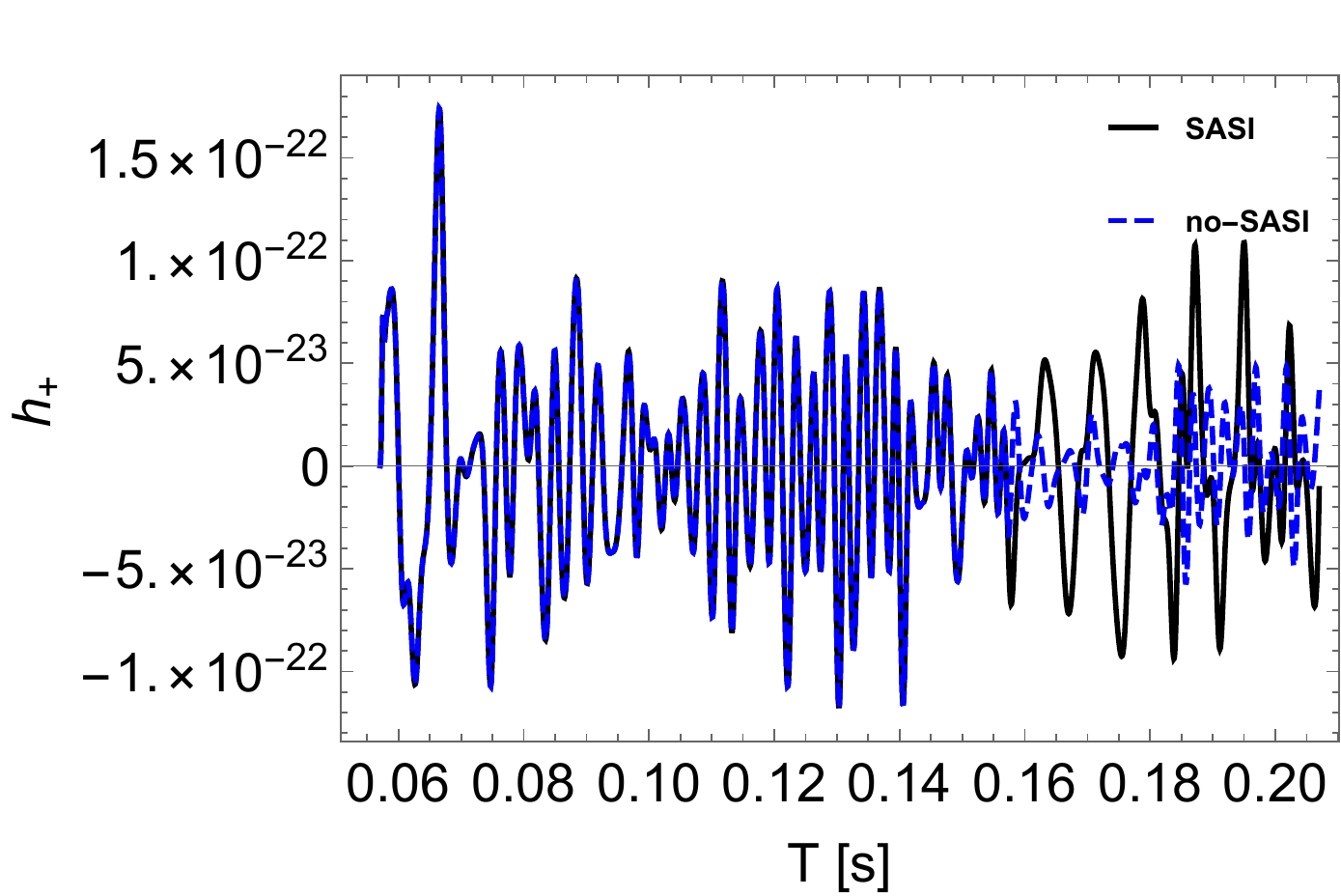}
    \caption{The cross and plus gravitational wave polarizations of the KKHT model \cite{Kuroda:2017trn} are plotted versus time in black solid. In blue dashed we plot the same quantities where a time frequency filter was applied to remove the GW SASI component according to the discussion in section \ref{sec:waveform}.}
    \label{fig:cross}
\end{figure}

   \begin{figure}[htp]
	\centerline{\includegraphics[width=0.4\textwidth]{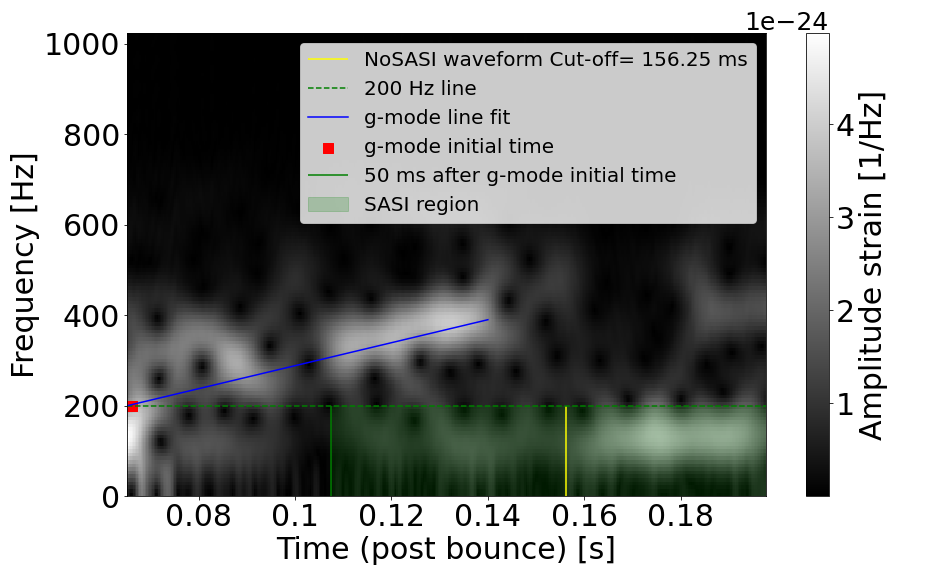}}
	\centerline{\includegraphics[width=0.4\textwidth]{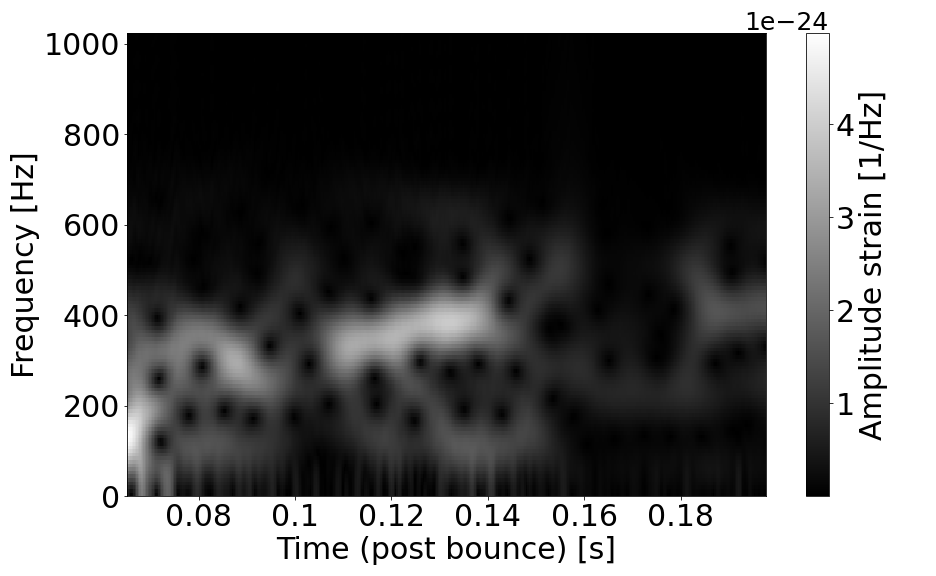}}
    \caption{S15 model Spectrogram for the plus polarization, with SASI (top) and without SASI (bottom). In the top spectrogram, the blue line indicates where the g-mode is located (with the red dot indicating  the time the g-mode is around 200 Hz). In green is the allowed time-frequency SASI region (to start after 50 ms of the g-mode initial time) and below the 200 Hz line. In yellow we indicate the temporal interval where the actual SASI oscillations (see the location of the SASI in the left plot of Fig. \ref{fig:KurodaT} and bottom plot of Fig. \ref{fig:cross}) are removed, after the yellow vertical line at 156.25 ms}.
    \label{fig:SASI_spec}
\end{figure}

\end{center}

 Although the methodology of this paper is illustrated for the results of a specific numerical simulation, we designed it to be applicable to the range of 3-D simulations existing in the literature, which is summarized below for the interested reader.

In \cite{Mezzacappa:2020lsn},  results from a 3D simulation of GW emission for a $15M_\odot$ star are presented. From the time-frequency plots, we can see that after the quiescent phase, which lasts until $t\approx 100 ~\mathrm{ms}$ post-bounce, a low frequency signal, with frequencies below $\approx 200~\mathrm{Hz}$ begins, after $\approx 50~ \mathrm{ms}$ of the g-mode component corresponding to $200~ \mathrm{Hz}$, which persists through the remainder of the simulation (until $\approx 425 ~\mathrm{ms}$ post-bounce), which is due to aspherical mass motions in the gain layer from neutrino-driven convection and the SASI. Also, in the interval $\sim 150-200 ~\mathrm{ms}$ post-bounce, an intermediate frequency emission ($\approx 400-600 ~\mathrm{Hz}$) joins the low frequency emission which is described as SASI-induced aspherical accretion. 

In \cite{Kuroda:2017trn}, 3D GR with $\nu$ transport simulation has been conducted for three non-rotating progenitors of 11.2, 15 and 40 $M_\odot$. Prompt convection can be seen in both low and high frequency regions in the early stage until $t \approx 50~\mathrm{ms}$ post-bounce of the simulation, with $S11.2$ showing the strongest prompt convection. In addition to the PNS g-mode, which is a relatively narrow-banded spectrum that can be seen for all models, in $S15.0(SFHx)$ a SASI-induced low frequency component is seen in $100 < f < 150 ~\mathrm{Hz}$ band for $t>150 ~\mathrm{ms}$ post-bounce, which is $\approx 75 ~\mathrm{ms}$ after the 
g-mode component.

In \cite{Andresen:2016pdt}, four models are discussed: s11.2, s20, s20s and s27. In s11.2, no growth of SASI is observed because of the large shock radius. The other models are SASI dominated. In s20, strong SASI activity (dominated by the spiral mode) is seen from 120-280 ms after core bounce which is the extended phase with peak from 200-250 ms. After a period of transient shock expansion, SASI (much weaker) continues. In s20s, prior to shock revival, post-shock flow is dominated by large-scale SASI sloshing motions between 120-280 ms post bounce. In s27, two episodes of pronounced SASI activity can be seen interrupted by a phase of transient shock expansion following infall of the Si/O interface. The first phase is from 120-260 ms post-bounce and the second phase from 410 ms post bounce to the end of the simulation (575 ms post bounce). Also, a low-frequency emission between 280-350 ms post bounce can be seen which is not associated with SASI. 

SASI produces a sizable l=1 mode in the range 50-100 Hz, and l=2 components in 100-200 Hz range. From the literature, we see that noticeable SASI components begin after a delay of $\approx$ 50 ms from the g-mode component corresponding to 200 Hz.

\section{Single messenger analysis: method}
\label{sec:method}

\begin{figure*}[htp]
	\includegraphics[width=0.8\textwidth]{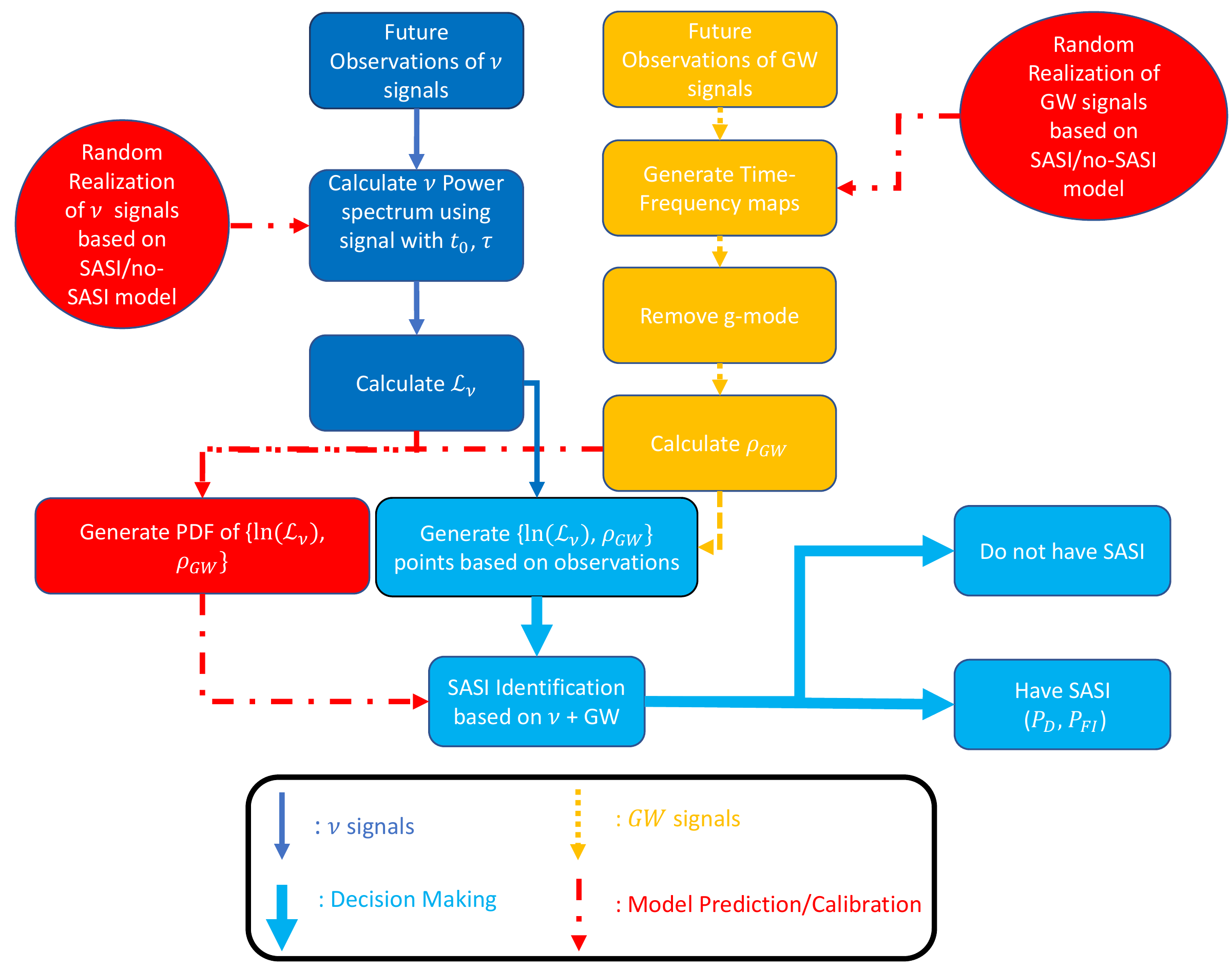}
	 \caption{Flowchart of the multimessenger SASI meter developed in this work. 
	 Due to the absence of current detections,  we characterize the pipeline with random realizations of reconstructed $\nu$ and GW signals for the test example. For the $\nu$, this means adding Poissonian fluctuations on 
	 signals that have SASI as well as signals where the SASI was removed. In the case of future detections, we can achieve the same result by taking a smoothed out version of the detected neutrino luminosities as the no SASI $\nu$ signature, and randomize it with Poissonian fluctuations to identify the threshold for the desired $P_\mathrm{FI}$ used 
	 as a reference in this work for single channel or multiple channel identification mode ($P_\mathrm{FI}$=0.1). For the GW channel,
	 in the test example used in this paper, we inject the GWs with and without SASI in real interferometric noise.
	 In a realistic scenario, the no-SASI injections can be used to tune the threshold on the identification metric 
	 for the desired single or multimessenger $P_\mathrm{FI}$ ($P_\mathrm{FI}$=0.1). The signals in the $\nu$ and GW channels with SASI are used
	 in this paper to characterize the performance of the GW-$\nu$ SASI meter.}
	\label{fig:flowchart}
\end{figure*}

We now illustrate the statistical method of analysis for each messenger separately. The flowchart in Fig. \ref{fig:flowchart} provides a compact summary of the single messenger analysis procedure, and its relationship with the multi-messenger analysis, which will be discussed in section \ref{sec:multimessenger}.

We assume that the noises in the neutrino and GW channels are statistically independent. 
The main reason is
that for the neutrino channel the noise is partly driven by the neutrino luminosity itself in the search for oscillatory signals in detected neutrino events, while in laser interferometers the noise is instrumental in origin.

 The signals in the \gw\ and \n\ channels have a different dependence on the distance to the CCSN, $D$. The 
  amplitude of the GW signal at an interferometer scales like $D^{-1}$, whereas the event rate in a \n\ detector is proportional to $D^{-2}$. This means that the dependence on $D$ of  the SASI-meter, as well as the SASI parameter estimation, could be different in the neutrino and GW channels, resulting in a non-trivial dependence of the combination of the two channels on the distance, which needs to be properly investigated, see sections \ref{sec:multimessenger}.

\subsection{Neutrino-only analysis 
}\label{subsec:nuonly}

In the present context, a \n\ detector like Hyper-K operates as a counting device, so the main observable is the number $N$ of neutrinos detected (``events") in the detector's volume in a given time bin. $N$ is a function of time, and it is affected by the physical fluctuations of the incoming  \n\ flux (some due to SASI) as well as statistical (Poissonian) fluctuations. The latter scale as $\sqrt{N}$, so their relative amplitude increases with the distance: $\sqrt{N}/N\propto D$.

In the following we discuss our likelihood ratio methodology for identifying SASI-induced neutrino signals in detail. This methodology is inspired by the Neyman Pearson detection method of a signal in Gaussian noise, and provides the probability to detect the SASI as a function of the corresponding false alarm probability. The method is presented extensively in \cite{Lin:2019wwm}; for completeness, here we briefly summarized its main points.

\subsubsection{Computing Receiver Operating Curves}\label{subsubsec:nuonlyMethod}

First, we establish two parametric templates in the time domain which characterize the main features of neutrino signals with and without the SASI activity respectively. For the case with SASI activity
we choose a single frequency function:
\begin{equation}
N_{\mathrm{S}}(t)=(A-n)(1+a \sin(2\pi  f_\mathrm{S} t))+n~, 
\label{eq:mod2}
\end{equation}
where $N_\mathrm{S}$ is the number of neutrino events collected in a unit time bin centered at $t$, $A$ is the time-averaged event rate (the ``DC component") in the detector including instrumental noise (after possible experimental cuts), $a$ is the relative SASI amplitude, $n$ is the mean value of the background events ($n\simeq 0$ for Hyper-K),  and $f_S$ is the nominal frequency of the SASI. 
The second template,  for the case without SASI, is a constant: 
\begin{equation}
N_\mathrm{nS}(t)=A~,
\label{eq:mod0}
\end{equation}
(with $A$ having the same meaning as in Eq. (\ref{eq:mod2})). 
In the above templates,  $f_S$ and $a$ are treated as free parameters, with respect to which the likelihood will be maximized. We note that there are two implicit variables in Eq. (\ref{eq:mod2}) and Eq. (\ref{eq:mod0}): the starting time $t_0$ and the duration $\tau$ of the neutrino time series of interest (which is typically a subset of the entire \n\ burst, where SASI is likely to be found). We consider these as fixed for the time being, as done in \cite{Lin:2019wwm}; varying them is discussed later in Appendix \ref{appendix:neutrino}. 
The quantity $A$ is treated as fixed as well, because it can be determined accurately by measuring the total number of neutrino events collected in a time series with $t_0$ and $\tau$ \cite{Lin:2019wwm}.

We consider the neutrino events that are recorded in a detector after an initial time $t_0$, in time bins of width $\Delta=1~{\rm ms}$. The  $j$-th time bin then corresponds to the time $t_j=t_0+j\Delta$. 
The observed number of events in the same bin will then be $N(t_j)$. The spirit of the method consists in establishing how well the time series $\{ N(t_j) \}$ matches the templates in Eqs. (\ref{eq:mod2}) and (\ref{eq:mod0}). 
Considering the oscillatory character of SASI, the matching is done in the frequency space. 

Following \cite{Lund:2010kh, Lund:2012vm}, we perform a discrete Fourier transform of the series $\{ N(t_j) \}$ over the time interval $[t_0,t_0+ \tau ]$, containing $N_{bins}=\tau/\Delta$ time bins. The discrete frequency resolution is then
\begin{equation}
	\delta =\frac{1}{\tau}~,
	\label{eq:tau}
\end{equation}
which represents the minimum width of frequency bins for which statistical independence between adjacent bins can be realized.
The Nyquist frequency is
\begin{equation}
	f_{\textit{Nyq}}=\frac{1}{2\Delta},
	\label{eq:fnyq}
\end{equation}
which corresponds to the frequency index
\begin{equation}
k_{\textit{Nyq}}=\frac{f_{\textit{Nyq}}}{\delta }=\frac{\tau}{2 \Delta}=\frac{1}{2}N_{bins}~.
\label{eq:knyq}
\end{equation}
 We define the discrete Fourier-transformed neutrino signal as:
\begin{equation}
 h(k\delta )=\sum_{j=0}^{N_{bins}-1}{N}(t_j)e^{i2\pi j \Delta k\delta }~,
\label{eq:h}
\end{equation}
and the one-sided power spectrum as:
\begin{equation}
P(k\delta )=\begin{cases}
2| h(k\delta )|^2/N_{bins}^2~~~ \text{for } 0<k\delta <f_{Nyq}~,\\
\\
| h(k\delta )|^2/N_{bins}^2 ~~~\text{for } k\delta =0. ~
\end{cases} 
\label{eq:power1}
\end{equation}

The factor of $1/N^2_{bins}$ is included in order to fix the normalization, so that at $k=0$ we have $ P(0)=( N_{ev}/N_{bins})^2$ (here $ N_{ev}= \sum^{N_{bins}-1}_{j=0} {N}(t_j)$). 

For a given quantity, a symbol
with tilde will be used when referring to an actual outcome of a measurement, which is affected by statistical fluctuations. The probability that an observed power at a specific frequency $k\delta$, ${\tilde P}(k\delta)$ is a realization of a certain hypotherical template with parameters set $\Omega$ is \cite{Lin:2019wwm}  :


\begin{equation}
\begin{split}
Prob(\tilde{P}, \Omega)&=\frac{N_{bins}^2}{4\sigma^2} \exp{ \left[ -\frac{N_{bins}^2}{4\sigma^2} \left(\tilde{P} + P \right)\right]}\\
&\times I_0\left( \frac{N_{bins}^2}{2\sigma^2} \sqrt{ \tilde{P}P} \right)~,
\label{eq:prob}
\end{split}
\end{equation}
where $I_0$ is the modified Bessel function of the first kind, $P$ is the power calculated by the template $\Omega$ at frequency $k\delta$, and 
\begin{equation}
\begin{split}
\sigma^2=\frac{N_{ev}}{2}~.
\end{split}
\label{eq:sigma2}
\end{equation}

Let us now define the likelihood that a given observed power series vector, $\mathcal{ \tilde{P}}= \{   \tilde{P}(k\delta) \}$ is a realization of a certain hypothetical template $\Omega$.
It is defined as:  
\begin{equation}
L(\mathcal{ \tilde{P}},\Omega)=\prod_{k=K_{min}}^{K_{max}}Prob({ \tilde{P}}(k\delta), \Omega)~,
\label{eq:likeli}
\end{equation}
where $K_{min}$ and $K_{max}$ represents the minimum and maximum frequencies, with typical values being $K_{min}\delta\approx 50$ and $K_{max}\delta\approx 150$ Hz, that covers the SASI frequencies predicted by state-of-art simulations\cite{Mezzacappa:2020lsn,Kuroda:2017trn,Mueller:2012sv}.

Given two hypotheses (i.e., two templates) with parameter sets $\Omega_\mathrm{S}$ and $\Omega_{\mathrm{nS}}$, and observed set $\mathcal {\tilde{P}}$, the likelihood ratio is:
\begin{equation}
\mathcal{L}(\mathcal{\tilde P})=\frac{Max_{\Omega_S}[L(\mathcal{\tilde P},\Omega_S)]}{Max_{\Omega_{nS}}[L(\mathcal{\tilde P},\Omega_{nS})]}~. 
\label{eq:likeR}
\end{equation}
In the numerator (denominator), the template corresponding to SASI(no-SASI) is used and $Max_{\Omega_{S(nS)}}[L(\mathcal{\tilde P},\Omega_{S(nS)})]$ is the maximized likelihood with extremal parameters ${\Omega}_\mathrm{S}$ (${\Omega}_{nS}$) with respect to the observed power spectrum $\mathcal{\tilde{P}}$.  
In this work, the templates in Eqs. (\ref{eq:mod2}) and  (\ref{eq:mod0}) will be used as representative of the SASI  and no-SASI cases.
Their parameters are $\Omega_{S}=\{a, f_S \}$ and $\Omega_{nS}=\{ Null\}$  respectively. 

It is intuitive to see how the likelihood ratio in Eq. (\ref{eq:likeR}) is sensitive to SASI. Since our templates $N_{S}$ (Eq. (\ref{eq:mod2})) and $N_{nS}$ (Eq. (\ref{eq:mod0})) capture well the main features of the neutrino event rates of the models with and without SASI respectively, as the SASI features in the data become more pronounced, the numerator Eq. (\ref{eq:likeR}) is likely to increase, while at the same time the denominator is likely to decrease (poorer fit for the $N_{nS}$ template), so $\mathcal{L}$ is likely to increase. Vice-versa,  $\mathcal{L}$ will take lower values if the SASI signatures in the data become weaker. 
Therefore, Eq. (\ref{eq:likeR}) serves as our \emph{``SASI-meter"} to identify the presence of SASI.  

\begin{figure*}
    \centering
    \includegraphics[width=0.3\textwidth]{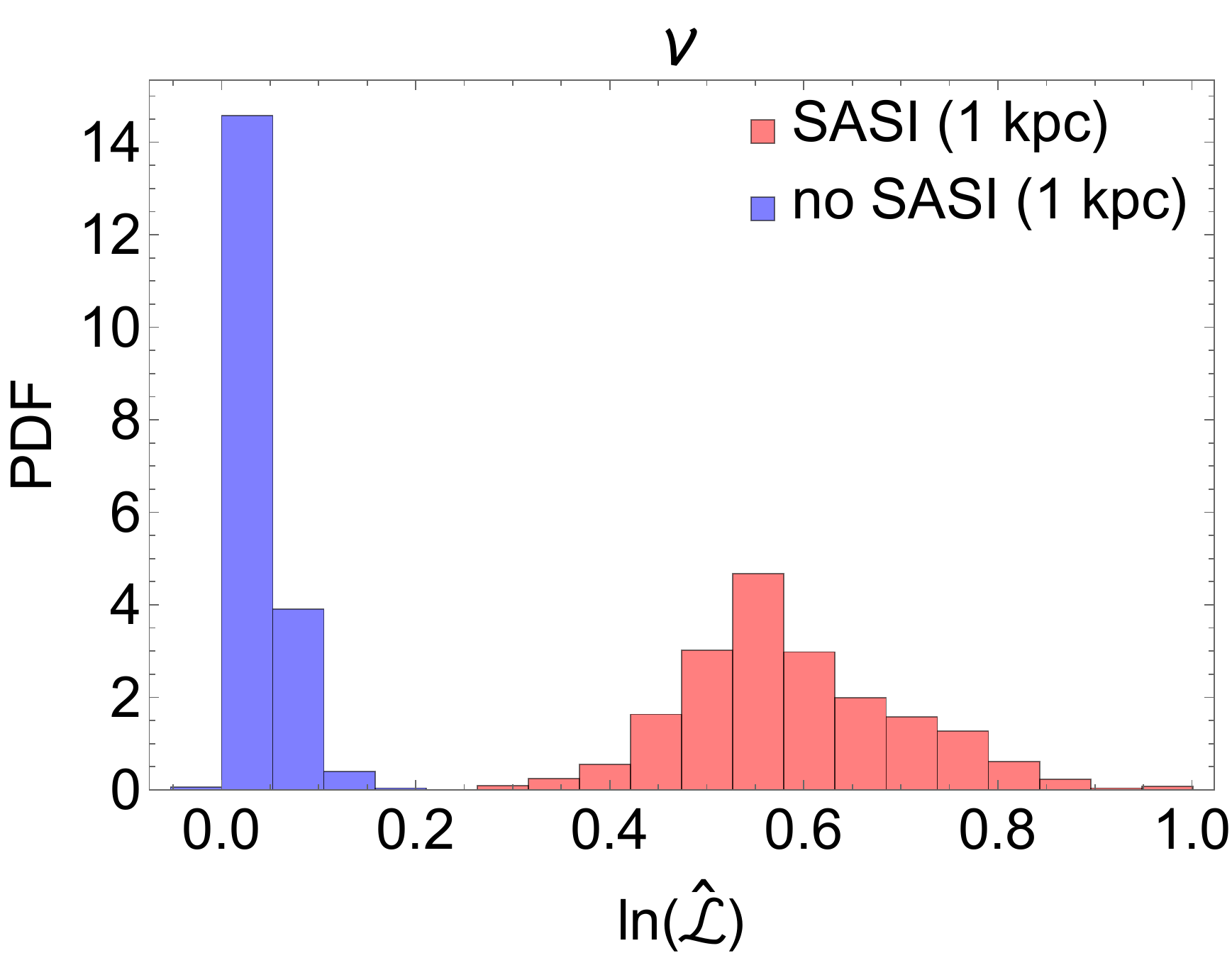}
    \includegraphics[width=0.3\textwidth]{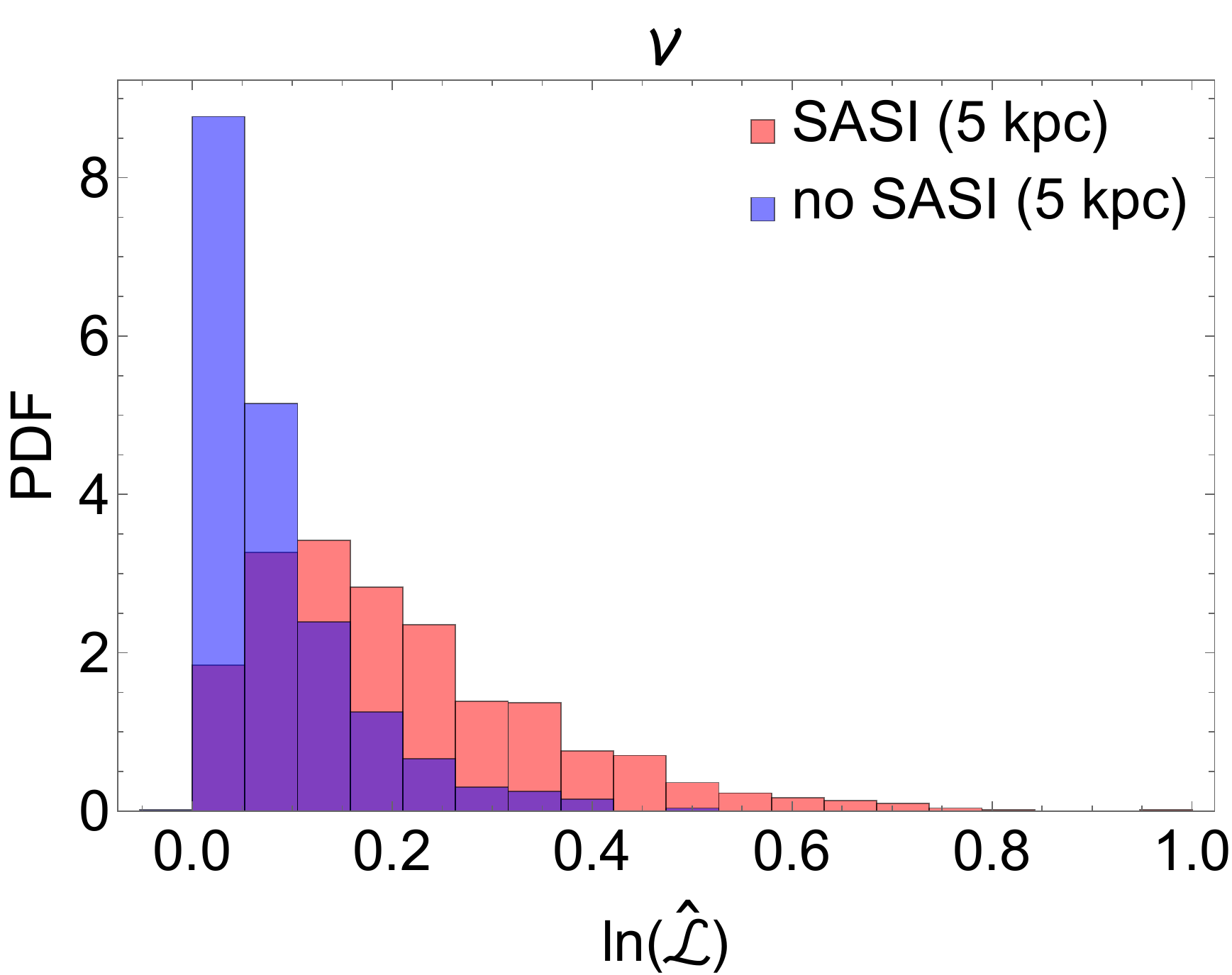}
    \includegraphics[width=0.3\textwidth]{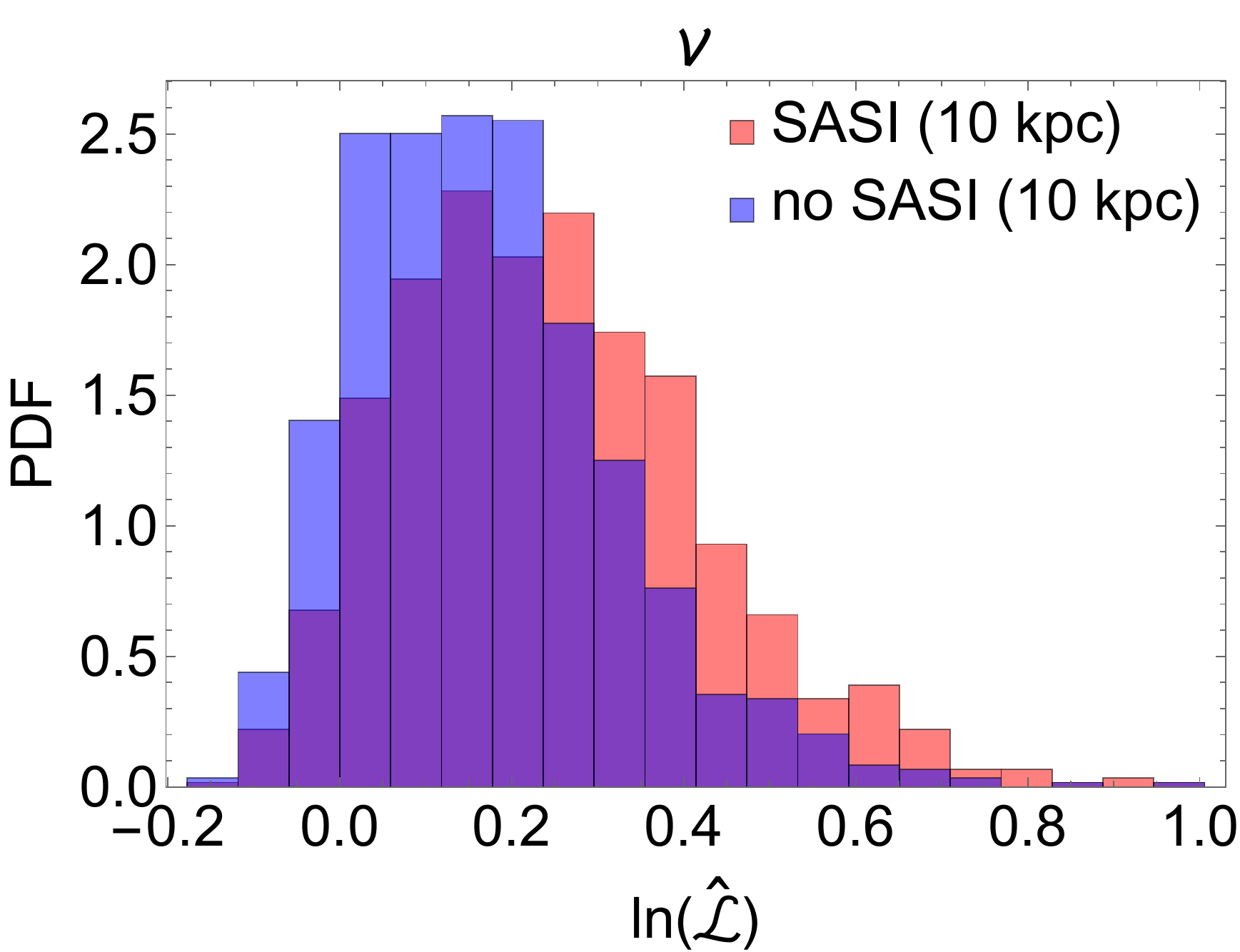}

    \includegraphics[width=0.3\textwidth]{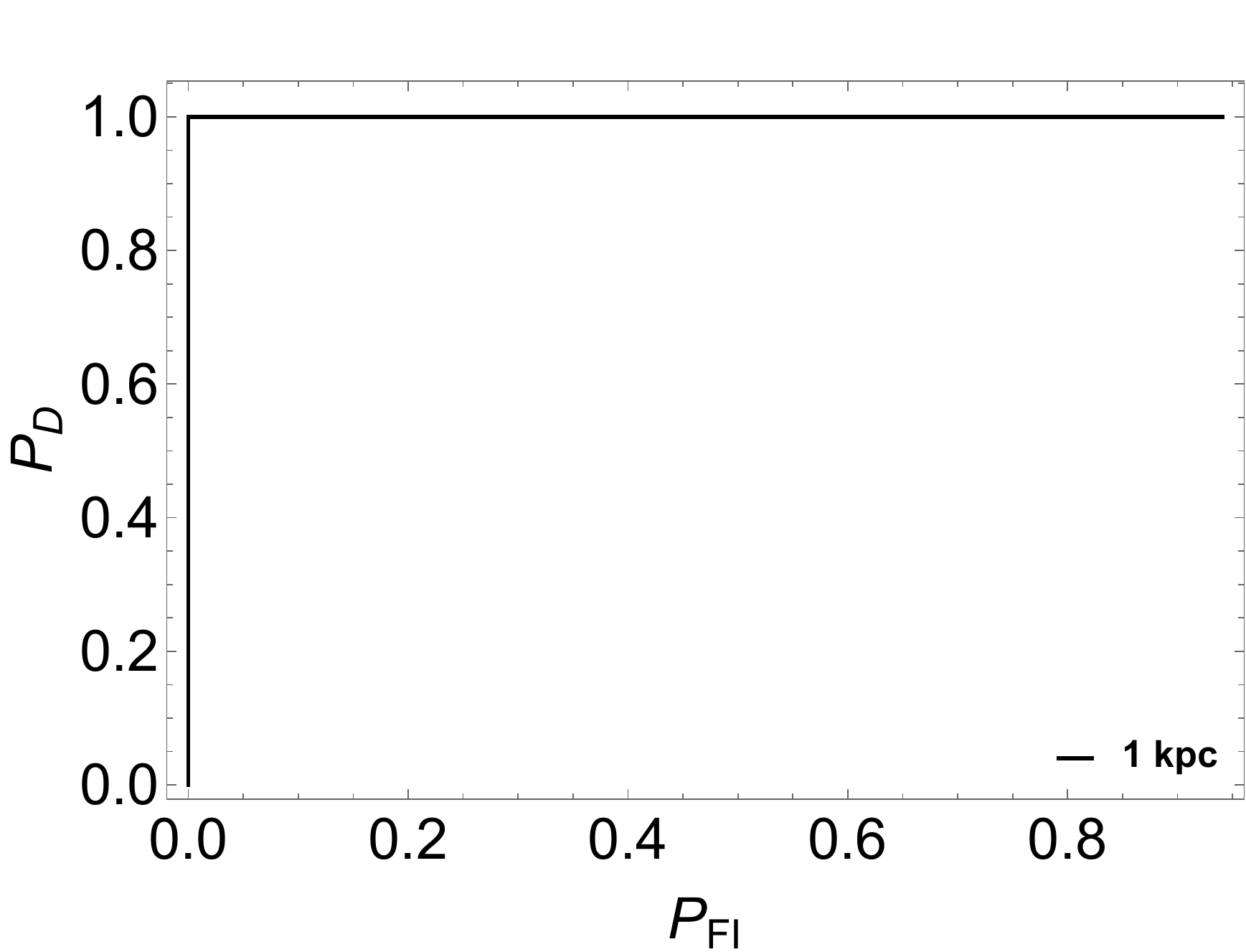}
    \includegraphics[width=0.3\textwidth]{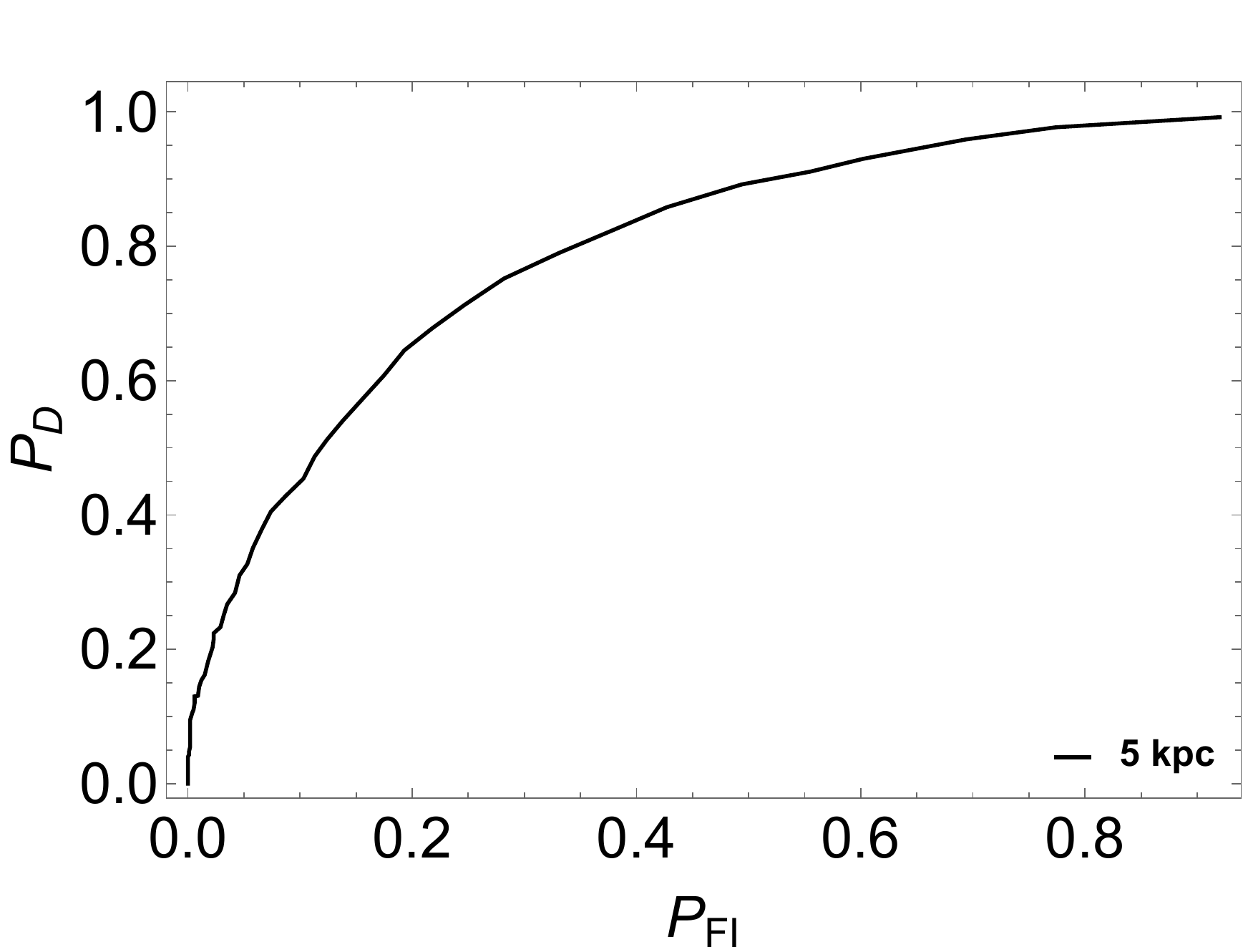}
    \includegraphics[width=0.3\textwidth]{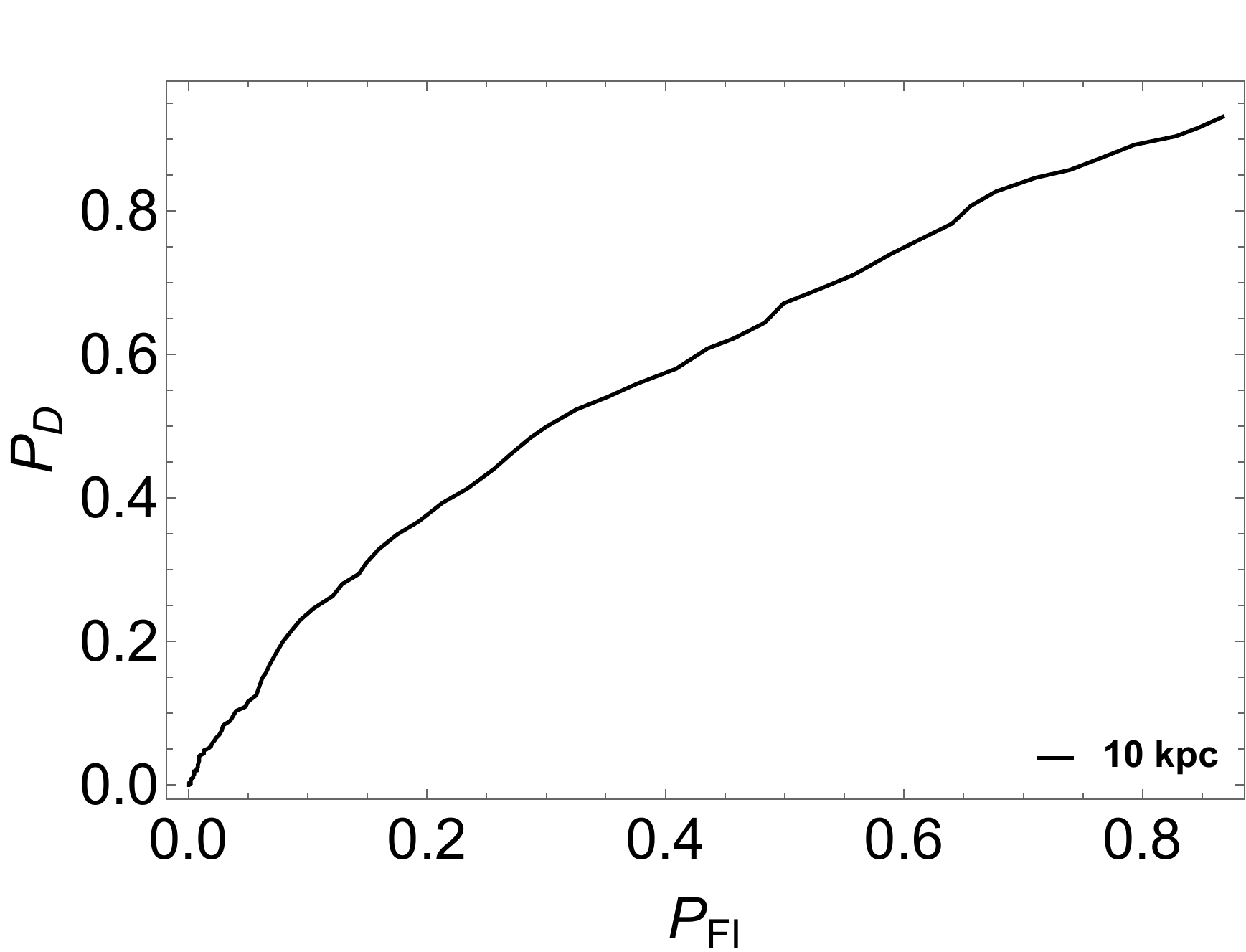}
    \caption{\emph{Top row:} examples of distributions of the test-statistics $ln(\hat{{\mathcal L}})$ obtained from Monte-Carlo generated neutrino data at Hyper-K with the starting time $t_0=150~\mathrm{ms}$ and duration $\tau=50~\mathrm{ms}$. The blue histograms correspond to the one from a simulation where the SASI component has been previously filtered out and the red histograms correspond to the one from a simulation where the SASI component has been kept. Here, $ln(\hat{\mathcal{L}})$ is the rescaled logarithmic likelihood ratio with its maximum in SASI case being 1. The three colums correspond to distances $D=1,5,10$ kpc to the \sn. \emph{Bottom row:} the corresponding Receiver Operating Curves describing the identification probability versus the false identification probability.}
    \label{fig:NuOnlyResults}
\end{figure*}

\begin{figure}[htp]
	\includegraphics[width=0.45\textwidth]{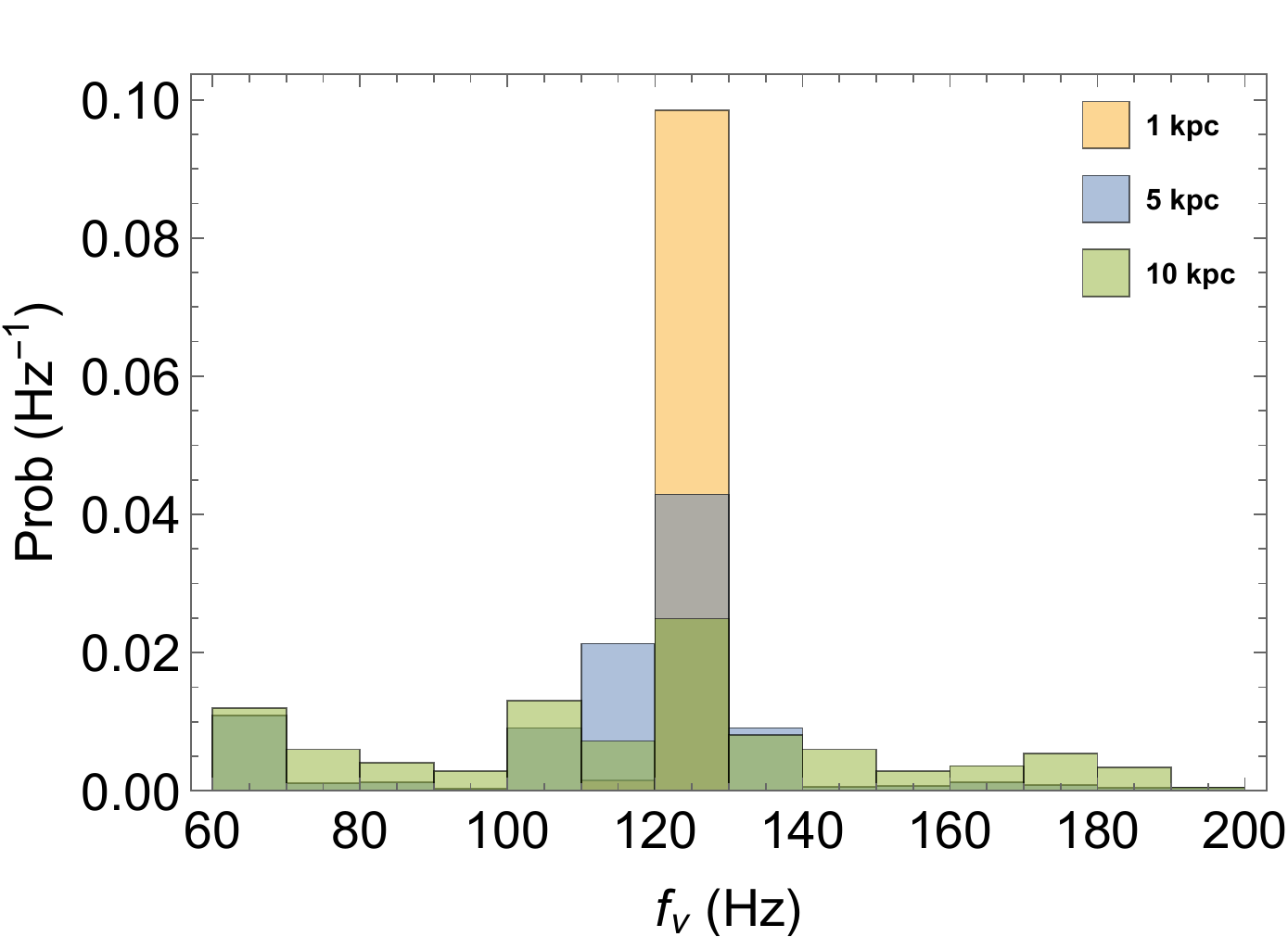}
	 \caption{Histograms of the estimated SASI frequency from neutrino signatures. As expected the variance decreases for closer distances since the amplitude of the poissonian fluctuations decreases with closer distances while the amplitude of the SASI fluctuations with respect to the DC component is distance independent.}
	\label{fig:fdistributionNu}
\end{figure}

To assess the robustness of the \si-meter method, 
we find the probability density distributions of $\mathcal{L}$ (or, equivalently, $\ln{\mathcal L}$) under the two hypotheses as testbed.  This was done by simulating (using a Monte Carlo method) $10^3$ sets of $  \mathcal{ P} $ with shot noise based on the \kur\ models with and without \si.
 
We then obtain the probability density distribution of $\mathcal{L}$ in the two scenarios,  $Prob_\mathrm{\nu,S}(ln(\mathcal{L}))$
and $Prob_\mathrm{\nu,nS}(ln(\mathcal{L}))$. A useful way to describe these two distributions, and compare them with one another, is to examine the probabilities that -- under the two hypotheses --   the likelihood ratio exceeds a certain threshold value, $\Lambda_\nu$: 

\begin{eqnarray}\label{eq:pdpfanu}
& P_\mathrm{D}^{\nu}=\int_{\Lambda_\nu}^\infty d ln(\mathcal{L})Prob_\mathrm{\nu,S}(ln(\mathcal{L})),\label{eq:pd}~\\
& P_\mathrm{FI}^{\nu}=\int_{\Lambda_\nu}^\infty d ln(\mathcal{L})Prob_\mathrm{\nu,nS}(ln(\mathcal{L}))~. \label{eq:pf}
\end{eqnarray}

$\Lambda_\nu$ represents a value of the likelihood ratio above which the SASI hypothesis is accepted as true (``detection"). Therefore, $P_\mathrm{D}^{\nu}$ takes the meaning of  SASI \emph{detection probability}, because it represents the probability that the method accepts the \si\ hypothesis as true  when the SASI is in fact true. $P_\mathrm{FI}^{\nu}$ then represents the \emph{false identification probability}, i.e., the probability that the SASI hypothesis is accepted when in fact the \nsi\ hypothesis is the true one. The curve of the points $(P_\mathrm{D}^{\nu},P_\mathrm{FI}^{\nu})$ for varying $\Lambda_\nu$ is the Receiver Operating Curve (ROC), which allows us to evaluate the effectiveness of the method at a glance.

In addition to assessing the detectability of SASI, our SASI-meter can also be used for parameter estimation. Indeed, in every realization of the Monte Carlo simulations based on the KKHT model, the extremal parameters $\tilde{\Omega}_{SASI}$ are found when searching for the maximized $\tilde{L}(\mathcal{P},\tilde{\Omega}_{SASI})$. Thus, the probability distribution of $\tilde{\Omega}_{SASI}$ is sampled by our Monte Carlo simulations. From the Monte Carlo simulations, one can find the uncertainty on the parameters $f_S$ and $a$ due to  the statistical fluctuations of neutrino events in the detector. These uncertainties increase as the CCSNe distance to the Earth increases, as a result of the decreased number of \n\ events \cite{Lin:2019wwm}.

\subsubsection{Results: neutrino Receiver Operating Curves}

Our results for the \n-only analysis are shown in Figs. \ref{fig:NuOnlyResults} and \ref{fig:fdistributionNu}. 
In particular, Fig. \ref{fig:NuOnlyResults} shows the statistical distributions of ${ln(\hat{\mathcal {L}})}$ for the SASI and no-SASI cases,  for different values of the distance $D$. The probability density function (PDF) of ${ln(\hat{\mathcal {L}})}$ with its maximum in SASI case being one is obtained by rescaling the distribution of ${ln(\mathcal {L}})$. The corresponding ROCs are shown as well. 
We notice the expected trends (see section \ref{subsubsec:nuonlyMethod}), namely the two distributions having increasingly large overlap as $D$ increases, which results in worsening (i.e., approaching the line $P^\nu_D = P^\nu_{FI}$) of the ROCs. 
For example,  at 5 kpc $P^\nu_D\approx0.5$ (for $P^\nu_{FI}=0.1$). At 10 kpc, $P_\mathrm{D}\approx0.2$, indicating that it would be difficult to identify SASI in HyperK. The methodology introduced here is model independent, namely that the the identification thresholds
for a $P^\nu_\mathrm{FI}$=0.1 are determined automatically from the smoothed detected luminosity. However the performance will be model dependent from the amplitude of the neutrino luminosity fluctuations and the mean neutrino luminosity.

We then discuss SASI parameter estimation using the neutrino SASI-meter. The estimation of parameters characterizing the SASI starting time and duration is discussed in Appendix \ref{appendix:neutrino}. We estimate the SASI frequency (oscillation amplitude) in KKHT model by calculating the mean and the uncertainties of the extremal frequency (extremal oscillation amplitude) in the time interval where the \emph{monochromatic} feature is observed (see Appendix \ref{appendix:neutrino} for a more detailed discussion of the \emph{monochromatic} feature). We plotted the probability density distribution of the estimated SASI frequency at various CCSNe distances in Fig. \ref{fig:fdistributionNu}. We further summarized the estimated SASI frequency and oscillation amplitude of the KKHT model (with uncertainties) at various CCSNe distance in Tab. \ref{tab:sasiIcePfa10}.   
 To conclude, the extremal frequencies in the monochromatic region indicated the SASI frequency (oscillation amplitude),  $f_\nu\approx120~\mathrm{Hz}$ ($a_\nu\approx0.05$) when the neutrino signals are simulated using the KKHT model.

\begin{table*}
	\caption{\label{tab:sasiIcePfa10} 
Estimated mean (median for GW SASI duration) and standard deviation of the SASI parameters in neutrino and GW data analysis and g-mode slope in GW data analysis. The SASI frequency $f_\nu$ and amplitude $a_\nu$ in neutrino analysis are estimated using neutrino events with starting time $t_0=150 ~\mathrm{ms}$ and duration $\tau=50 ~\mathrm{ms}$, where the $P^\nu_\mathrm{D}$ is maximized at different CCSNe distances. The estimated SASI starting time $t_0^\nu$ ($t_0^\mathrm{GW}$) and duration $\tau^\nu$($\tau^\mathrm{GW}$) in neutrino (GW) analysis are also provided. See the appendix \ref{appendix:neutrino} for detailed discussion of determining $t_0^\nu$ and $\tau^\nu$ using neutrino signals. See section \ref{sec:method} for detailed discussion of determining $\tau^\mathrm{GW}$ using GW signals. The reasons for the non-monotonic behaviour of the duration estimation with the distance are discussed in section \ref{sec:conclusion}.
}  
	\begin{ruledtabular}
		\begin{tabular}{ccccc}
			G-mode slope &SASI&$10$ kpc&$5 $ kpc &$1$ kpc \\
			\hline
			 &&&&\\
			 &$f_\nu$(Hz)& 113.38& 111.03 &119.85\\
			 &$\delta f_\nu$(Hz)&32.9& 22.6  &1.22 \\
			 &$a_\nu$& 0.063& 0.047 &0.044\\
			 &$\delta a_\nu$&0.022& 0.013 & 0.005  \\
			 &$f_{GW}$(Hz)& 120.08& 120.42 &122.36\\
			 &$\delta f_{GW}$(Hz)&18.65& 13.80  &5.48 \\
			 &$t_0^\nu$(ms)& N/A (Due to large $\delta f_\nu$) & $>150$& $>150$\\
			 &$\tau^\nu$(ms)&N/A (Due to large $\delta f_\nu$) &$>50$ &$>50$\\
			 &$\tau^\mathrm{GW}$(ms)& 259& 494&166\\
			 &$\delta \tau^\mathrm{GW}$(ms)& 347& 552&261\\
			 
			$m_{opt}^{GW}(s^{-2})$& & 2564.84&2645.02&3190.68\\
			$\delta m_{opt}^{GW}(s^{-2})$& & 1301.08&1132.72&929.62\\
		\end{tabular}
	\end{ruledtabular}
\end{table*}

\subsection{Gravitational waves-only analysis} 
\label{subsec:GWonly}
In this section, we discuss the GW SASI identification ROC and estimate some of its parameters. We assume the existence of a detected GW event from the CCSN of interest in coincidence with a neutrino  detection. 

We restrict the analysis to the frequency  range $16 \leq f \leq 2000 ~\mathrm{Hz}$, and assume SASI starting times later than $50$ ms after the beginning of the g-mode related turbulence in GW analysis, based on the literature review performed at the end of section \ref{sec:waveform}. 
   
\subsubsection{Receiver operating curves and parameter estimation}\label{sec:GW}
\begin{figure*}
\centering
  \includegraphics[width=0.4\textwidth]{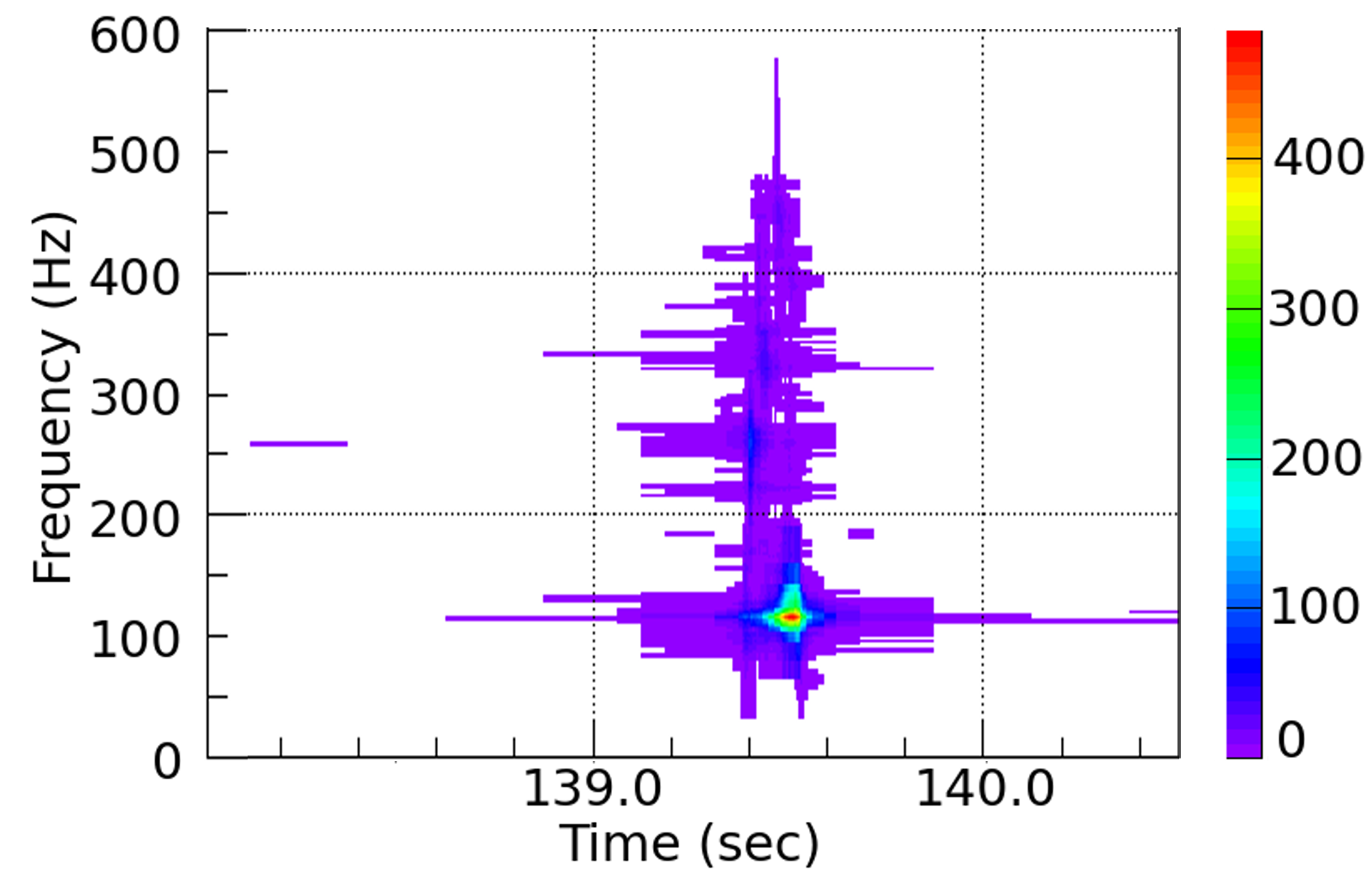}
 \includegraphics[width=0.45\textwidth]{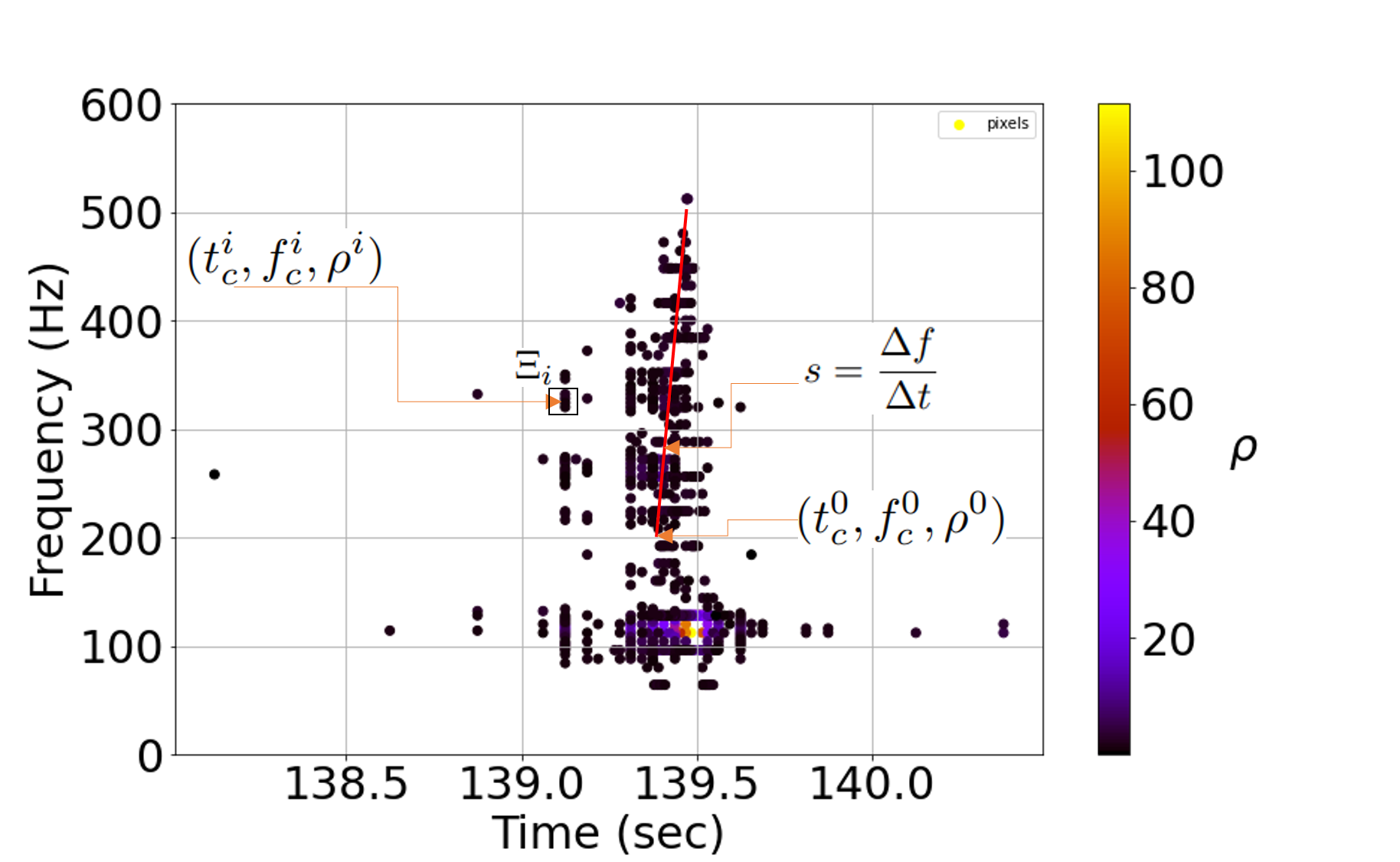}
  \caption{\emph{Left:} Example of a standard cWB scalogram for a GW event. Each wavelet components involved in the reconstructed event is represented with a rectangle of sides equal to the frequency and temporal resolution. The darker blue regions highlight the reconstructed SASI and g-mode features. The duration resolution of tens of milliseconds of some of the wavelets below 200 Hz indicate that some impact on the duration reconstruction is to be expected. \emph{Right:} The same event is displayed with a single dot for each wavelet, the coordinates representing the central time and frequency of the wavelets. While the color bar to the right corresponds to single pixel likelihood, the color bar to the left (not used in the calculations of this paper), corresponds the likelihood for a fixed size time frequency region.
For each wavelet, we use/define different parameters for the $\chi^2$ localization of the g-mode frequency evolution (red) with slope ($s=\frac{\Delta f}{\Delta t}$): the wavelet central time ($t_c$), wavelet central frequency ($f_c$), wavelet likelihood ($\rho$) and the density of wavelets ($\Xi$) in a given time frequency box. For details on these parameters and their usage, see appendix \ref{appendix:GW1}.}
\label{fig:scalogram}
\end{figure*}

In detection and identification problems, a very important aspect is the choice of the metric to be used as an identification tool. The metric we suggest here allows us to leverage on the relative importance
of the recorded energy in the time-frequency region of the
SASI with respect to the overall energy of the candidate. 
Two visualizations that illustrate the time frequency layout of the events
(in particular the location of the SASI and g mode), are presented in Fig. \ref{fig:scalogram}. One is an example of the standard cWB scalogram and the other is the equivalent pixelization adopted for the estimates of the identification metrics in this paper. The difference in the scaling of the likelihood values is because of the presence of overlapping pixels in the left plot, whose likelihood values (that are all positive) add up, causing the upper limit of it's likelihood color bar to be higher than the one on the right plot. In the second plot we also illustrate the relevant parameters used in each pixel (pixel central time, pixel central frequency, pixel SNR) as well as collective measures like local density of pixels, the estimated g-mode start time and slope.

The performance of applying thresholds on a certain metric should be tested on a large number of test cases. Here we produce the test cases by injecting the simulated gravitational waves into different instances of real laser interferometric noise from the LIGO O3 scientific run. In each case, cWB was used to process the data according to the configuration described in Appendix B. 
 The normalized likelihood, for a given trigger, is defined as the sum of the likelihood values of all the pixels surviving in the SASI region (see appendix \ref{appendix:GW1}) w.r.t the total sum of likelihoods in the trigger: 
 \begin{equation}
 \label{eq:SNR_norm}
     \rho_{norm}=\frac{\sum_{i \in SASI} \rho^i}{\sum_{j \in trigger}\rho^j}~,
 \end{equation}
where the $i^{th}$ pixel $\in$ SASI region and $j^{th}$ pixel $\in$ whole trigger. If there are no pixels in the SASI region, the normalized likelihood of the trigger is 0. In this case our recommendation is not to use the gravitational wave data but instead use only the neutrino SASI-meter for the identification and parameter estimation of the SASI.  
In this analysis, the situation with no pixels in the SASI-region occurs with probability of $\lesssim 1\%$ at 1 Kpc and $\lesssim ~20\%$ at 10 Kpc.

 We prepare distributions of the normalized likelihood by repeating the analysis for injections at different times in the noise. The fraction of events above a given threshold can be considered as an estimate of the probability that the normalized likelihood is above the threshold. The detection probability ($P_\mathrm{D}^{\mathrm{GW}}$) and false identification probability ($P_{\mathrm{FI}}^{\mathrm{GW}}$) are calculated from the probability density function (PDF) of $\rho_{\mathrm{norm}}$ based on the simulated GW waveforms with and without SASI activities. The $P_\mathrm{D}^{\mathrm{GW}}$ ($P_\mathrm{FI}^{\mathrm{GW}}$) are the ratio of cumulative area under the SASI (no-SASI) PDF curve of $\rho_{\mathrm{norm}}$ with $\rho_\mathrm{norm}>\Lambda_\mathrm{GW}$ to the total area under the SASI (no-SASI) PDF curve (see the three distributions in the top row of Fig. \ref{fig:GWOnlyResults}). The ROC is the plot of $P_\mathrm{D}^{\mathrm{GW}}$ as a function of $P_{\mathrm{FI}}^{\mathrm{GW}}$, with varying $\Lambda_\mathrm{GW}$:
 
 \begin{eqnarray}\label{eq:pdpfaGW}
& P_\mathrm{D}^{\mathrm{GW}}=\int_{\Lambda_{\mathrm{GW}}}^\infty d\rho Prob_{\mathrm{GW,S}}(\rho),\label{eq:pdgw}~,\\
& P_\mathrm{FI}^{\mathrm{GW}}=\int_{\Lambda_{\mathrm{GW}}}^\infty d\rho Prob_{\mathrm{GW,nS}}(\rho)~. \label{eq:pfgw}
\end{eqnarray}

 The ROC with $P_\mathrm{D}^{\mathrm{GW}}=P_\mathrm{FI}^{\mathrm{GW}}$ corresponds to the 50-50 classification/detection scenario which is equivalent to flipping a coin. The operating point 
 depends on the maximum $P_\mathrm{FI}$ that we would like to use. Similarly to \cite{Lin:2019wwm}, we take the 
$P_\mathrm{FI}^\mathrm{GW}=0.1$.

        The two SASI parameters we're calculating for each trigger are its central frequency and duration. The central frequency of the SASI $f_\mathrm{GW}$  
         is estimated as the weighted mean of the frequencies of the pixels belonging to the SASI region ($f_c^i$) in the time-frequency pixel map presented in Fig.~\ref{fig:scalogram}. The weights are given by the likelihoods of corresponding pixels ($\rho^i$) where, the $i^{th}$ pixel $\in$ SASI region. The $f_\mathrm{GW}$ is:
           \begin{eqnarray}
f_{\mathrm{GW}}=\frac{\sum_{i\in SASI} \rho^if_c^i}{\sum_{i\in SASI} \rho^i}~,
  \end{eqnarray}
where the summation is done over all the pixels remaining in the area of interest of the trigger (SASI region) and, $\rho^i$ is the likelihood value of the $i^{th}$ pixel.

The GW duration of the SASI ($\tau^{\mathrm{GW}}$) is estimated here by the difference in time coordinates of the two extreme pixels in the SASI region. Since the pixels correspond to wavelet components from multi-resolution, the time resolutions of the two extreme pixels are also used in the estimate according to
  \begin{eqnarray}
\tau^{GW}= t_{max}+\frac{\delta t_{max}}{2}-t_{min}-\frac{\delta t_{min}}{2},
  \end{eqnarray}
where $t_{max}$ and $t_{min}$ are the time coordinates of the rightmost pixel (with $\delta t_{max}$ as its time resolution) and the leftmost pixel (with $\delta t_{min}$ as its time resolution), respectively.\\

\subsubsection{Results: GW receiver Operating Curves}

\begin{figure*}
    \centering
    \includegraphics[width=0.3\textwidth]{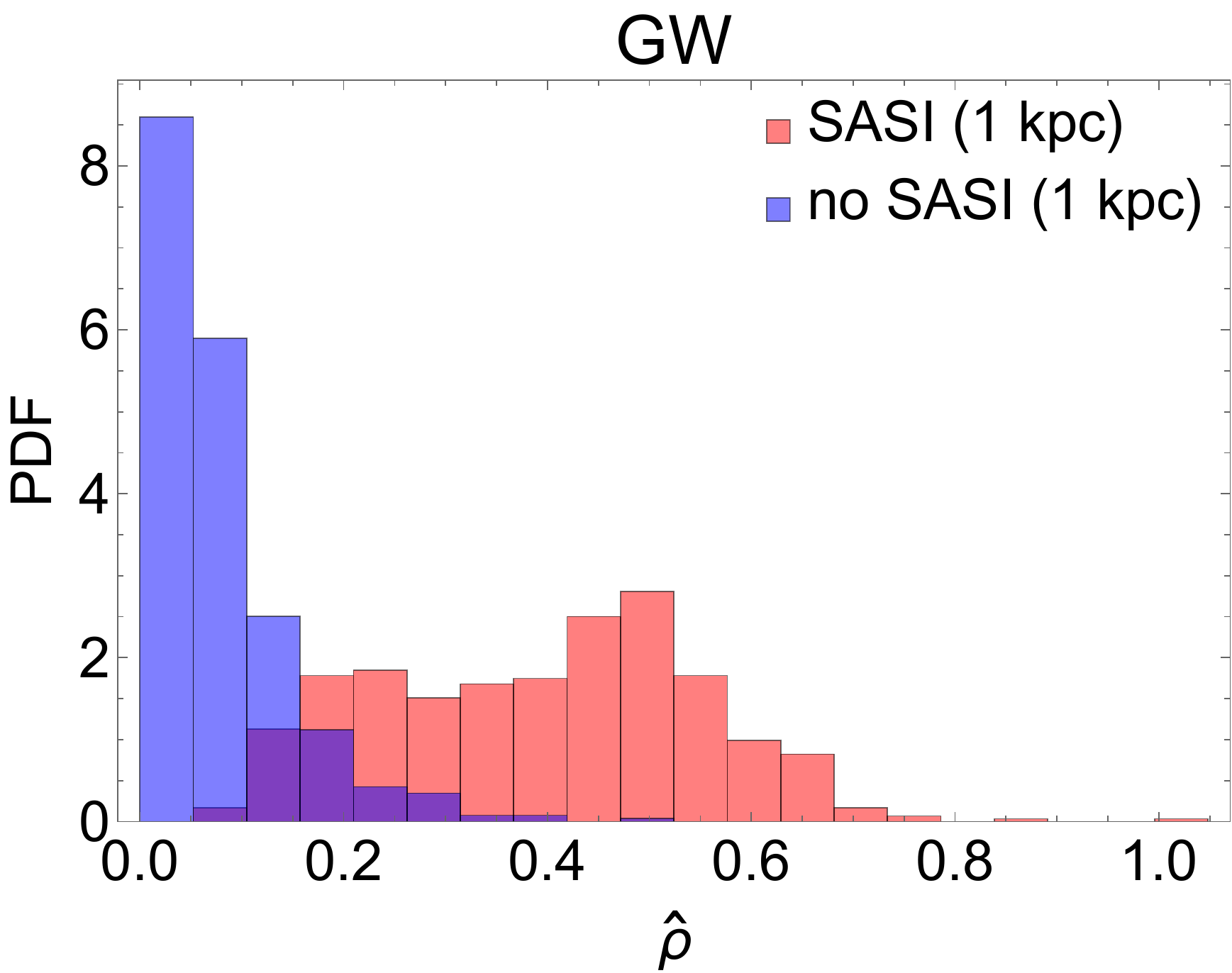}
    \includegraphics[width=0.3\textwidth]{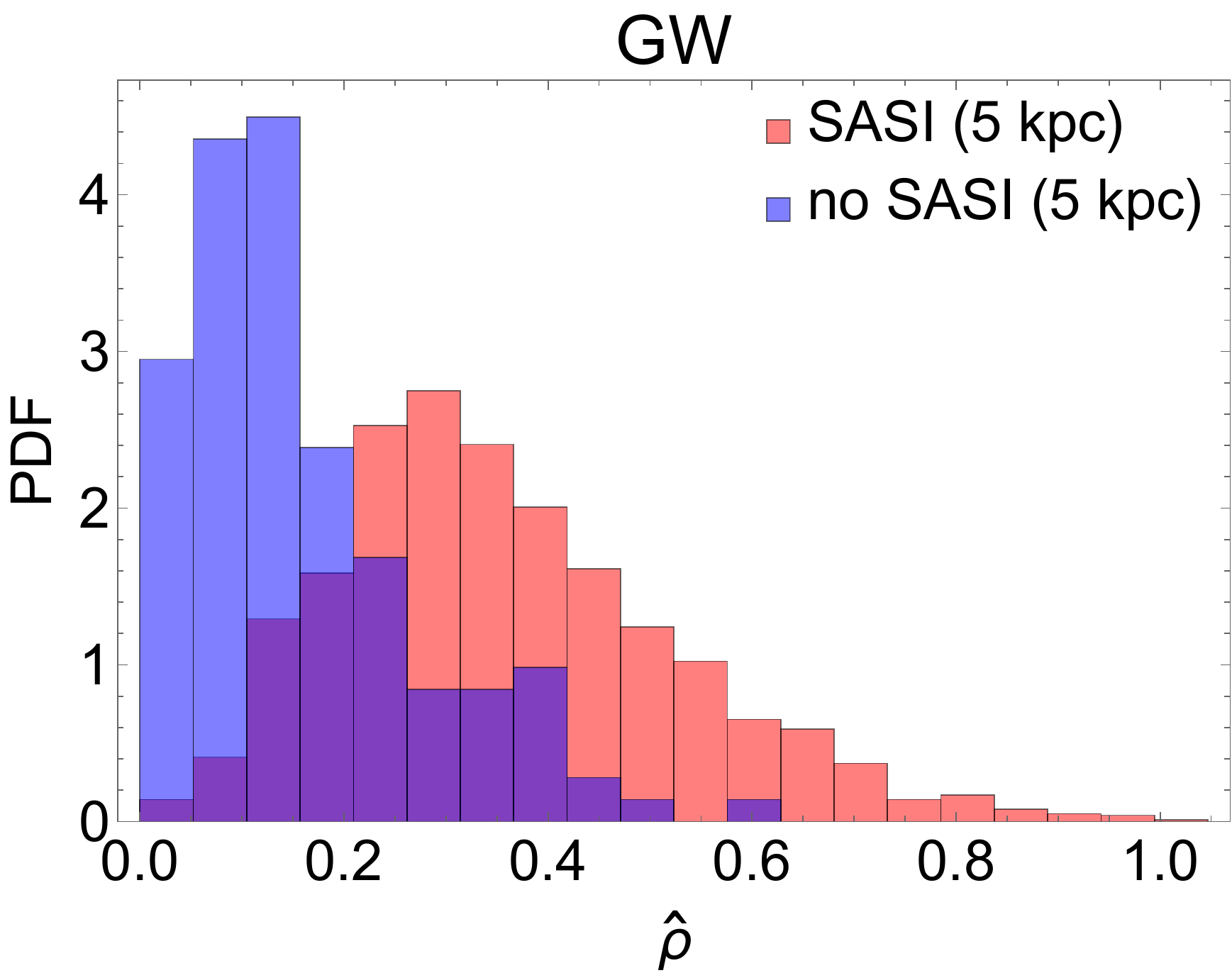}
    \includegraphics[width=0.3\textwidth]{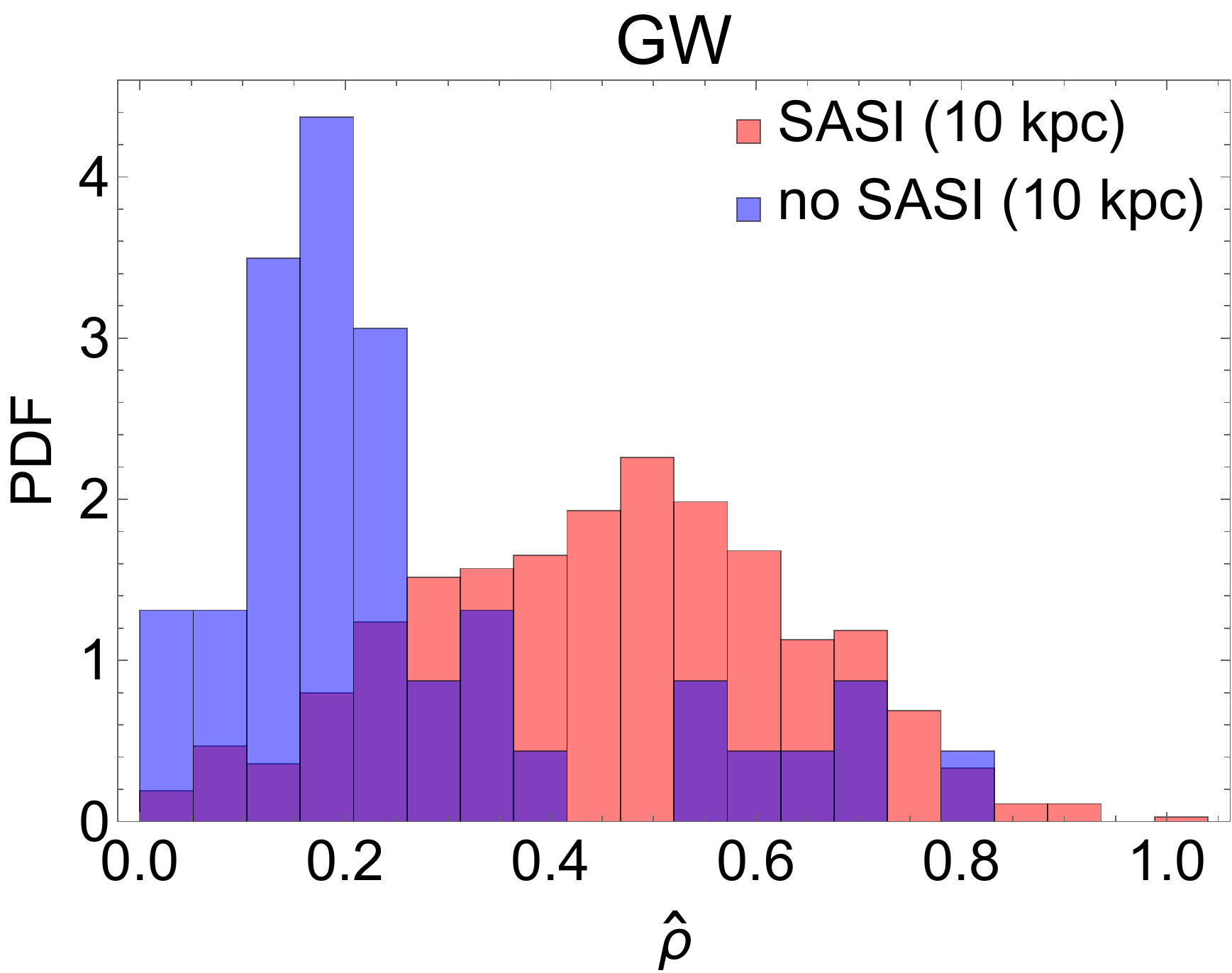}

    \includegraphics[width=0.3\textwidth]{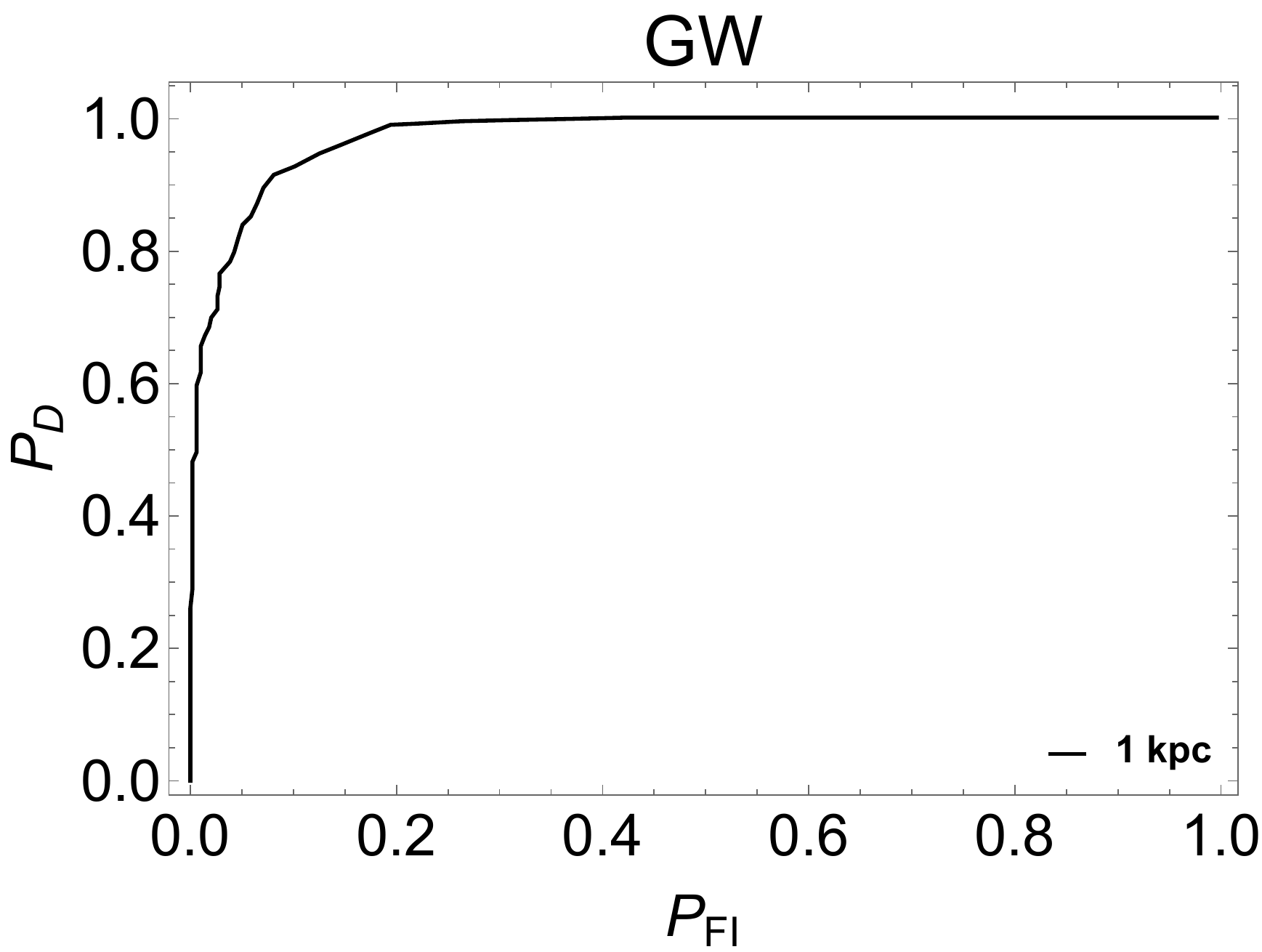}
    \includegraphics[width=0.3\textwidth]{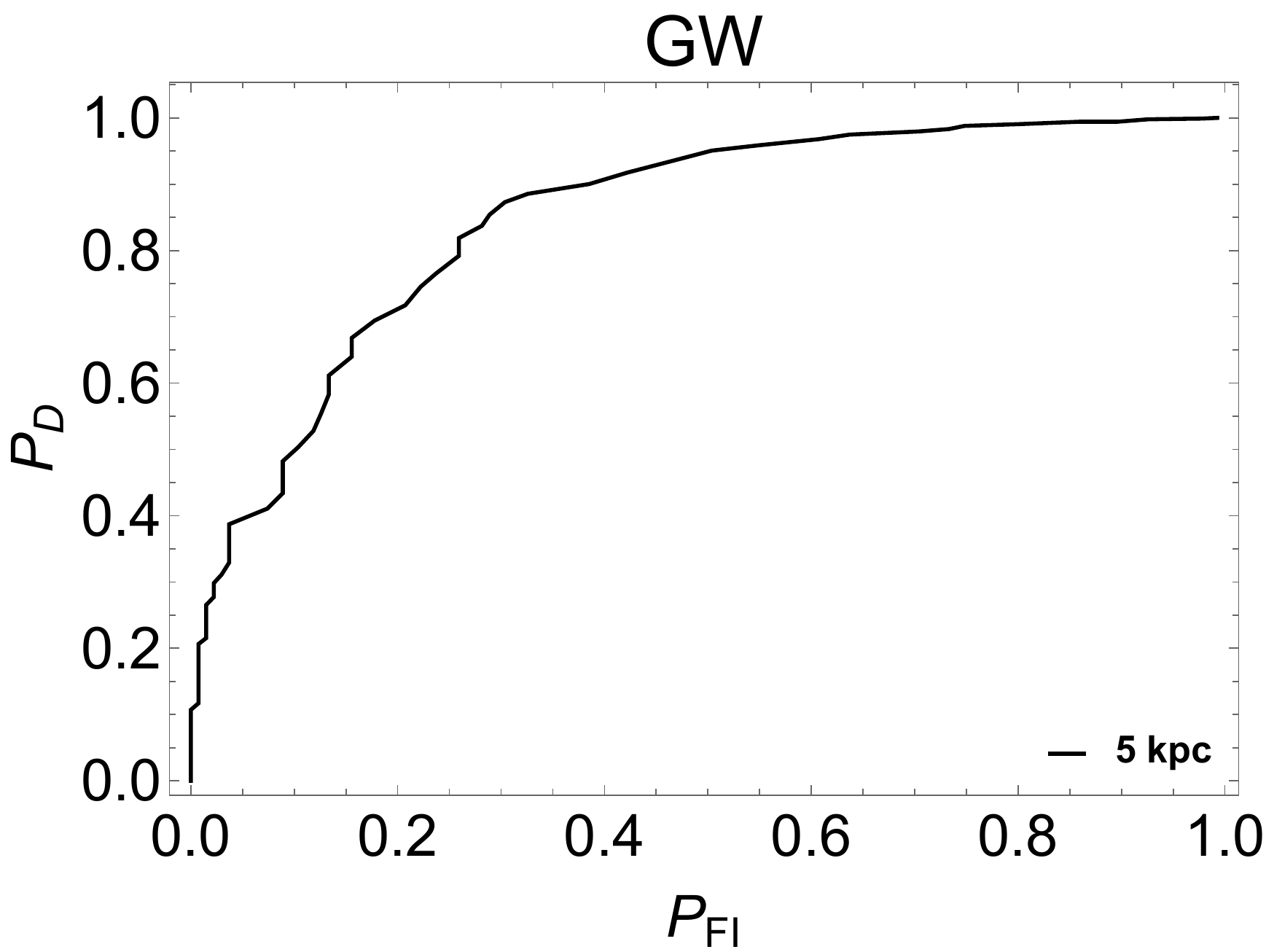}
    \includegraphics[width=0.3\textwidth]{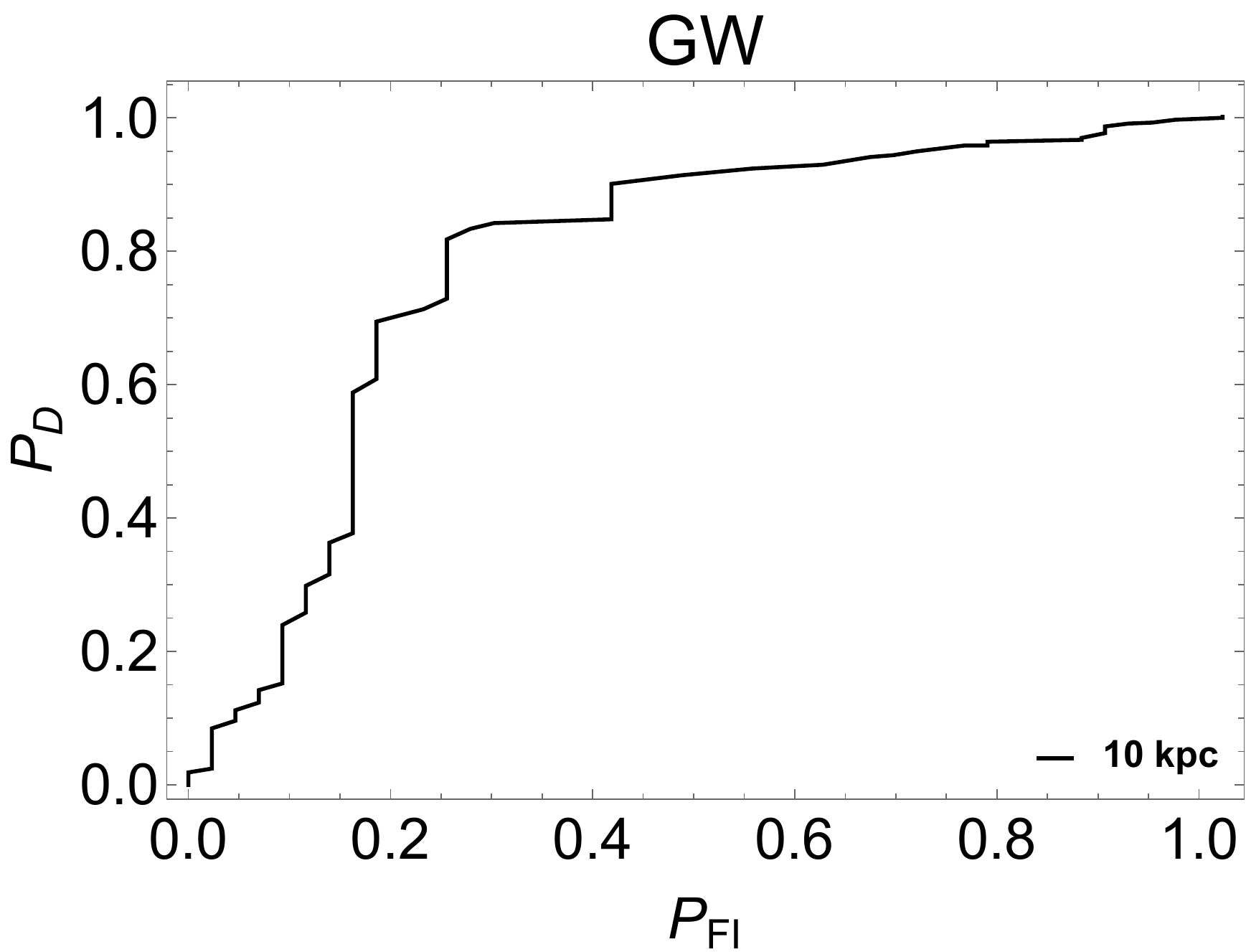}
    \caption{\emph{Top row:} examples of distributions of the test-statistics $\hat{\rho}$ obtained from simulated GW data, for distances $D=1,5,10$ kpc to the \sn. Here, $\hat{\rho}$ is the rescaled logarithmic likelihood ratio of GW signals with its maximum in SASI case being 1. \emph{Bottom row:} the corresponding Receiver Operating Curves.}
    \label{fig:GWOnlyResults}
\end{figure*}

Fig. \ref{fig:GWOnlyResults} displays the PDFs as well as the ROCs of the test-statistics $\hat{\rho}$  for the SASI and no-SASI cases,
for different values of CCSNe distance. The PDF of $\hat{\rho}$ with its maximum in SASI case being one is obtained by rescaling the distribution of $\rho_\mathrm{norm}$. As expected, the value of $P^{\mathrm{GW}}_{\mathrm{D}}$ (for fixed $P^{\mathrm{GW}}_{\mathrm{FI}}$) decreases with increasing distance. The decline is noticeably slower than the one observed in the \n\ channel, reflecting the slower scaling of the GW signal with $D$. For $D=10$ kpc and $P^\mathrm{GW}_\mathrm{FI}=0.20$, we have $P^\mathrm{GW}_\mathrm{D}\sim 0.60-0.65$, which  is larger than the corresponding neutrino result ($P^{\nu}_\mathrm{D}\simeq 0.35$ for $P^{\nu}_{\mathrm{FI}}=0.20$).  

\begin{figure}[ht!]
	\includegraphics[width=0.45\textwidth]{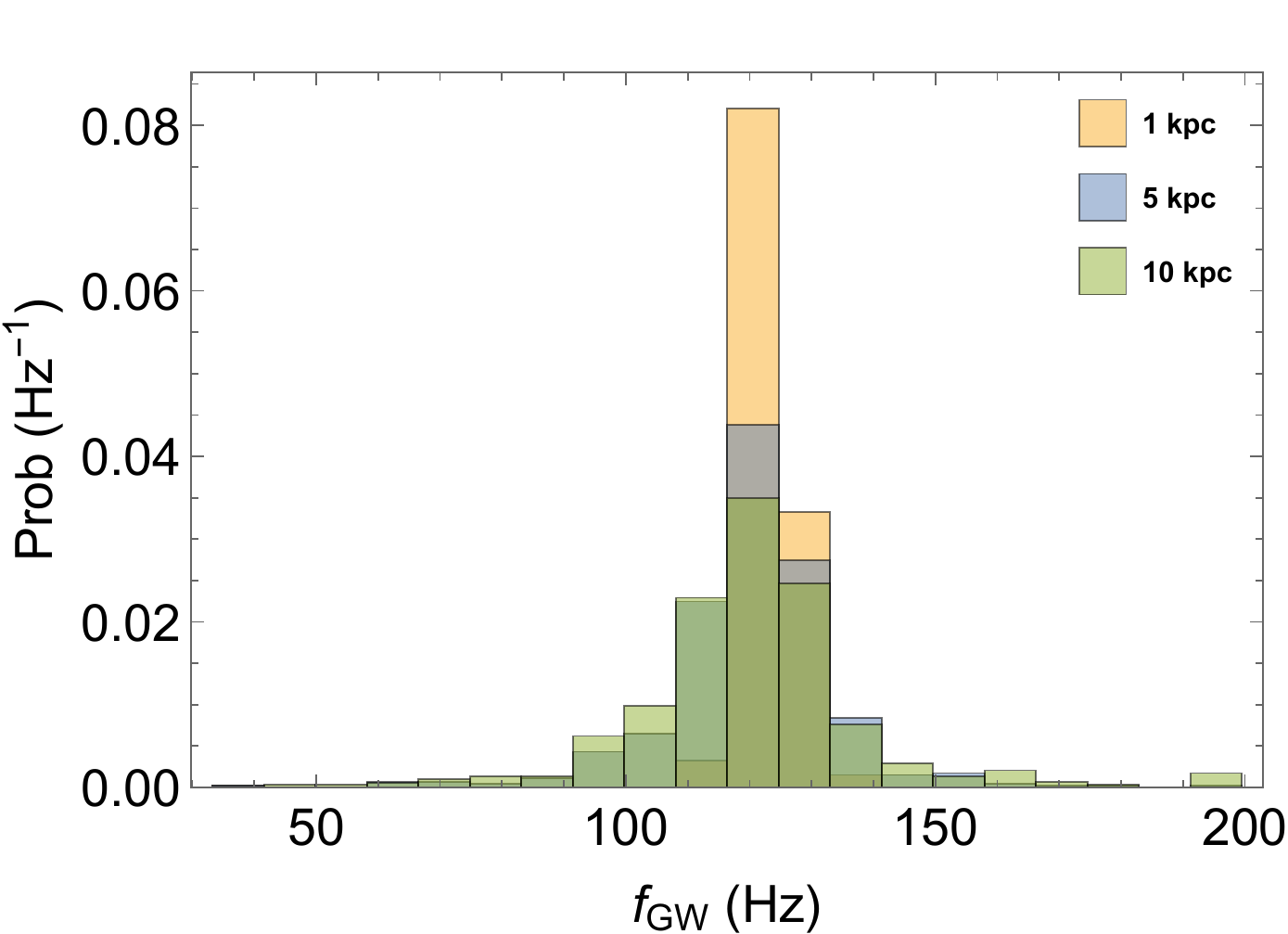}
	\includegraphics[width=0.45\textwidth]{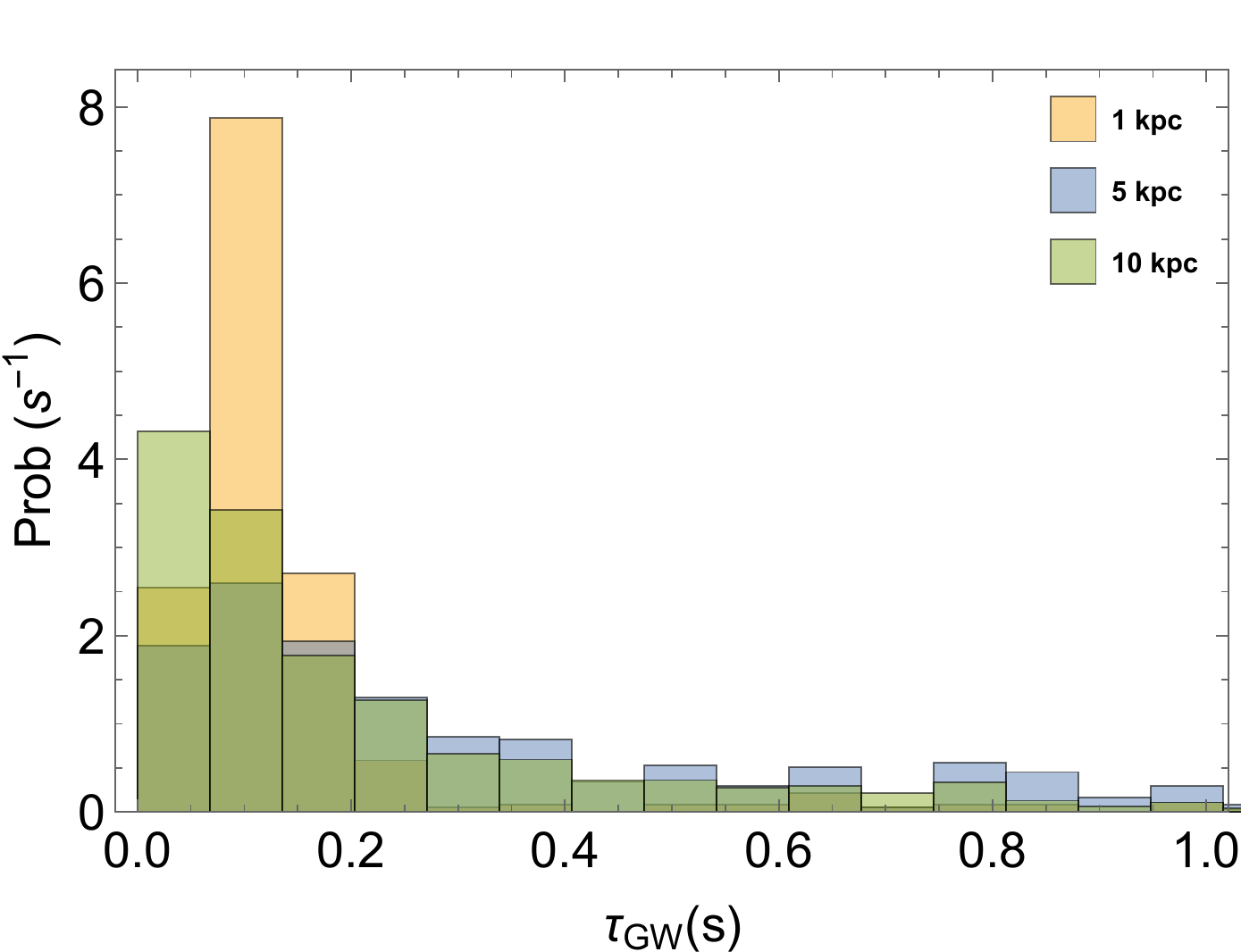}
	 \caption{Probability density distribution of the SASI central frequency (upper) and the SASI duration (lower) estimations based on GW signals. }
	\label{fig:fre_tau_GW}
\end{figure}

In Fig. \ref{fig:fre_tau_GW} and Tab. \ref{tab:sasiIcePfa10}, we present the parameter estimation results. 
Fig. \ref{fig:fre_tau_GW} shows the probability density distribution of the SASI starting time as well as the SASI duration in GW signals at various CCSNe distance.

In Tab. \ref{tab:sasiIcePfa10}  we show the estimated values and uncertainties for the frequency $f_{\mathrm{GW}}$, and duration $\tau^{\mathrm{GW}}$, of the SASI.   The results for $f_{\mathrm{GW}}$ can be directly compared to those obtained from the \n-only analysis (also shown in the Table). Here we notice the different dependence on $D$ in the \n\ and GW results: for $D=1$ kpc the uncertainties are comparable in the two channels, with the performance being slightly better in the \n\ channel. However, as the distance grows, 
the performance in the two channels decrease at a different rate.

The parameter estimation performance is poorer for $\tau^{GW}$, for which the uncertainty is comparable or larger than the central value \footnote{The fact that the estimated SASI duration and/or its uncertainty exceed the duration of the simulated signal is due to having combined the signal with a stream of background data that extend in time beyond the signal itself. }. This result  could be a limitation of our definition of duration. In the future we might employ instead a definition based on a SNR weighted duration.

The results discussed here show the possibility to identify the presence of the SASI for a galactic CCSN candidate at current interferometers. Our results can be used to forecast the performance of future interferometers that will have order of magnitude better sensitivity.  Because GW signal amplitudes scale as $D^{-1}$, these detectors are expected to have ROCs similar to those shown here for larger distances (proportional to their sensitivity).

\section{Multi-messenger analysis}\label{sec:multimessenger}

\begin{figure*}[htp]

	\includegraphics[width=0.45\textwidth]{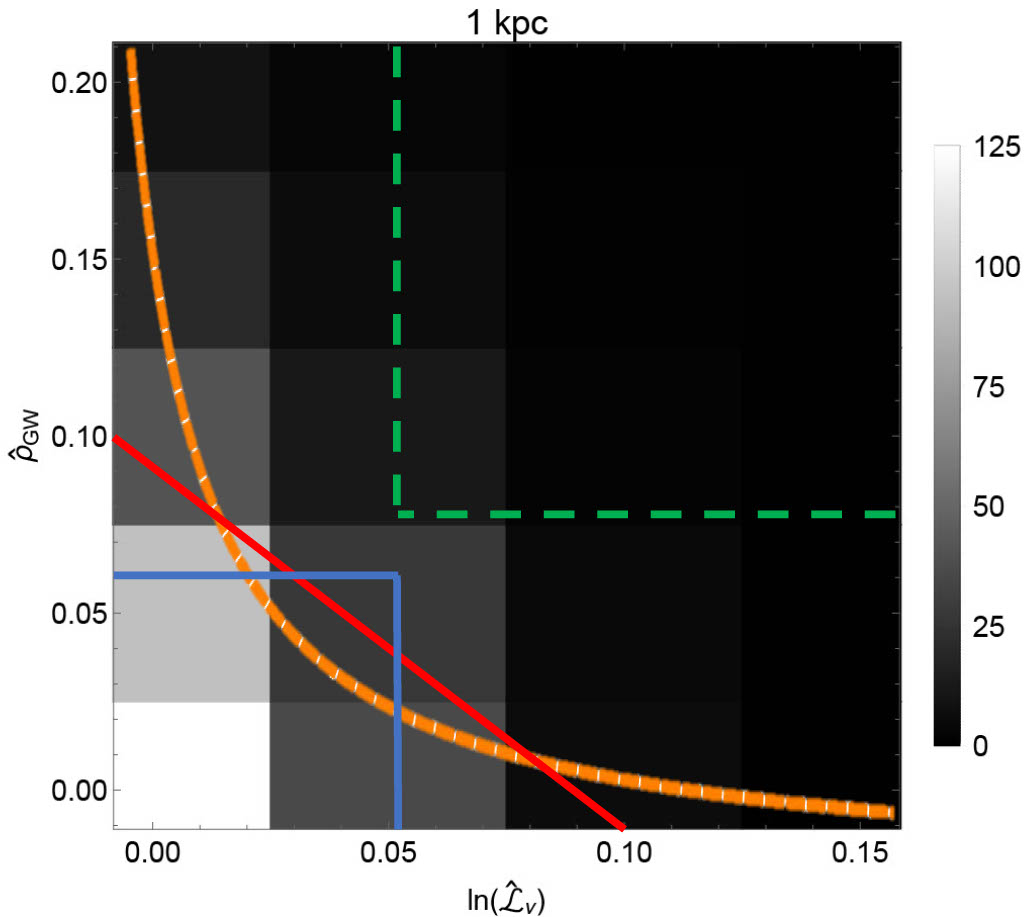}
	\includegraphics[width=0.44\textwidth]{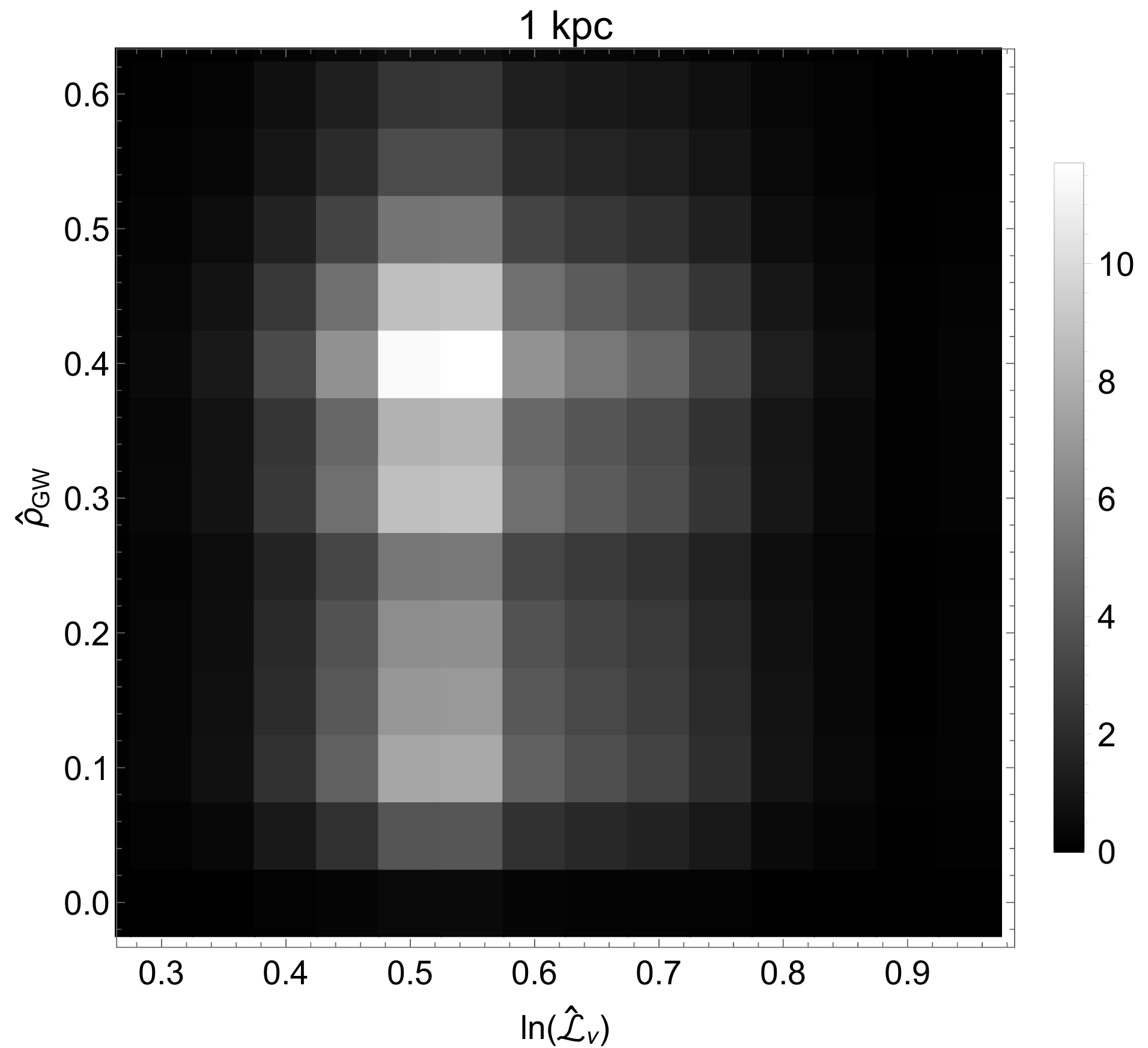}
	\includegraphics[width=0.45\textwidth]{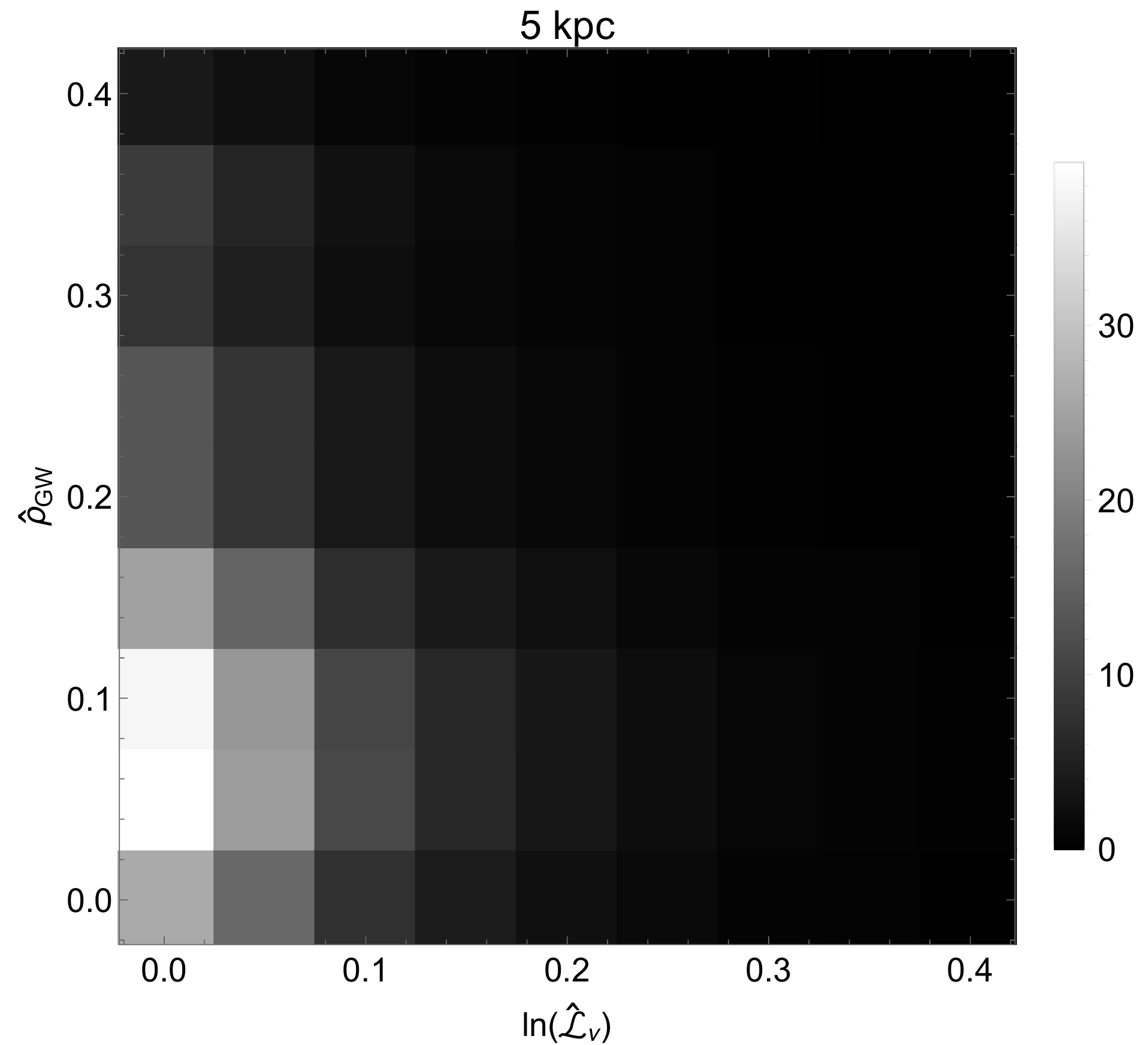}
	\includegraphics[width=0.45\textwidth]{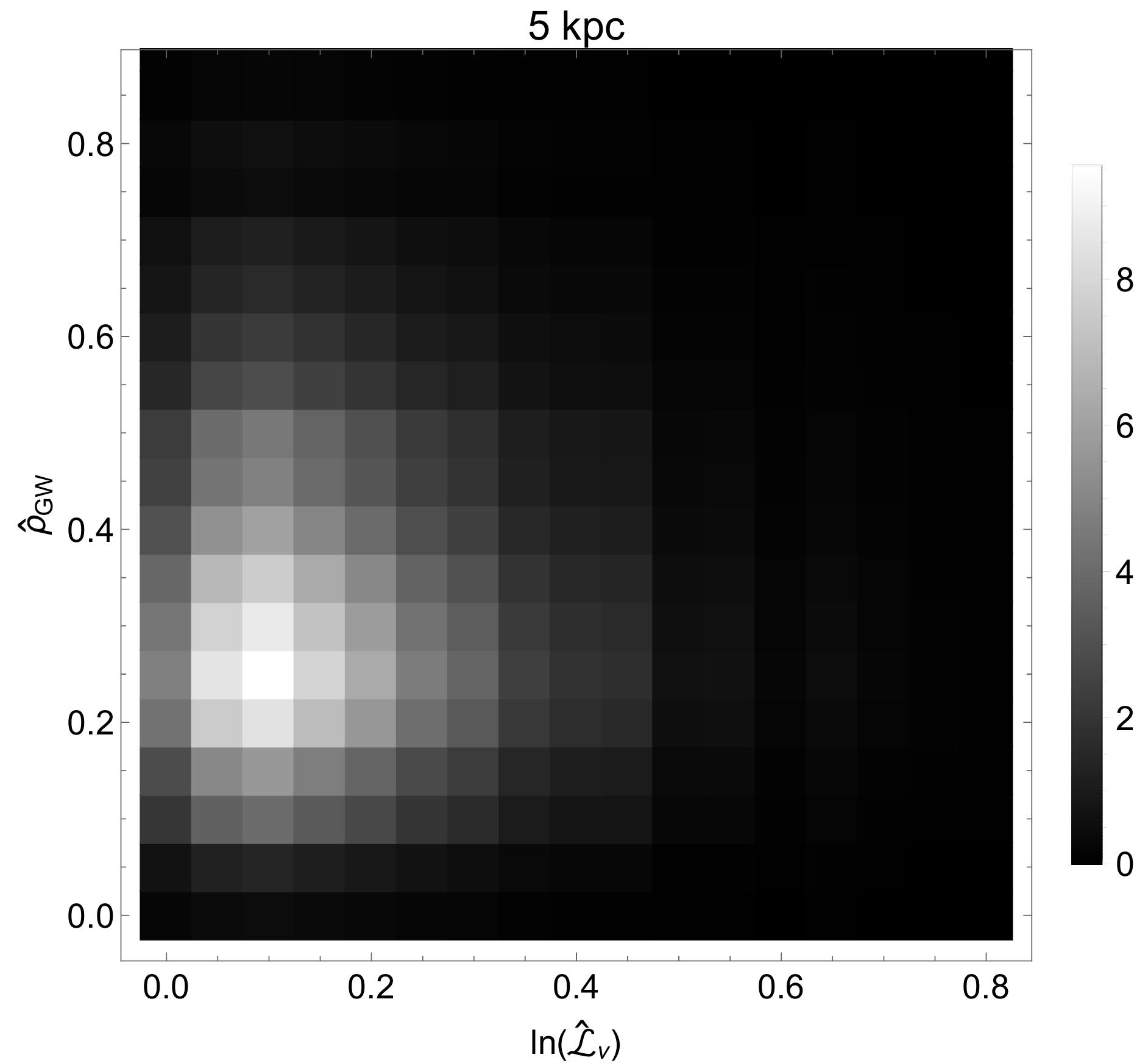}
	\includegraphics[width=0.45\textwidth]{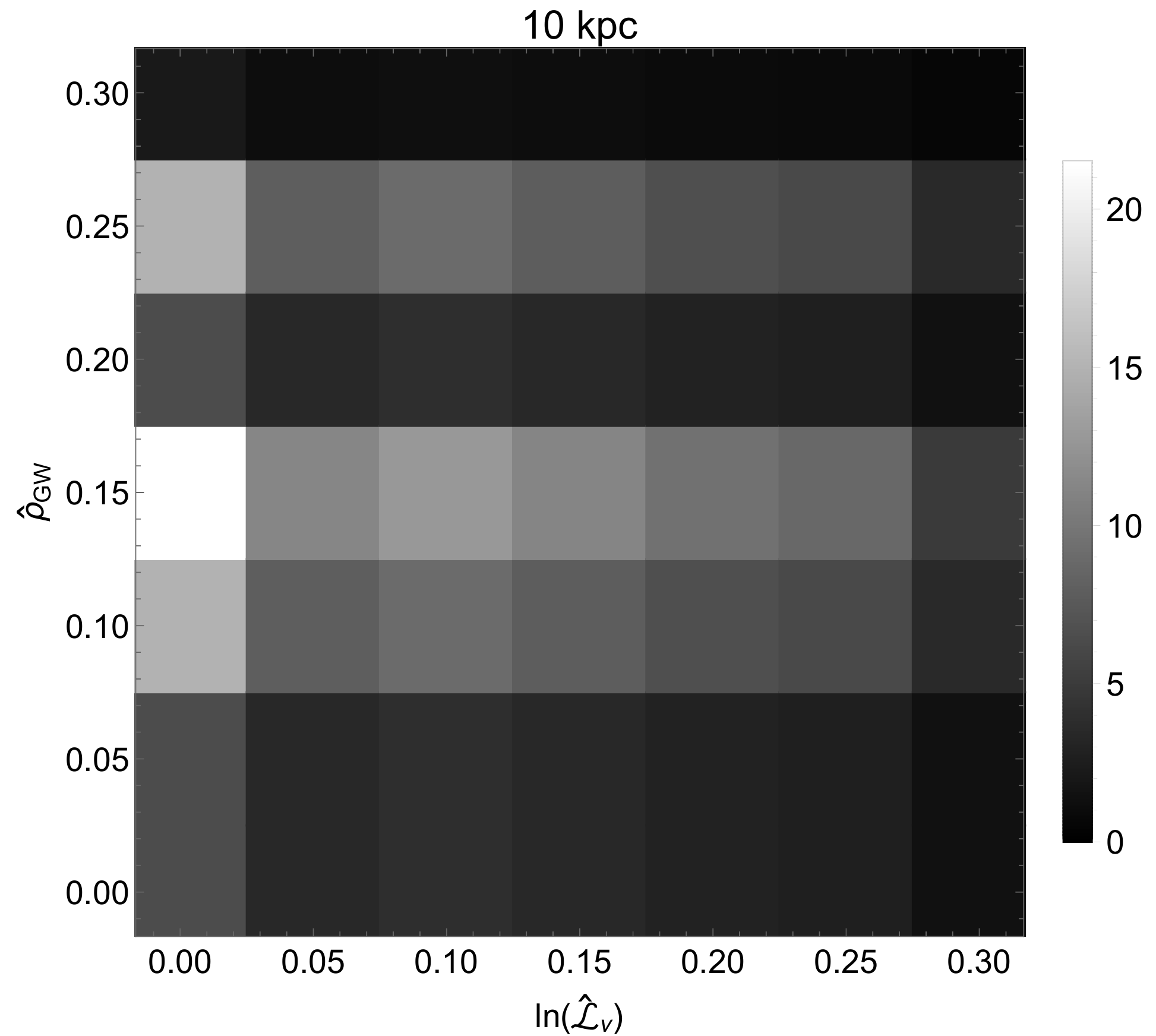}
	\includegraphics[width=0.44\textwidth]{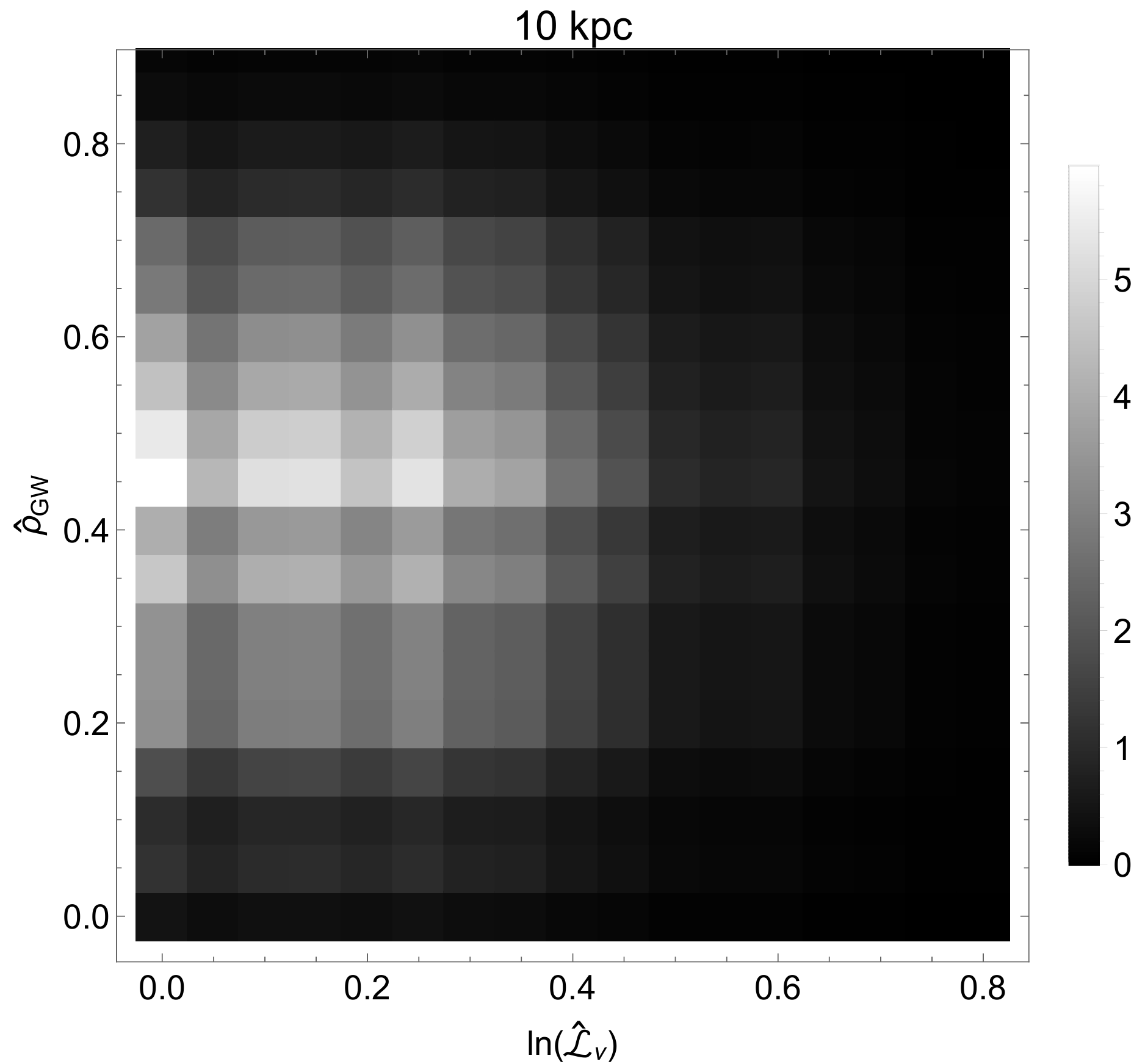}
	
	\caption{2D probability density distribution of  $\{ln(\hat{\mathcal{L}}_{\nu}),\hat{\rho}_{GW}\}$ at 1 kpc (upper panels), 5 kpc (middle panels) and 10 kpc (lower panels), where the distribution in the left(right) panels are based on neutrino and GW signals without(with) SASI activities. Here, $ln(\hat{\mathcal{L}})$ is the rescaled logarithmic likelihood ratio with its maximum in SASI case being 1. And $\hat{\rho}$ is the rescaled logarithmic likelihood ratio of GW signals with its maximum in SASI case being 1. In the upper left panel, the curves in color illustrate the different integration thresholds that can be used to identify the presence of the SASI, as discussed in Sec. \ref{subsec:comprob}. Specifically, the region inside the green dashed rectangle is for the ``logical And" ; the region outside the blue solid rectangle is for the ``logical Or". The region outside the red solid triangle corresponds to the ``x + y = const” case, and region at the upper right of the orange dashed curve represent the ``x $\times$ y = const” case. }  
	\label{fig:combinehistogram}
\end{figure*}

\begin{figure*}[htp]

	\includegraphics[width=0.45\textwidth]{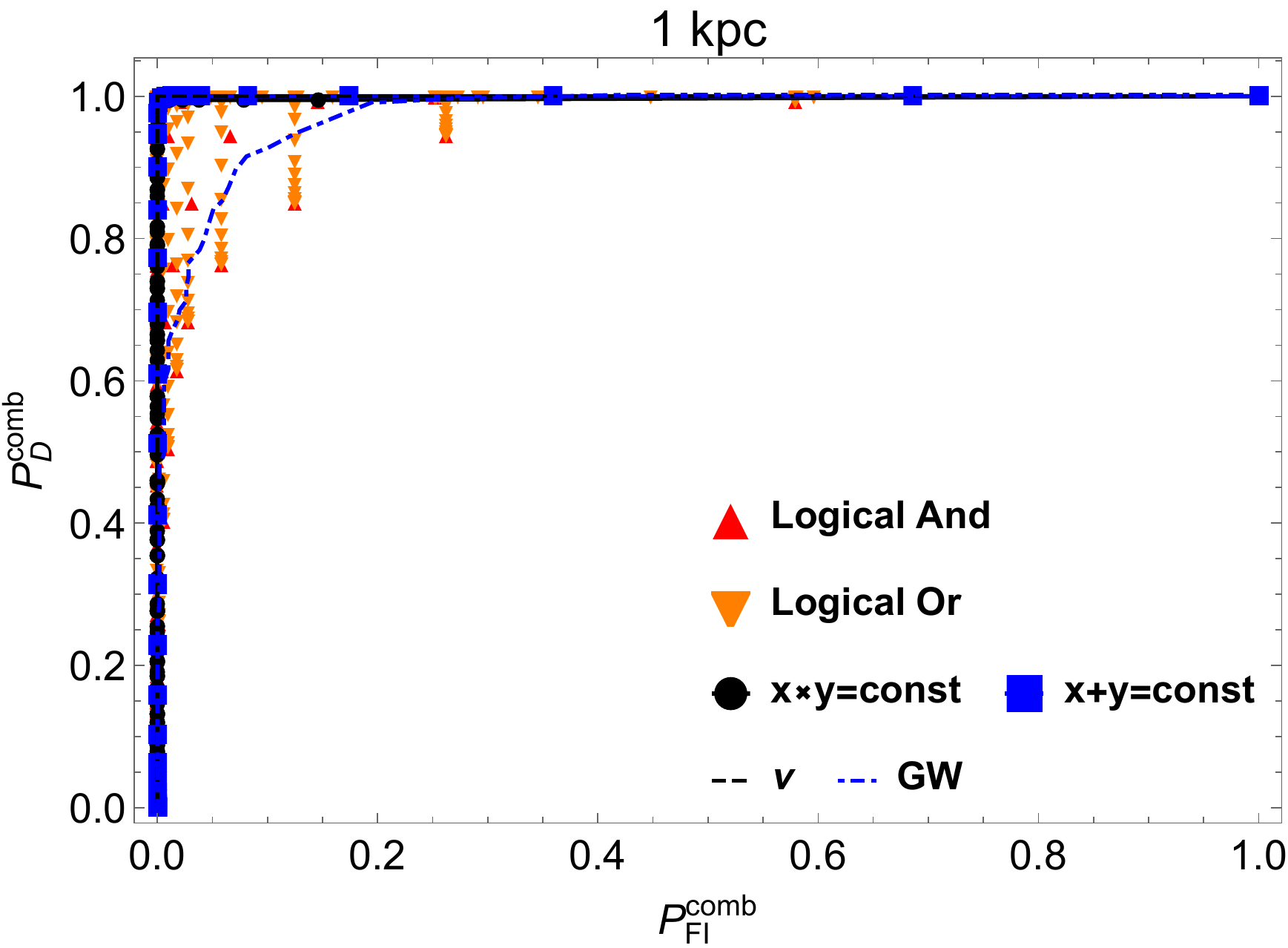}
	\includegraphics[width=0.45\textwidth]{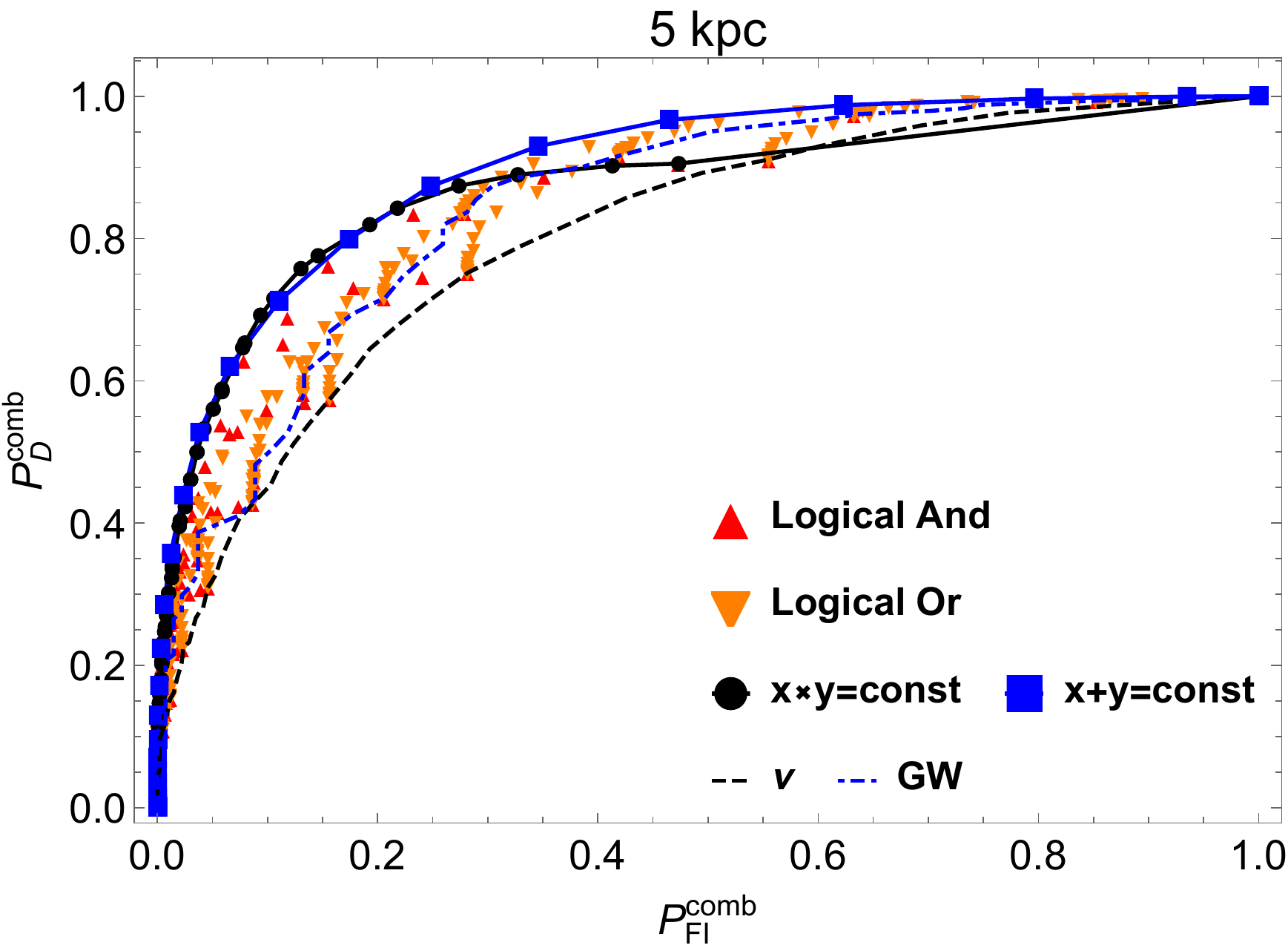}
	\includegraphics[width=0.45\textwidth]{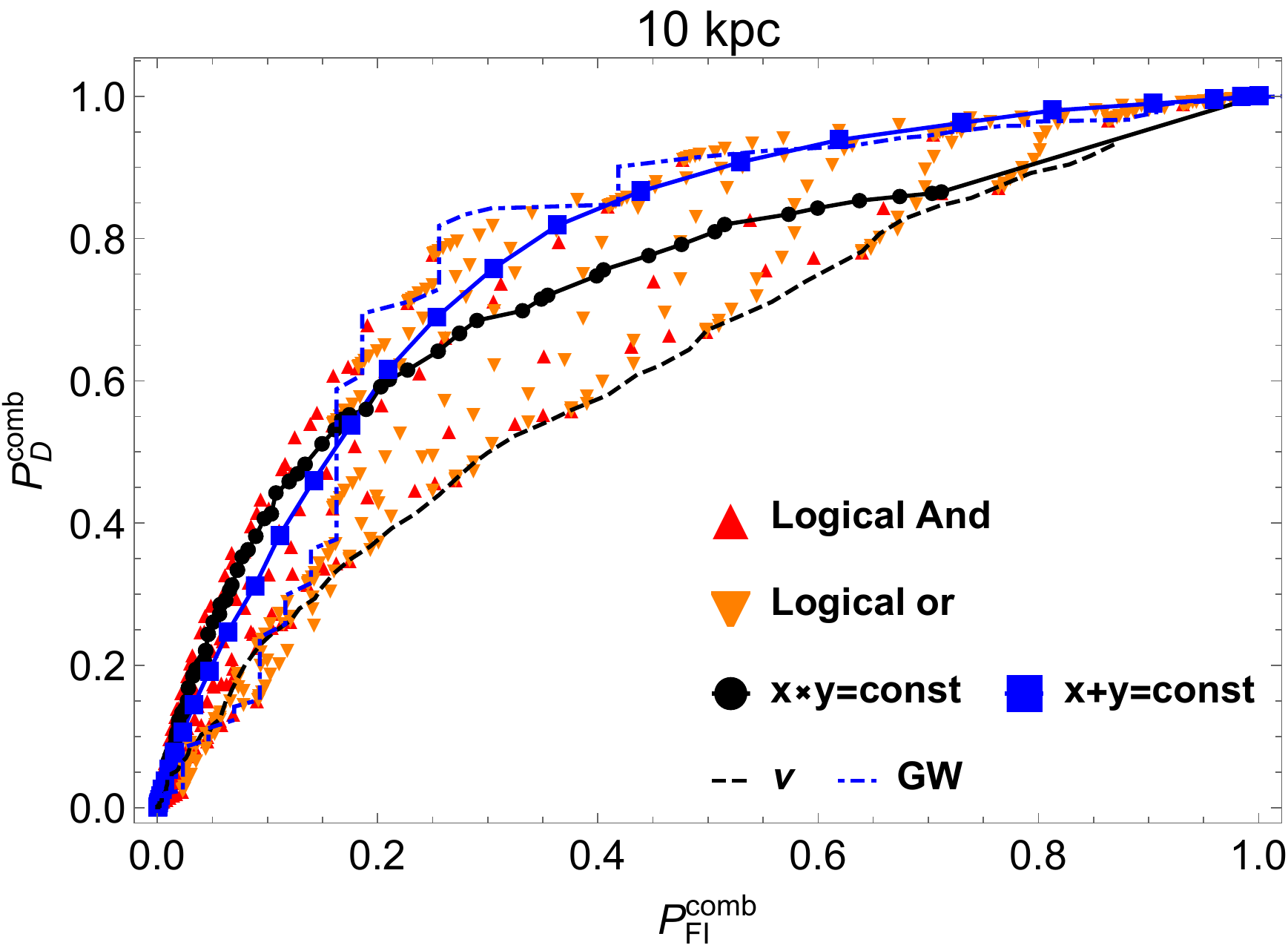}
	
	\caption{Combined receiver operating curve at 1 kpc (upper left) , 5 kpc (upper right) and 10 kpc (lower)  based on Fig. \ref{fig:combinehistogram}, by using different selection thresholds.}
	\label{fig:combineROC}
\end{figure*}
In this section we discuss a novel methodology analyzing the SASI using \emph{both} the neutrino and the GW signals combined. To begin with, we present a flow chart illustrating the main procedure, see Fig. \ref{fig:flowchart}. In the following, the generation of combined receiver operating curves for SASI identification is discussed. 

\subsection{Combining probabilities}
\label{subsec:comprob}

The identification of SASI activity can happen in two scenarios, where (usable) data are available either (i) only in a single messenger (either neutrino or GW signatures); or ~(ii) in multi messengers.
Each scenario is expected to happen with a specific probability. Realistically, the detection probability of neutrino signals from a galactic or nearby extragalactic CCSN is unity (when the detectors are active), and the neutrino data can always be used for the detection of the SASI activity. The probability of having usable data in the GW channel,  $P^\mathrm{D}_\mathrm{GW}$, depends on the probability to detect the GW waveform itself (for a recent discussion see \cite{Szczepanczyk:2022lni}) and the probability that the GW presents surviving wavelet components in the SASI time frequency region.

In the case where only neutrino data are available, the detection probability of SASI activities is $P_\mathrm{D}=P_\mathrm{D}^{\nu}$. Similarly, $P_\mathrm{FI}=P_\mathrm{FI}^{\nu}$. In the second case, where both usable neutrino and GW data are present, we define joint detection and false identification probabilities as follows.

We can define the 2-dimensional probability density distributions for the the SASI and no-SASI cases as: 
\begin{eqnarray}
    Prob_\mathrm{S}(ln(\mathcal{L}),\rho)=Prob_\mathrm{\nu,S}(ln(\mathcal{L}))Prob_\mathrm{GW,S}(\rho)~,\nonumber \\
Prob_\mathrm{nS}(ln(\mathcal{L}),\rho)=Prob_\mathrm{\nu,nS}(ln(\mathcal{L}))Prob_\mathrm{GW,nS}(\rho)~.
\label{eq:probcomb}
\end{eqnarray}

Similarly to the single-messenger case, this distribution can be integrated to obtain detection and false identification probabilities. 
In this case, however, the integration threshold for SASI identification can be chosen in more than one way. The choice is driven by the goal of finding the optimal ROC curve (i.e., maximize the joint detection probability for fixed false identification probability) for the identification of the SASI.

For example the identification of the presence of the SASI can be pursued in the three following ways: 

(1) \emph{Logical And: } SASI is established when $\mathcal{L}>\Lambda_\nu$ \emph{and} $\rho>\Lambda_{GW}$. 
The combined probability of detection and the combined false identification probabilities are:  
\begin{equation}\label{eq:pdlogicAnd}
\begin{split}
    P_\mathrm{D}^{comb}&=P_\mathrm{D}^{\nu}\times P_\mathrm{D}^{GW}\\&=\int_{\Lambda_\nu}^\infty\int_{\Lambda_{GW}}^{\infty}dln(\mathcal{L})d\rho Prob_\mathrm{\nu,S}(ln(\mathcal{L}))Prob_\mathrm{GW,S}(\rho),
\end{split}
\end{equation}
and 
\begin{equation}\label{eq:pfalogicAnd}
\begin{split}
    P_\mathrm{FI}^{comb}&=P_\mathrm{FI}^{\nu}\times P_\mathrm{FI}^{GW}\\&=\int_{\Lambda_\nu}^\infty\int_{\Lambda_{GW}}^{\infty}dln(\mathcal{L})d\rho Prob_\mathrm{\nu,nS}(ln(\mathcal{L}))Prob_\mathrm{GW,nS}(\rho).
\end{split}
\end{equation}
In a 2-D probability density distribution of $(\mathcal{L},\rho)$, the SASI threshold defined above appears like a rectangle (see illustration in Fig. \ref{fig:combinehistogram}), with its accepted GW component lying between $\Lambda_{GW}$ to $\infty$ and its accepted neutrino component lying between $\Lambda_\nu$ to $\infty$. And only the $(\mathcal{L},\rho)$ lying inside of the defined rectangle would be identified as a detection of the SASI. 

(2) \emph{Logical Or: } SASI is established when at least one of $\mathcal{L}>\Lambda_\nu$ and $\rho>\Lambda_{GW}$ is verified;    

The combined probability of detection and the combined false identification rate are then given by:
\begin{equation}\label{eq:pdlogicOr}
\begin{split}
    P_\mathrm{D}^{comb}&=1-(1-P_\mathrm{D}^{\nu})\times (1-P_\mathrm{D}^{GW})\\&=1-\int_{0}^{\Lambda_\nu}\int_0^{\Lambda_{GW}}dln(\mathcal{L})d\rho Prob_\mathrm{\nu,S}(ln(\mathcal{L}))\\&\times Prob_\mathrm{GW,S}(\rho),
\end{split}
\end{equation}
and 
\begin{equation}\label{eq:pfalogicOr}
\begin{split}
    P_\mathrm{FI}^{comb}&=1-(1-P_\mathrm{FI}^{\nu})\times (1-P_\mathrm{FI}^{GW})\\&=1-\int_{0}^{\Lambda_\nu}\int_0^{\Lambda_{GW}}dln(\mathcal{L})d\rho Prob_\mathrm{\nu,nS}(ln(\mathcal{L}))\\&\times Prob_\mathrm{GW,nS}(\rho).
\end{split}
\end{equation}
Here, the SASI threshold appear like a ``rectangular hole" (shown in Fig. \ref{fig:combinehistogram}, with its rejected GW component lying between 0 to $\Lambda_{GW}$ and its rejected neutrino component lying between 0 to $\Lambda_\nu$. And only the $(\mathcal{L},\rho)$ lying outside of the defined rectangle hole would be identified as a detection of the SASI imprints. 

 (3) \emph{Mixed: } SASI is established when $f(\mathcal{L},\rho)>\Lambda$, where $f$ is a function of $\mathcal{L}$ and $\rho$. 
Here, the SASI threshold can be defined in several ways. For example, as $\mathcal{L}\times\rho>\Lambda$, which is denoted as ``$x\times y=const$" threshold. Or, the SASI threshold can be defined as $\mathcal{L}+\rho>\Lambda$, which is denoted as ``$x+y=const$" threshold (see Fig. \ref{fig:combinehistogram} for illustration). And the $P_\mathrm{D}^{comb}$ based on the above thresholds are:
\begin{equation}\label{eq:pdlogicPlus}
\begin{split}
    P_\mathrm{D}^{comb,x\times y}&=\int_{\mathcal{L}\times\rho>\Lambda}^\infty dln(\mathcal{L})d\rho Prob_\mathrm{\nu,S}(ln(\mathcal{L}))Prob_\mathrm{GW,S}(\rho),
\end{split}
\end{equation}
and 
\begin{equation}\label{eq:pdlogicCross}
\begin{split}
    P_\mathrm{D}^{comb,x+y}&=\int_{\mathcal{L}+\rho>\Lambda}^\infty dln(\mathcal{L})d\rho Prob_\mathrm{\nu,S}(ln(\mathcal{L}))Prob_\mathrm{GW,S}(\rho).
\end{split}
\end{equation}
The false alarm probability is defined similarly, with $Prob_\mathrm{\nu,S}\rightarrow Prob_\mathrm{\nu,nS}$ and $Prob_\mathrm{GW,S}\rightarrow Prob_\mathrm{GW,nS}$.

We will discuss the relationships between the above combined $P_\mathrm{D}$ in more details in the next section. Note that the strength of the SASI-imprints  may depend on the distance $D$ differently for neutrinos and GW. So a combination of relatively high $\mathcal{L}$ and relatively low $\rho$ (or vice versa)  from a CCSN is possible. The 2-D probability distribution map may provide a way to account for possible tensions between neutrino and GW SASI identifications, and provide a statistical interpretation for a combined data set including both neutrino and GW observations.

\subsection{Results: the joint SASI-meter}

\begin{table*}
	\caption{\label{tab:pd} 
$P_\mathrm{D}$ corresponding to $P_\mathrm{FI}$ of 10 \%, which we take as the operating point, and 20 \%, where the difference between $\nu$ and GW channel becomes more pronounced, for the distances of 1 kpc, 5 kpc and 10 kpc in the single messenger (GW and $\nu$) and combined cases.
}  
	\begin{ruledtabular}
		\begin{tabular}{ccccc}
			$P_\mathrm{FI}$ &Channel& $P_\mathrm{D}$($10$ kpc)&$P_\mathrm{D}$ ($5 $ kpc) &$P_\mathrm{D}$ ($1$ kpc) \\
			\hline
			 &&&&\\
			 0.10&GW& $\approx 0.15-0.25$& $\approx0.40-0.50$ &$0.90$\\
			 0.10&$\nu$&$ 0.25$& $ 0.40$  &1.00 \\
			 0.10&Logical And&$\approx 0.19-0.47$ &$\approx 0.47-0.68$  &$\approx 0.80-1.00$\\
			 0.10&Logical Or&$\approx 0.18-0.25$ & $\approx 0.48-0.60$ &$\approx 0.83-1.00$   \\
			 0.10&$x*y=$const.& 0.41 & 0.71 &$ 1.00$\\
			 0.10&$x+y=$const.& 0.35& 0.70   &$ 1.00$ \\
			 \hline
			 &&&&\\
			 0.20&GW& $\approx 0.60-0.65$& $ 0.70$ &$ 1.00$\\
			 0.20&$\nu$&$0.35$& $0.65$  &1.00 \\
			 0.20&Logical And&$\approx 0.40-0.69$ &$\approx 0.69-0.82$  &$\approx 0.90-1.00$\\
			 0.20&Logical Or&$\approx 0.36-0.65$ &$\approx 0.66-0.76$  &$\approx 0.92-1.00$   \\
			 0.20&$x*y=$const.& 0.59& 0.83 &$ 1.00$\\
			 0.20&$x+y=$const.& 0.61& 0.83  &$ 1.00$ \\
		\end{tabular}
	\end{ruledtabular}
\end{table*}

Results are presented in Fig. \ref{fig:combinehistogram}, where 2D Probability density distribution of  $\{ln(\hat{\mathcal{L}}_{\nu}),\hat{\rho}_{GW}\}$ are shown for various CCSNe distances.

We first compare the 2D probability density distributions of the joint indicator for SASI and no-SASI, for fixed $D$. As expected, for all the 3 distances investigated here, the joint SASI indicator in SASI scenario is likely to locate in the upper right region of the panel, while the indicator in no-SASI scenario is likely to reside in the lower left corner. For example, for $D=1$ kpc, the indicator is most likely to locate in the region of $\{0.5, 0.4\}$ when SASI appears, while it is most likely to be near $\{0.0,0.0\}$ when SASI is absent in the signatures. 

We then consider how the 2D probability density distributions of the joint SASI indicator vary with the distance. As $D$ increases, the PDF of the SASI indicator becomes broader, due to the increasing importance of the statistical fluctuations, in both neutrino and GW signatures, which result in larger uncertainties of both $ln(\hat{\mathcal{L}}_{\nu})$ and $\hat{\rho}_\mathrm{GW}$. Consequently, the PDFs for SASI and no-SASI have increasing overlap, meaning that the detectability of SASI activities in this joint analysis decreases with increasing distance. Fig. \ref{fig:combinehistogram} shows another interesting trend: as $D$ increases, the maximum of the 2D PDF for the SASI scenario moves from  the $ln(\hat{\mathcal{L}}_{\nu})>\hat{\rho}_{\mathrm{GW}}$ region to the $ln(\hat{\mathcal{L}}_{\nu})<\hat{\rho}_{\mathrm{GW}}$ one. Such behavior reflects that the faster decline of the sensitivity in \ns\ with the distance compared to GW signatures. 

Given the 2D PDFs of $\{ln(\hat{\mathcal{L}}_{\nu}),\hat{\rho}_{\mathrm{GW}}\}$, we perform quantitative SASI-/no-SASI- scenario identification by constructing the receiver operating curves ($P^\mathrm{comb}_\mathrm{D}$ as a function of $P^\mathrm{comb}_{\mathrm{FI}}$) at different CCSNe distances, according to the prescriptions discussed in Sec. \ref{sec:method} (Eqs. \cref{eq:pdlogicAnd,eq:pfalogicAnd,eq:pdlogicOr,eq:pfalogicOr,eq:pdlogicPlus,eq:pdlogicCross}). 

The resulting curves are shown in Fig. \ref{fig:combineROC}. Note that for the ``logical And" and ``logical Or" prescriptions the 2D SASI thresholds are composed of two independent thresholds $\Lambda_\nu$ and $\Lambda_{GW}$. By varying $\Lambda_\nu$ and $\Lambda_{GW}$ independently, multiple ROC curves are found (because one specific $P^\mathrm{comb}_\mathrm{D}$ correspond to multiple $P^\mathrm{comb}_{\mathrm{FI}}$s), which form a ``receiver operating band" in the $P^\mathrm{comb}_\mathrm{D}$---$P^\mathrm{comb}_{\mathrm{FI}}$ plane. For comparison, in Fig. \ref{fig:combineROC} we also show the single-messenger ROCs (labeled $\nu$ and $GW$), from eqs. (\ref{eq:pdpfanu}) and  (\ref{eq:pdpfaGW}). 

From  Fig. \ref{fig:combineROC}, one can see the dependence of the ROCs on $D$. For $D=1$ kpc, the performance of the $\nu$-only ROC  is much better than the GW-only one, with the multi-messenger ROCs being intermediate between the two. However, as the CCSNe distance increases, the decline of the $\nu$-only ROC performance is much faster compared to GW-only. As $D$ increases, interestingly, we find that the strategy of jointly using neutrino/GW information could provide ROCs that perform better than both the single-messenger ROCs. For example, at 5 kpc, we observe that the $P_\mathrm{D}\approx 71\%$ at $P_\mathrm{FI}=10\%$ using the ``$x * y=const$" threshold in multi-messenger analysis, while the $P_\mathrm{D}\approx40-50\%$ at $P_\mathrm{FI}=10\%$ in single-messenger analysis. 

We summarized the $P_\mathrm{D}$ based on single-messenger as well as multi-messenger methods at $P_\mathrm{FI}=0.1$ and $P_\mathrm{FI}=0.2$ in Table. \ref{tab:pd}. As shown in Table. \ref{tab:pd}, when using the  ``$x * y=const$" threshold and the  ``$x + y=const$" threshold, $P_\mathrm{FI}=0.1(0.2)$ corresponds to a range of $P_\mathrm{D}$s, rather than a single $P_\mathrm{D}$. This is because when using these types of SASI-meter threshold, we obtained a ``receiver operating band", as explained above. Finally, the $P_\mathrm{D}$ at 10 kpc (with $P_\mathrm{FI}$=0.1, 0.2) and at 5 kpc (with $P_\mathrm{D}$=0.1) in GW channel cannot be determined accurately because of the difficulty of obtaining enough GW triggers and the resulting ``zig-zags" on GW ROC.

\section{Discussion and Conclusion}\label{sec:conclusion}

We have introduced a novel multi-messenger methodology (\emph{``SASI-meter"}) to identify and characterize the presence of SASI activities in a future core collapse supernova event that is detected with \ns\ and gravitational waves. 
 
For each messenger, the SASI-meter indicates the presence of SASI with a desired maximum false identification probability.
We study the effectiveness of the procedure with Receiver Operating Curves (ROC), which give the probability of establishing the presence of SASI for a  fixed false identification probability, that are tuned on generic properties of the GW and Neutrino CCSNe signatures
when the SASI is not present. 
The results are produced using numerical simulations with and without SASI-induced signatures at different distances from ground-based detectors.

The method also performs parameter estimation, by characterizing the features of the SASI oscillations such as oscillation amplitude, frequency, duration and starting time.

More explicitly, we characterize the pipeline with random realizations of reconstructed $\nu$ and GW signals for the test example. For the $\nu$, this means adding Poissonian fluctuations on 
	 signals that have SASI as well as signals where the SASI was removed. In the case of future detections, we can achieve the same result by taking a smoothed out version of the detected neutrino luminosities as the no-SASI $\nu$ signature, and randomize it with Poissonian fluctuations to identify the threshold for the desired $P_\mathrm{FI}$ used 
	 as a reference in this work for single channel or multiple channel identification mode (here $P_\mathrm{FI}$=0.1). For the GW channel,
	 in the illustrative example used in this paper, we inject the GWs with and without SASI in real interferometric noise.
	 In a realistic scenario, the no-SASI injections can be used to tune the threshold on the identification metric 
	 for the desired single or multimessenger $P_\mathrm{FI}$ (here $P_\mathrm{FI}$=0.1). The signals in the $\nu$ and GW channels with SASI are used
	 in this paper to characterize the performance of the GW-$\nu$ SASI meter.

We anticipate and observe a different scaling with distance in the frequency estimation uncertainties in the GW and neutrino channels because of the different dependence of the signal amplitude with respect to the source distance. The method is capable of accounting for an intrinsic uncertainty of the SASI frequency
due to, for example, the shock radius fluctuations of a progressive frequency drift like in
the case of spiral SASI. The frequency estimate in the neutrino channel uses the peak
frequency in the spectrum. In the GW channel, the estimated SASI frequency is a weighted average frequency among the
wavelet components in the SASI time frequency region. 
This concept of average is well defined even when the frequency drifts because of
shock radius fluctuations or other reasons. We also show estimates of the slope of the g mode in the GW channel. The estimation of the SASI duration in the GW channel might not be optimized yet. Similar identification performances are expected to happen at order of magnitude larger distances for third generation GW detectors.

For \ns\ we have elaborated on a previously presented maximum likelihood method. Our single messenger results show that for a galactic event the SASI can be identified with high confidence, and its main parameters can be estimated.

The single-messenger methods have been combined into a fully consistent multi-messenger SASI-meter, where a joint Receiver Operating Curve is found. The results confirm the power of multi-messenger astrophysics: they show that, for a typical galactic supernova (distance $D\gtrsim few$ kpc), a joint analysis can be more sensitive than each of the single-messenger ones, depending on the degree of optimization of the integration domain of the multi-dimensional probability distribution curves.

Given future galactic CCSNe observations, the SASI-meter calculates the quantities characterizing the strength of SASI-induced oscillations, namely $\{ln(\hat{\mathcal{L}}_\nu),\hat{\rho}_\mathrm{GW}\}$, based on observed neutrino and GW data in time-frequency domain. The $\{ln(\hat{\mathcal{L}}_\nu),\hat{\rho}_\mathrm{GW}\}$ based on observations can be compared with theoretical predictions. In this way, the results of the SASI-meter could be used to validate and interpret future CCSNe numerical simulations. 
 The detailed calibration procedure is not discussed in this work and will be left for future investigations.

In this work we use the results of a representative but specific numerical simulation. As more simulations become available with both neutrino and GW signatures, we will repeat the SASI-meter analysis. We expect that the thresholds to achieve at given $P_\mathrm{FI}$, will change weakly when more models are included in the analysis, while the identification range will change depending on the relative amplitude of the SASI.

Future works may include: 1) generalizing the GW SASI-meter to other wavelets bases (in the future the method could be tested in those directions as well); 2) applying the SASI-meter method on a model with fast rotating progenitor (note that KKHT model has a non-rotating progenitor). The PDF of $\{ln(\hat{\mathcal{L}}_\nu),\hat{\rho}_\mathrm{GW}\}$ in SASI scenario might be sensitive to the observational direction with respect to the rotating axis of CCSNe. A direction-dependent SASI activity indicator $\{ln(\hat{\mathcal{L}}_\nu),\hat{\rho}_\mathrm{GW}\}$  from simulations can then be compared with observations. In this way, the joint SASI-meter may help to identify the rotating axis of the CCSN in a quantitative way.  

\section*{Acknowledgments}

ZL acknowledges support from the National Science Foundation grant number PHY 21-16686 and AST 19-09490 and from DOE Scidac Grant No. DE-SC0018232. MZ acknowledges support from  the National Science Foundation grant number 2110555. CL's work was supported by the National Science Foundation grant number PHY-2012195. We acknowledge LIGO and the Gravitational-Wave Open Science Center (GWOSC) for the open source LIGO noise data from $O3$ run \cite{LIGOScientific:2022enz} and the computing resources that were used.
We are grateful to K.~Kotake, A.~Mezzacappa and M.~Szczepańczyk for many fruitful discussions.

\appendix
\section{Neutrino analysis: optimization of the SASI initial time and duration}\label{appendix:neutrino}

In the SASI-meter method illustrated above, the duration $\tau$ and starting time $t_0$ are fixed before the process of identifying the existence and analyzing the features of the SASI. Let us now discuss an aspect which is new of this work, namely that we do not assume prior values of $t_0$ and $\tau$ (as was done in \cite{Lin:2019wwm}), but rather analyze the output of the \kur\ simulation for their extremal values. Following the method suggested in \cite{Lin:2022lck}, the process of SASI identification described above is performed repeatedly for various time series $[t_0,t_0+\tau]$. The extremal interval $[t_0,t_0+\tau]$ for the detection of the SASI can be found by exploring all the possible pairs  $(t_0,\tau)$.  
The starting time $t_0$ is varied in the range  $50-210$ ms, with a time step of $20$ ms. The duration $\tau$ is varied within $20-70$ ms, with a time step of $10$ ms. For each pair  $(t_0,\tau)$, the interval $[t_0,t_0+\tau]$ is  used to analyze the detectability as well as the features of SASI\footnote{Note that Eq. (\ref{eq:tau}) implies that longer $\tau$ will give finer resolution of the frequency, which means when we are varying $\tau$ the $\delta$ correspondingly changes.}. The detection probability, $P_\mathrm{D}$, is calculated (for a fixed $P_\mathrm{FI}$). 
The optimal values of $t_0$ and $\tau$ are identified as those for which $P_\mathrm{D}$ is maximum. Indeed, if the SASI-induced fluctuations predicted by the \kur\ SASI model are present in $[t_0,t_0+\tau]$, then the PDF of $\mathcal{L}$ from \kur\ SASI model and that from \kur\ no-SASI model would deviate from each other. Thus, the resultant $P_\mathrm{D}$ would be high. Otherwise, the $P_\mathrm{D}$ would be low.

\begin{figure*}[htp]
	\includegraphics[width=0.45\textwidth]{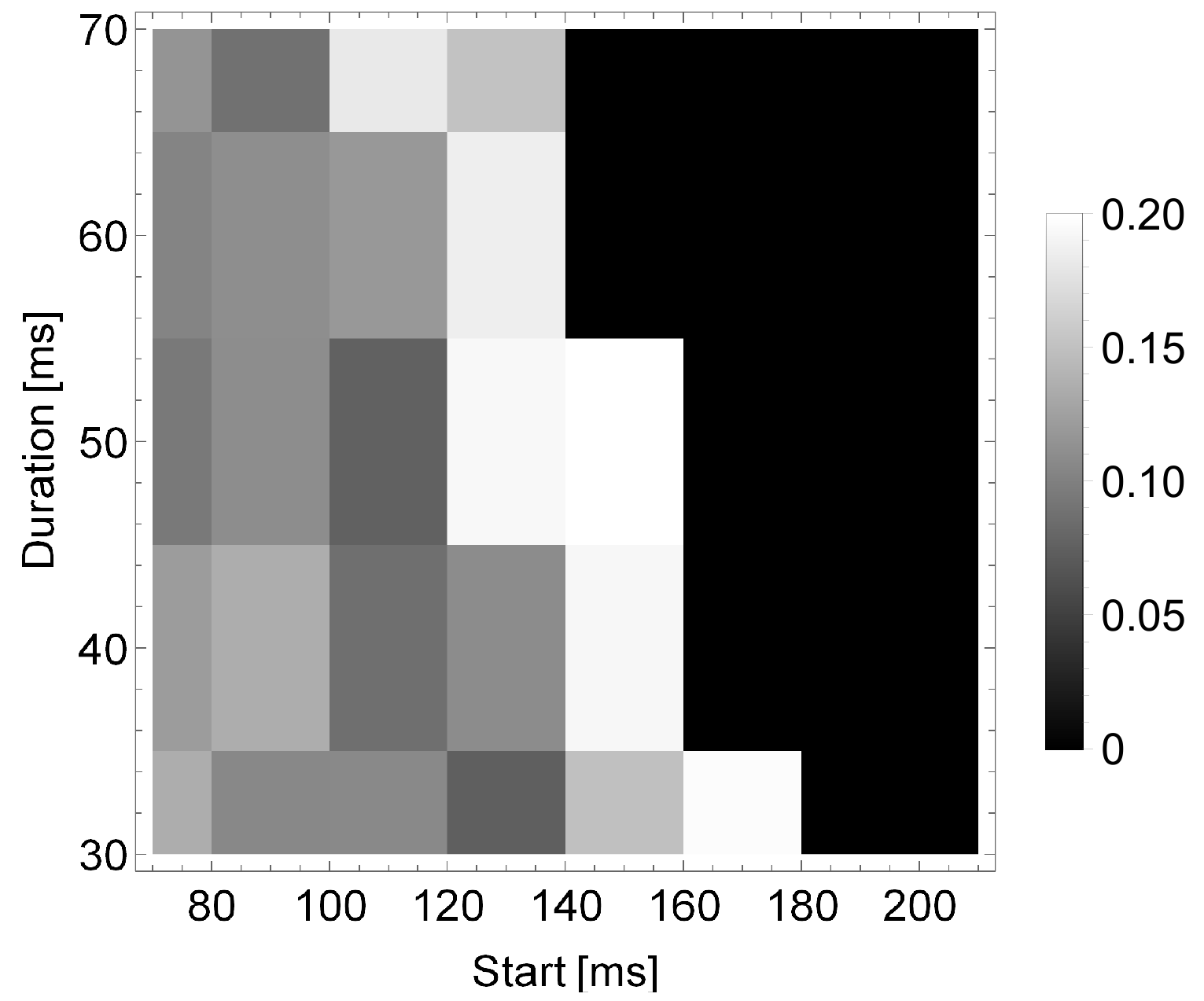}
	\includegraphics[width=0.45\textwidth]{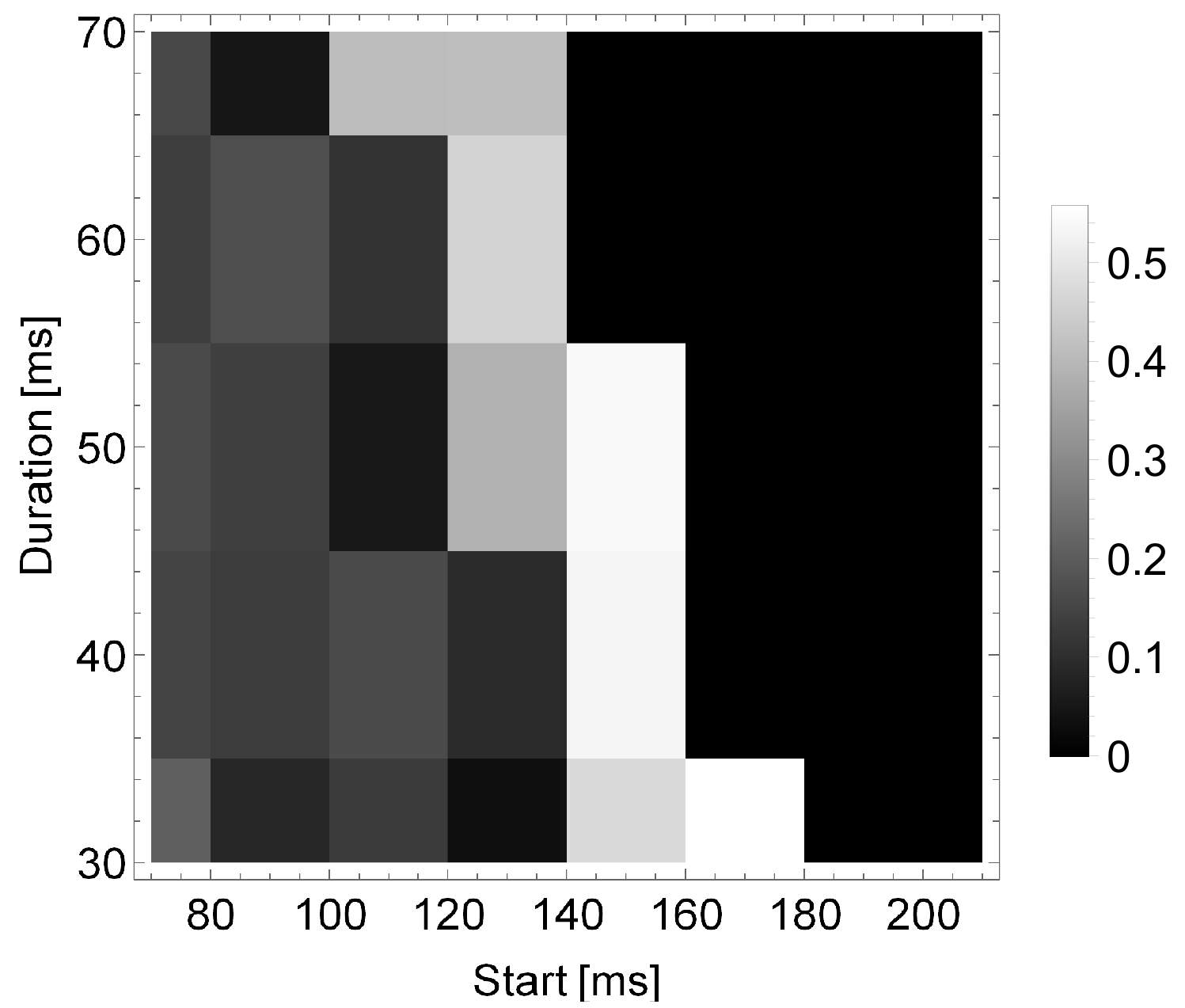}
	\includegraphics[width=0.45\textwidth]{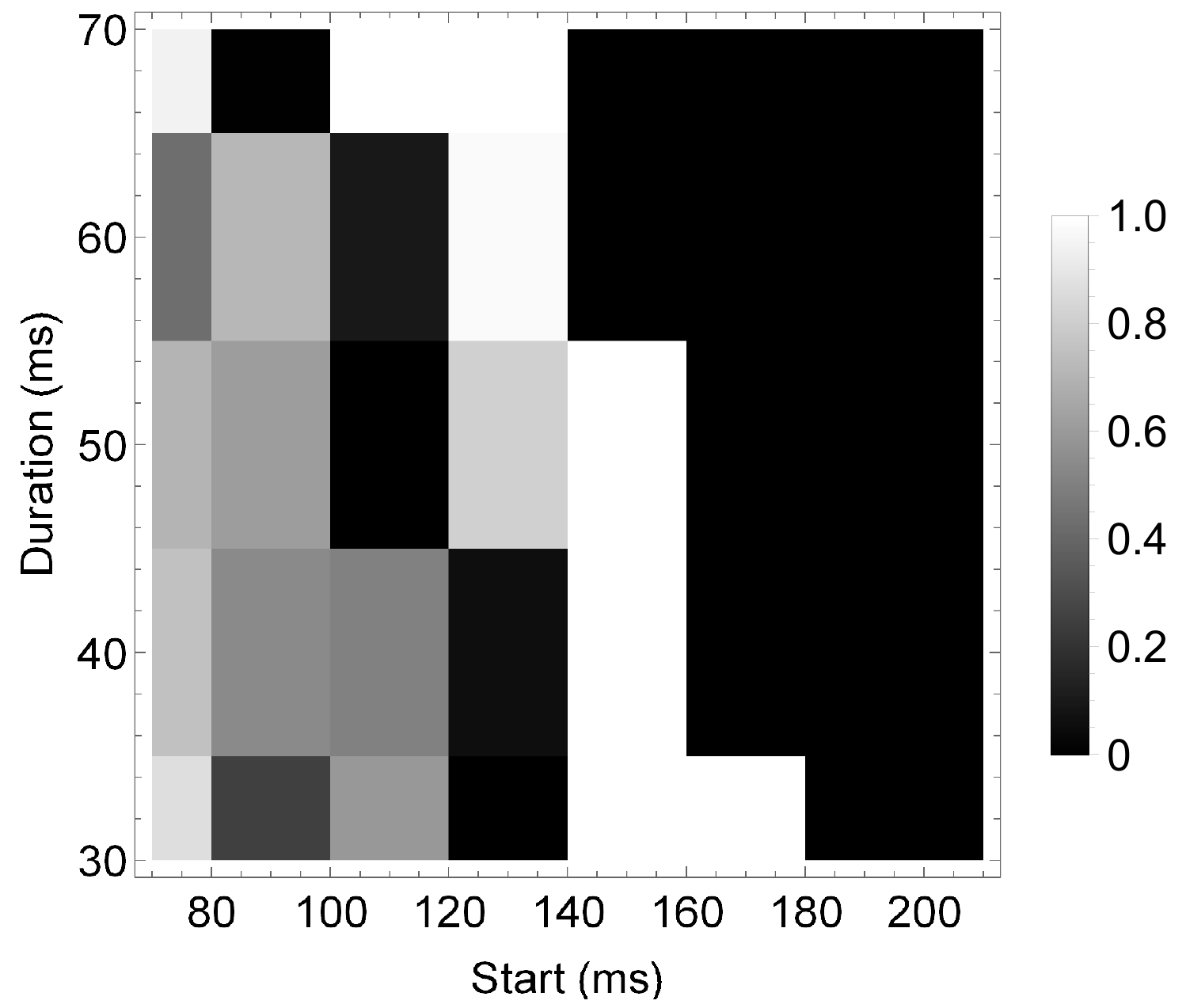}
	
	\caption{Probability of detection $P_\mathrm{D}$ of SASI at false alarm rate $P_\mathrm{FI}=10\%$. The $P_\mathrm{D}$s are evaluated in neutrino time series with different duration and starting time. The upper left, upper right and lower panels are $P_\mathrm{D}$s with SNe distance being 10, 5 and 1 kpc.   }
	\label{fig:pdmap}
\end{figure*}

By applying the SASI-meter on neutrino signatures of $[t_0,t_0+\tau]$, the $P_\mathrm{D}$ and the $P_{\mathrm{FI}}$ in a specific neutrino time series is found. In Fig. \ref{fig:pdmap}, at fixed $P_{\mathrm{FI}}=10\%$ we plot $P_\mathrm{D}$ in neutrino times series of $[t_0,t_0+\tau]$, where $t_0$ ($\tau$) varies from 80 (30) ms to 200 (70) ms. We found the $P_\mathrm{D}$ obviously increases when the neutrino time series of $[t_0, t_0+\tau]$ (partly) overlaps the SASI region predicted by KKHT model. For example, at CCSNe distance $D=1~ \mathrm{kpc}$, the $P_\mathrm{D}\approx 1.0$ in neutrino time series of $[110 ~\mathrm{ms}, 110+70 ~\mathrm{ms}]$, $[130 ~\mathrm{ms}, 130+60 ~\mathrm{ms}]$, $[130 ~\mathrm{ms}, 130+70 ~\mathrm{ms}]$, $[150 ~\mathrm{ms}, 150+30 ~\mathrm{ms}]$, $[150 ~\mathrm{ms}, 150+40 ~\mathrm{ms}]$, $[150 ~\mathrm{ms}, 150+50 ~\mathrm{ms}]$, and $[170 ~\mathrm{ms}, 170+30 ~\mathrm{ms}]$, and the SASI region predicted by KKHT model is approximately in $[150 ~\mathrm{ms}, 150+50~\mathrm{ms}]$. In neutrino time series that ends after 200 ms, the $P_\mathrm{D}=0.0$. This is because the simulation of CCSNe in KKHT model was truncated after 200 $\mathrm{ms}$. Finally, for neutrino time series ending before 150 ms, $P_\mathrm{D}\ll 1.0$, indicating that the SASI-induced neutrino oscillations cannot be identified in these time series. The $P_\mathrm{D}$ distributions at different distances are qualitatively similar.

As one may notice, the extremal $(t_0,\tau)$ defines the region where the SASI activity in KKHT model has highest probability to be verified against observations. 
Naturally, this interval may not adequately represent the starting time and the duration of the SASI activities residing in \emph{observed} neutrino events. In fact, the comparison with the true, observed, values of $t_0$ and $\tau$ is a test of the model. Such true values can be roughly measured in the SASI-meter method, as follows. 
First, note that the SASI modulation on neutrino and GW emissions is predicted as a quasi-periodic signature and its oscillation frequency remains almost constant for tens of $\mathrm{ms}$. Thus, when exploring the $\tilde{\Omega}_{SASI}$ for various points $(t_0, \tau)$ in the parameter space we find that for all the time intervals $[t_0,t_0+\tau]$ that include SASI-induced oscillations at least partially, the estimated SASI frequency, $\tilde{f}_S$, is approximately the same, representing the \emph{``monochromatic"} feature of SASI-induced oscillations. By identifying the region of the parameter space where  $\tilde{f}_S$ stays constant, we can find the SASI region in time domain and estimate $t_0$ and $\tau$, without relying on a model. In this way, the SASI-meter can give an approximate estimation of SASI duration and starting time. Such estimation is necessarily rough, less accurate than the measurement for $f_S$ and $a$, since the SASI duration/starting time are features in time domain, while the SASI-meter is designed mainly for analyzing SASI features in frequency domain.

\begin{figure*}[htp]

	\includegraphics[width=0.45\textwidth]{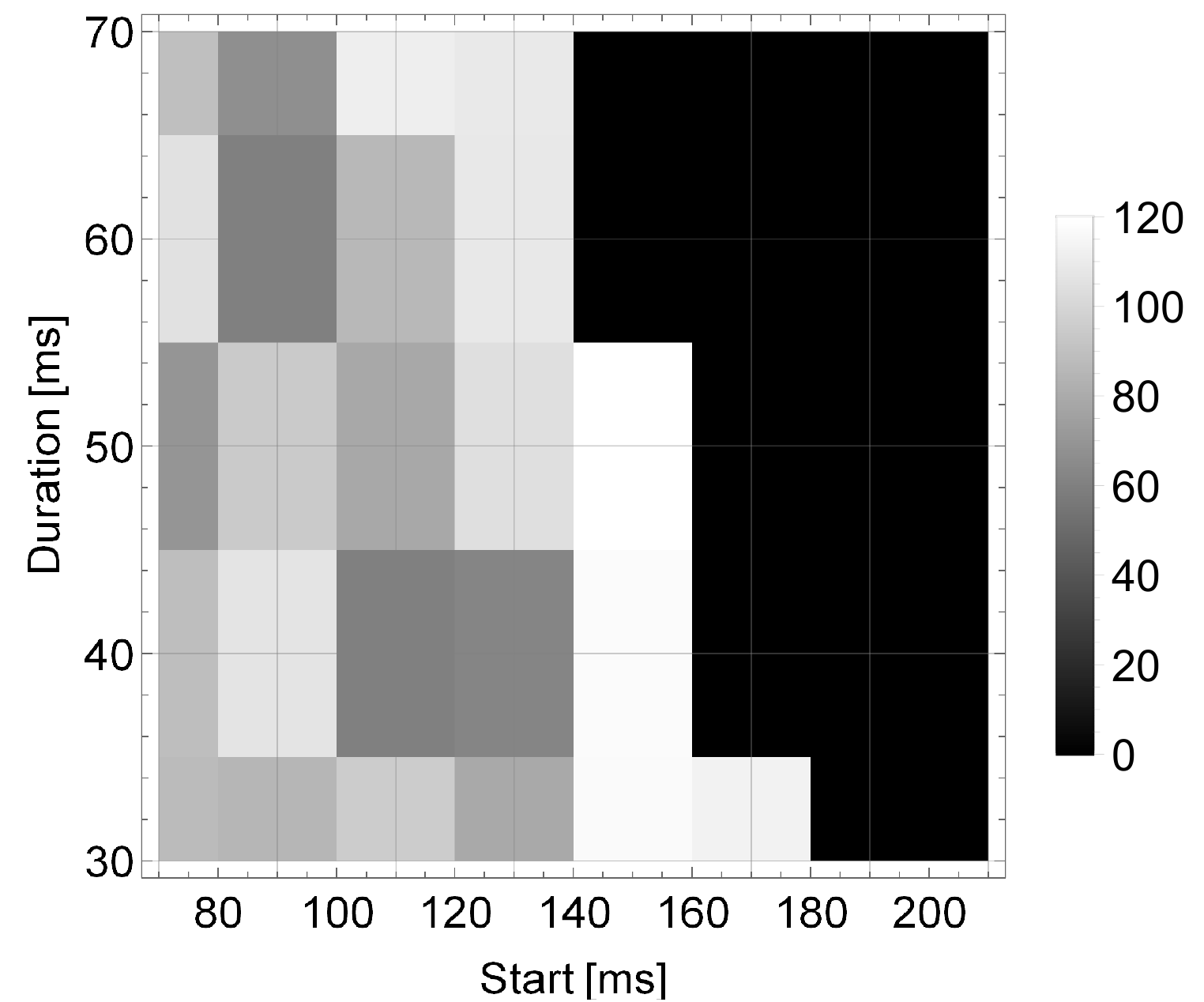}
	\includegraphics[width=0.45\textwidth]{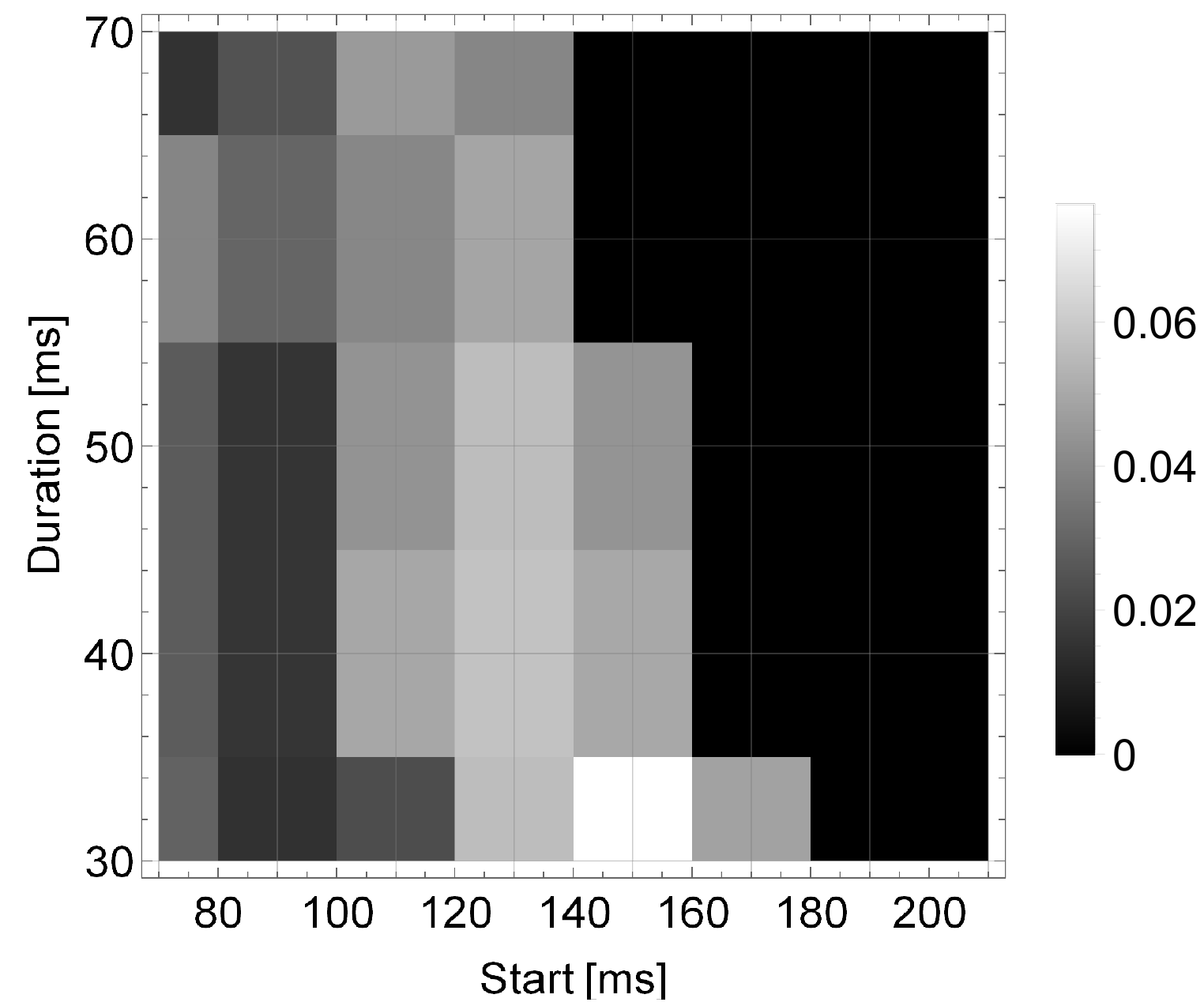}
	
	\caption{Averaged optimal frequency(left) and amplitude (right) evaluated in neutrino time series with different duration and starting time. The corresponding distance to CCSNe is 1 kpc.  }
	\label{fig: parametermap}
\end{figure*}

In Fig. \ref{fig: parametermap}, the extremal frequency that maximizes the $\mathcal{L}$ in various neutrino time series were plotted. Note that for time series of $[150 ~\mathrm{ms}, 150+30 ~\mathrm{ms}]$, $[150 ~\mathrm{ms}, 150+40 ~\mathrm{ms}]$, $[150 ~\mathrm{ms}, 150+50 ~\mathrm{ms}]$, the corresponding optimal frequencies are approximately identical, indicating that a monochromatic SASI oscillation exists in these periods. Since the determination of optimal SASI oscillation parameters does not require a \emph{prior} PDF of $\mathcal{L}$, we can \emph{model-independently} identify the neutrino time series with SASI-induced oscillations and measure the corresponding oscillation frequency given the observation of neutrino signatures. For example, at 1 kpc, the fact that the optimal frequency in $[t_0,t_0+\tau]$ with $t_0<150~\mathrm{ms}$ is different from those in $[150 ~\mathrm{ms}, 150+30 ~\mathrm{ms}]$, $[150 ~\mathrm{ms}, 150+40 ~\mathrm{ms}]$, and $[150 ~\mathrm{ms}, 150+50 ~\mathrm{ms}]$ suggest that the SASI activity happens no sooner than $\approx150~\mathrm{ms}$. Additionally, the fact that the optimal frequencies in $[150 ~\mathrm{ms}, 150+30 ~\mathrm{ms}]$, $[150 ~\mathrm{ms}, 150+40 ~\mathrm{ms}]$, and $[150 ~\mathrm{ms}, 150+50 ~\mathrm{ms}]$ are approximately the same indicate that the SASI activity lasts for $\gtrapprox50~\mathrm{ms}$.

\section{GW analysis: initial processing and g-mode removal}\label{appendix:GW1}
 In this appendix, we define the different GW parameters and the method used by cWB to produce the likelihood time-frequency maps, describe the GW data we're using, the re-tuning of some cWB parameters we performed, and the g-mode slope estimation for the removal of the g-mode region in order to define the SASI dominant region (or the SASI region) where we perform the SASI parameter estimation.

 \subsection{Generation and processing of the time-frequency (TF) maps}

For a single detector (with equal arms of length $l$), the response $X(t)$ is the sum of detector noise $n(t)$ and the GW signal contribution $\xi(t)$; 
\begin{equation}
    X(t)=n(t)+\xi(t)
\end{equation}
where, $\xi(t)$ depends on the absolute difference in the change in length of the two arms $\delta l_x(t)$ and $\delta l_y(t)$ relative to the original length:

    $$\xi(t)=\frac{|\delta l_x(t)-\delta l_y(t)|}{l}=F_+ h_+(t)+F_\times h_\times (t)$$
where, $h_+(t)$ and $h_\times (t)$ are the plus and cross polarization components of the GW and $F_+$ and $F_\times$ are the respective antenna patterns. $\xi$ can also be expressed as
\begin{equation}
\xi= \zeta\cdot \tilde{A}+\tilde{\zeta}\cdot A
\end{equation}
where $\zeta=h_++ih_\times$ and $A=\frac{1}{2}(F_++iF_\times)$.

Since the response of the interferometer, or the detector data, is in the form of a time series $X=\{x[0], x[1], ..., x[I]\}$ that may or may not contain GW signal, a decision has to made for the presence (hypothesis $H_1$) or absence (hypothesis $H_0$) of the GW signal described by the two probability densities $p(x|H_1)$ and $p(x|H_0)$, respectively. Any decision rule would then be based on a threshold applied to these densities. For this, we define the likelihood ratio $\Lambda(x)$ as

\begin{equation}
    \Lambda(x)=\frac{p(x|H_1)}{p(x|H_0)}
\end{equation}

In GW data analysis, due to the standard whitening procedure performed by cWB, we can assume the noise to be a zero mean temporarily uncorrelated Gaussian process with standard deviation $\sigma$ (non-Gaussian noise components are managed separately).   
The probability density functions associated to the two hypotheses $H_0$ and $H_1$
become
\begin{equation}
    p(x|H_0)=\prod_{i=1}^I \frac{1}{\sqrt{2\pi}\sigma}e^{-\bigg(\frac{x^2[i]}{2\sigma^2}\bigg)}
\end{equation}
\begin{equation}
    p(x|H_1)=\prod_{i=1}^I \frac{1}{\sqrt{2\pi}\sigma}e^{-\bigg(\frac{(x[i]-\xi[i])^2}{2\sigma^2}\bigg)}~.
\end{equation}

Then, the logarithmic value of the likelihood ratio, which we simply call likelihood, is

\begin{equation}  
\begin{split}
    \rho&=ln(\Lambda(x))=ln\bigg(\prod_{i=1}^I e^{\bigg(\frac{1}{\sigma^2}(x[i]\xi[i] - \frac{1}{2}\xi^2[i])\bigg)}\bigg)\\&=\sum_{i=1}^I \frac{1}{\sigma^2}\bigg(x[i]\xi[i] - \frac{1}{2}\xi^2[i]\bigg)~.
\end{split}
\end{equation}
For $N$ detectors, $\sigma=\{\sigma_1, \sigma_2,...,\sigma_N\}$, 
    $\xi=\{\xi_1, \xi_2,...,\xi_N\}$, 
    $F_+=\{F_{+1}, F_{+2},...,F_{+N}\}$, 
    $F_\times=\{F_{\times 1}, F_{\times 2},...,F_{\times N}\}$, and
    $A=\{A_1,A_2,...,A_N\}$ and the total likelihood for the $N$ detectors becomes
\begin{equation}
\label{eq:Likelihood}
    \rho=\sum_{k=1}^{N}\sum_{i=1}^I \frac{1}{\sigma_k^2}\bigg(x_k[i]\xi_k[i] - \frac{1}{2}\xi_k^2[i]\bigg)
\end{equation}

If we introduce

\begin{eqnarray}
    f_+=\bigg\{\frac{F_{+1}}{\sigma_1}, \frac{F_{+2}}{\sigma_2},...,\frac{F_{+N}}{\sigma_N}\bigg\}\\
    f_\times=\bigg\{\frac{F_{\times 1}}{\sigma_1}, \frac{F_{\times 2}}{\sigma_2},...,\frac{F_{\times N}}{\sigma_N}\bigg\}\\
    A_\sigma=\bigg\{\frac{A_1}{\sigma_1},\frac{A_2}{\sigma_2},...,\frac{A_N}{\sigma_N}\bigg\}\\
    g_c=\sum_{k=1}^{N} \frac{A_k^2}{\sigma_k^2}
\end{eqnarray}
where, $g_c$ is the network antenna pattern, the transformation $g_c \rightarrow g_c'$ which makes the imaginary part of $g_c'$ to vanish transforms to the Dominant Polarization Frame (DPF). If $g_c = |g_c|e^{2i\gamma}$, the
transformation of $A_k$ is, $A_k' = A_ke^{-i\gamma}$ and thus the normalized antenna patterns transform as 
\begin{equation}
    f_{k+}'=f_{k+}cos(\gamma)+f_{k\times}sin(\gamma)
\end{equation}
\begin{equation}
    f_{k\times}'=-f_{k+}sin(\gamma)+f_{k\times}cos(\gamma)~.
\end{equation}

The unitary vectors of the DPF are expressed as; $e_+'=\frac{f_+'}{|f_+'|}$ and $e_\times'=\frac{f_\times'}{|f_\times'|}$. It can be shown, after transforming to the DPF, where the plus and cross antenna patterns are orthogonal, and assuming they have the same magnitude, the maximum likelihood value, after applying the conditions $\frac{\delta \rho}{\delta h_+}=0$ and $\frac{\delta \rho}{\delta h_\times}=0$, is (see \cite{drago}):
\begin{equation}
\label{eq:Likelihood_mn}
    \rho_{max}=(X\cdot e_+')^2+(X\cdot e_\times')^2= \sum_k \frac{\xi_k^2}{\sigma_k^2}=\sum_k SNR_k^2 ~,
\end{equation}
where the SNR of the $k$th detector, calculated from detector
response $\xi_k$ and noise variance $\sigma_k$, is expressed as $SNR_k^2=\frac{\xi_k^2}{\sigma_k^2}$.
We can see that the maximum likelihood 
is the sum of squared SNR of the detectors.

\subsection{GW data:}
For the processing of the interferometric data, we apply the CCSNe configuration of the coherent WaveBurst (cWB) algorithm, in the 16 to 2048 Hz band, to detect and reconstruct the CCSNe GW signals, after injecting those signals on LIGO noise from $O3$ run \cite{LIGOScientific:2022enz} using data from detectors H1 and L1 (GPS times 1256652800 to 1269563392). In order to account for the time variability of the noise, we prepare hundreds of injected events (triggers), separately for each distances of 1 Kpc, 5 Kpc and 10 Kpc. We use our prepared SASI and no SASI waveforms (injections), with amplitudes scaled according to the distance, and the injections are added to the noise at different times.  

\subsection{Cluster formation, likelihood time-frequency maps and cWB parameter tuning:}
     Coherent WaveBurst is an excess-power search algorithm for detecting and reconstructing GWs based on a constrained likelihood formalism \cite{Klimenko:2015ypf}. The analysis of GW strain data is performed in a wavelet
domain \cite{Necula:2012zz} using the Wavelet Transform, a tool that transforms the signal into time-frequency domain. First, cWB performs data conditioning on the calibrated strain data by applying a Linear Prediction Error (LPE) filter to remove
“predictable” components from the time series, for instance lines of stationary
noise. The LPE filter and whitening is applied in the wavelet domain individually for each wavelet
layer\cite{Klimenko:2015ypf}. 

The cWB algorithm uses Wilson-Daubechies wavelets, resulting in 2D maps that are formed by wavelet components (pixels) with different time-frequency resolutions.

Since the wavelet decomposition is performed through seven wavelet basis in parallel, the time and frequency resolution of each pixel is in general different, as seen in Fig. \ref{fig:scalogram}. The cWB 
performs wavelets/pixels selection. 
Wavelets with amplitudes above a threshold, designed to spare only a small percentage of the noise induced ones (black pixel probability, (bpp)), are retained in each frequency. 
Each wavelet component is defined by its time, frequency, likelihood, and time-frequency resolution. The frequency resolution is inversely related to the time resolution.

For the time-frequency position $i,j$ in the time-frequency plane (after wavelet decomposition), with time resolution '$\delta t$', the resulting signal is expressed as $x(i,j,\delta t)$. From equation \ref{eq:Likelihood}, the Likelihood thus needs to be defined over both time and frequency as (see \cite{drago})

\begin{equation}
\begin{split}
     \rho_c&=\sum_{ij}\sum_{k=1}^N \frac{1}{\sigma_k^2(i,j)}\bigg(x_k(i,j,\delta t_k)\xi_k(i,j,\theta,\phi) - \frac{1}{2}\xi_k^2(i,j,\theta,\phi)\bigg)\\&=\sum_{ij}\rho(i,j,\theta,\phi)
\end{split}
\end{equation}
where, the likelihood functional $\rho(i,j,\theta,\phi)$, defined over time-frequency positions $i,j$ and angles $\theta, \phi$ referring to the source coordinates, and expressed as

\begin{equation}
\begin{split}
     \rho(i,j,\theta,\phi) &=\sum_{k=1}^N \frac{1}{\sigma_k^2(i,j)}\\&\times\bigg(x_k(i,j,\delta t_k)\xi_k(i,j,\theta,\phi) - \frac{1}{2}\xi_k^2(i,j,\theta,\phi)\bigg)~.
\end{split}
\end{equation}

The maximum likelihood statistic, for a given location $(i, j)$, is then determined by maximizing the
likelihood functional over the source coordinates $\theta, \phi$ ($\rho_m(i, j)$). The likelihood
time-frequency map (LTF), a pixel map, is then obtained from the set of these maximum likelihood values for each time and frequency. Coherent clusters are formed from these pixels, which are formed by selecting pixels with the maximum likelihood $\rho_m(i, j)$ greater than a chosen threshold, and it is
composed of pixels belonging to all detectors involved in the network.
Final clusters are used to reconstruct the gravitational wave signal.

 The cWB pipeline is divided into two stages: the coherent event generator and the post-production analysis. After the event generation, as explained above, the resulting data is stored as the output trigger files and the post-production stage deals with the selection of the optimal set of statistics. Some of the post-production metrics are explained below.\\ 

The likelihood is a quadratic form that can be expressed in a matrix $[\rho_{mn}]$. And according to eq. \ref{eq:Likelihood_mn} we have
\begin{equation}
   \rho_{max}=(X\cdot e_+')^2+(X\cdot e_\times')^2~.
\end{equation}
Expanding the dot product in N-dimensional space (where the m, n indices refer to the detector number), we get :

\begin{equation}
\begin{split}
    \rho_{max}&=\sum_{mn}\{(X_m e_{+m}')(X_n e_{+n}')+(X_m e_{\times m}')(X_n e_{\times n}')\}\\&=\sum_{mn}\rho_{mn}~.
\end{split}
\end{equation}

We define coherent network energy, or simply coherent energy ($E_c$), as the sum of the off-diagonal terms of the Likelihood matrix:  

\begin{equation}
    E_c=\sum_{m\neq n} \rho_{mn}
\end{equation}
  
  The null energy ($N_{ull}$), which is the total reconstructed energy of noise, is defined as; 
  
  \begin{equation}
      N_{ull}=|X-\xi|^2
  \end{equation}

  The network correlation coefficient ($CC$), which is an estimate of  coherence among different interferometers, is defined as;
  
  \begin{equation}
  \label{CC_eq}
      CC=\frac{E_c}{E_c+N_{ull}}
  \end{equation}

  Real gravitational-wave events are expected to have $CC$ closer to $1$, and noise events are expected to have $CC << 1$.
  
  Thresholds on the event metrics are applied. For example, we applied a threshold on the correlation coefficient $CC$ 
  and a threshold on $Z$, which is the effective correlated 
  SNR of a trigger based on $e_c$, see \cite{Klimenko:2015ypf}. The relationship between $Z$ and $CC$ is given by
  
  \begin{equation}
      Z=\sqrt{\frac{e_c}{N}CC}~,
  \end{equation}
  
  where the reduced correlated energy $e_c$ is defined as
  
  \begin{equation}
    e_c=\sum_{m\neq n} \rho_{mn}|r_{mn}|~,
\end{equation}
  
   where $m,n=1,2,...,N$, refer to the detector number. 
   The network correlation coefficient $r_{mn}$, obtained by cross-correlating detectors data, is defined as
  \begin{equation}
      r_{mn}=\frac{\rho_{mn}}{2\sqrt{\rho_{mm}}\sqrt{\rho_{nn}}}~.
  \end{equation}

\begin{figure*}[htp]
	\includegraphics[width=0.45\textwidth]{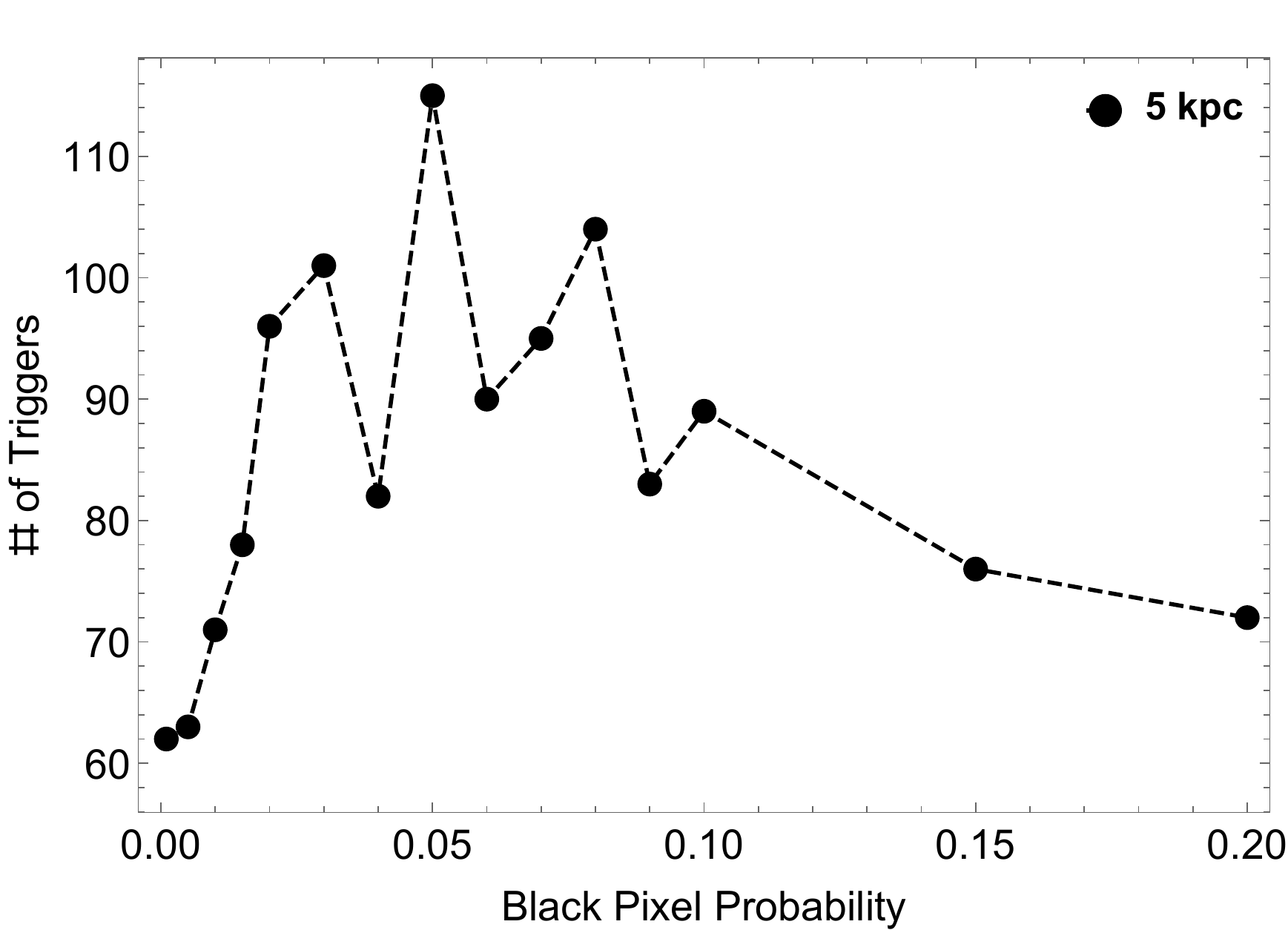}
    \includegraphics[width=0.45\textwidth]{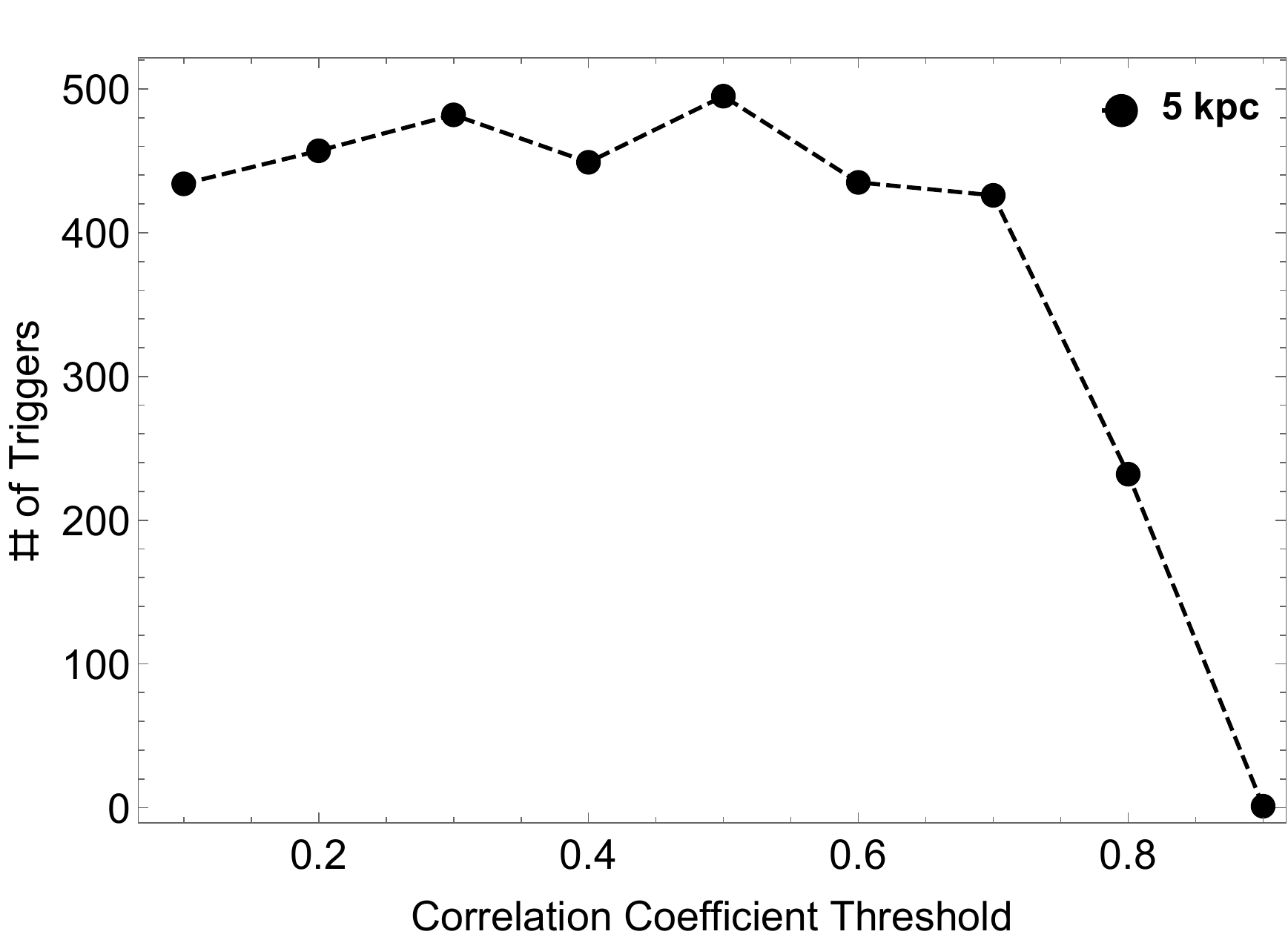}
    \caption{Showing the optimal choice of bpp(left) and CC threshold(right) at 5 Kpc in terms of maximization of the number of triggers produced by cWB, while other cWB parameters remain unchanged. cWB arranges a number of triggers into job files that were executed individually. Here, the analysis was carried out for 10 such jobs for each bpp value and 50 jobs for each CC value shown in the graph.}
    \label{fig:bpp}
\end{figure*}

The traditional tuning process of cWB focused on producing the best ROC for the detection of the overall signal and not the identification of a specific feature in the signal, like the SASI. For this reason, we redid some of the tuning. Here, the focus is to maximize the chances an event is reconstructed since otherwise it is not possible to identify the presence of the SASI, when present. For the cWB internal parameters optimization, we have executed different cwb jobs at different bpp and CC thresholds to find the set of values that give us the maximum number of triggers. As we can see from fig \ref{fig:bpp}, the optimal bpp to be used is 0.05 and the optimal cc threshold to be used is 0.5.  However, exploring a systematic optimization of all the cWB thresholds is beyond the scope of this study.

\subsection{g-mode location, parameter estimation and removal}  

The growth of the frequency of the fundamental mode of oscillation of the PNS in the GW spectrogram is the main feature that all the CCSNe numerical simulations display. Estimating the g-mode slope has merit by itself, but for the detection of the SASI its contribution in the GW event is not necessarily useful as its contribution to the overall SNR does not make the total SNR a good indicator for the presence of SASI. In this regard, the g-mode location and estimation in this work is considered sufficiently well-performed in terms of benefits to the SASI ROC curves. The optimization of the g-mode parameters is the topic of future publications. We locate the g-mode region, in terms of start time and initial slope, and later, remove the pixels from that part of the event to focus on the SASI pixels. It is also possible that the statistically more significant wavelet components induced by noise
are scattered in the time frequency plane (but in proximity of the GW event). A second mechanism where energy could percolate in the SASI region is an impulsive stimulation of the PNS. This could happen, for example from an unusually large accretion funnel of material, 
 inducing a broad band GW spike that also contains lower frequencies \cite{Mueller:2012sv, Murphy_2009}. Also, there is a possibility that 
 a CCSN explosion could have turbulent components containing some GW energy as well in the SASI region. Nevertheless, the metric introduced in equation (\ref{eq:SNR_norm}) allows to produce probability distributions where the two scenarios can be distinguished (see for example fig  \ref{fig:GWOnlyResults}).

We apply a (Python) code to process wavelet maps for all the reconstructed triggers to remove the g-mode contribution in those triggers, detect SASI and estimate its frequency and duration.  For the g-mode parameter estimation, we remove pixels below 200Hz, because there's no significant energy of the g-mode component there but the SASI component maybe present which can affect our g-mode parameter estimation. Since the g-mode slope estimation is affected by the noise induced pixels, we discard pixels with likelihood below the event-dependent arithmetic mean of likelihood of the event. Next, as the g-mode is one of the most energetic features in the GW event, from the surviving pixels, we only choose pixels in an interval of 0.2 seconds (roughly twice the visible g-mode duration for most waveforms) identified as the 0.2 seconds interval containing the most energetic pixels for the g-mode parameter estimation. Such an interval is selected by comparing the sum of likelihood of the pixels, as the measure of total energy, within all the 0.2 seconds intervals in the event and choosing the one with the maximum value. 
         
If less than two pixels survive, we do not perform slope estimation and thus the following steps for g-mode slope estimation will be bypassed and the next steps for SASI parameter estimation will be implemented (see \ref{sec:GW}). 
 The preliminary estimation of the slope and intercept of the g-mode line is performed using linear regression on the remaining pixels. For the triggers with estimated slope outside the physical range $(500, 5000) s^{-2}$ identified from the literature review, the slope-intercept optimization process mentioned below is bypassed, like before, and the next steps for SASI parameter estimation will be implemented (see \ref{sec:GW}). The statistics of the number of triggers for which g-mode slope estimation was not bypassed was observed to be 807 out of 1081 at 1 Kpc, 671 out of 2867 at 5 Kpc, and 357 out of 2764 at 10 Kpc. The estimation of the slope is performed in two steps. The initial one involves a simple linear regression. The estimation is then refined by defining intervals around those values (200 points were taken around the initial estimates, with a relative variation of $<33\%$ with respect to the initial points). In this way, we construct a grid in the slope-intercept plane and use this set of values to minimize the following  weighted $\chi^2$ function:

          \begin{eqnarray}
          \chi^2 = \sum_i \rho^i(f_c^i)^2\Xi^i\bigg\{t_c^i - \frac{(f_c^i-c)}{m}\bigg\}^2~.
\end{eqnarray}

Here, $t_c^i$ is the centre time of $i^{th}$ pixel. The $f_c^i$ is the central frequency of the $i^{th}$ pixel. The $(f_c^i)^2$ is a weight function used to compensate for the LIGO noise curve. It is necessary because the LIGO noise increases with frequency, resulting in worse SNR at higher frequencies. The
$\rho^i$ is another weight function and is the likelihood value of the $i^{th}$ pixel. By using $\rho^i$, we take the energy of the pixels into consideration. This is because the pixels corresponding to signal would be more energetic in general than those corresponding to noise. Finally, since the pixels related to signals are most likely to be clustered together and the pixels related to noise are most likely to be isolated in the t-f maps, we introduce a quantity $\Xi^i$ in the weights,  
 which measures the density of pixels around the $i^{th}$ one. It is calculated as the sum of the likelihoods in the neighbourhood of the pixel in a window of 0.01s time interval and 25 Hz frequency interval centred around the given pixel in the t-f map.

The summation is done over all the pixels remaining in the area of interest of the trigger for all the elements in the grid of possible slopes i.e, $m$ and intercepts i.e, $c$.

\begin{figure}[ht]
	\centerline{\includegraphics[width=0.45\textwidth]{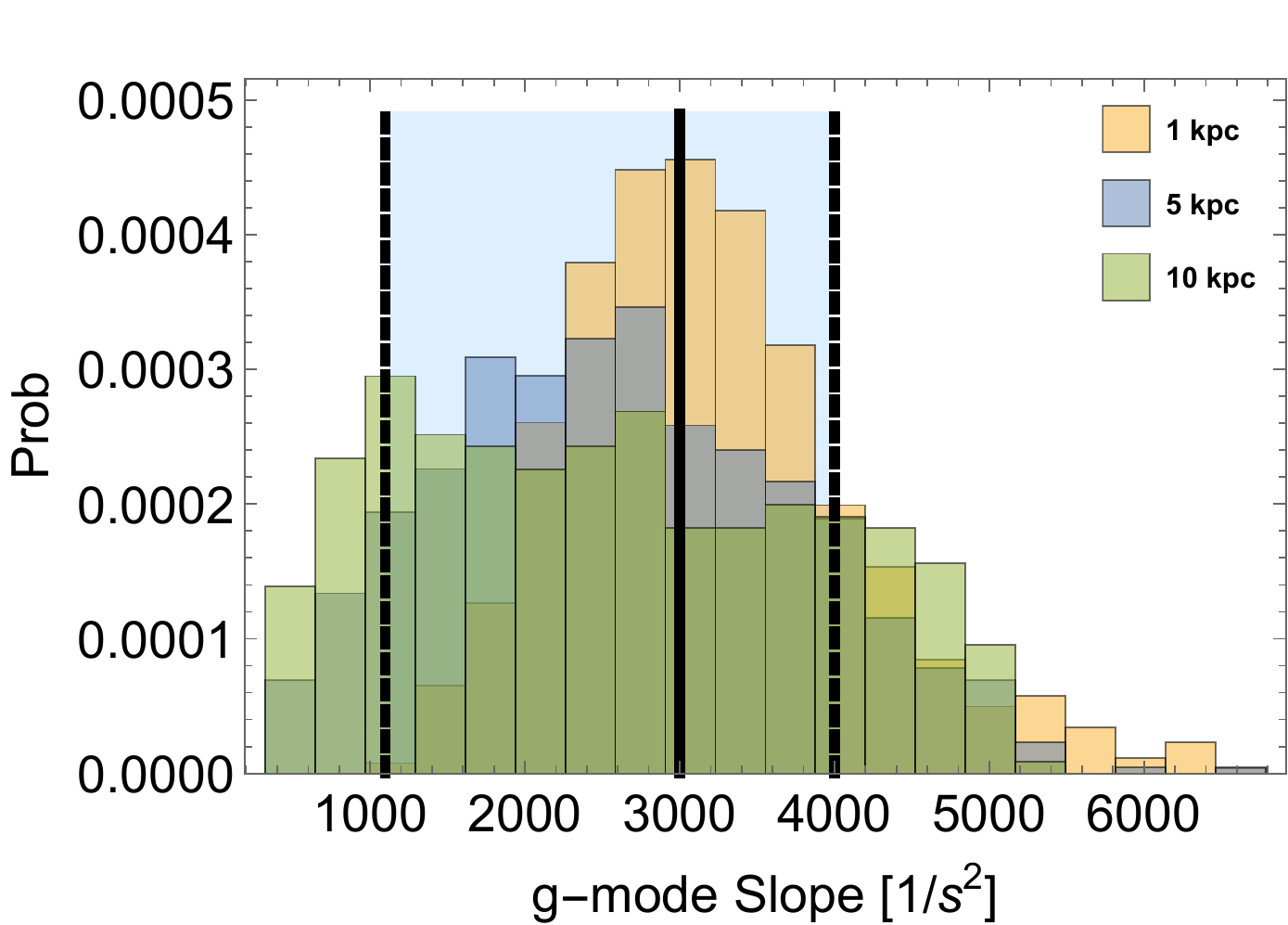}}
    \caption{G-mode slope histograms for $D = 1, 5, 10$ Kpc. The center solid vertical line identifies the slope of the g-mode evolution estimated from the spectrogram in fig.  \ref{fig:SASI_spec}, which is 3000 $\mathrm{s}^{-2}$ (which is the g-mode slope of our illustrative example, estimated visually from \ref{fig:SASI_spec}). The left and right dashed vertical line, represent the range of slopes for slowly rotating progenitors observed in the literature (derived from \cite{Warren_2020, Murphy_2009}), which are 1100 $\mathrm{s}^{-2}$ and 4000 $\mathrm{s}^{-2}$ respectively. The variance of the distribution shrinks with closer distances.}
    \label{fig:slope_1}
\end{figure}

 Minimization of the multi-variate $\chi^2$ function gives the optimal slope and intercept point (average of all the slopes and intercepts with minimum chi squared values) on the slope-intercept grid. Thus, the g-mode frequency ($f^g$) evaluation becomes;
 
\begin{eqnarray} 
    f^{g}=m^{GW}_{opt}t+c^{GW}_{opt}
\end{eqnarray}
            
where, $c^{GW}_{opt}$ and $m^{GW}_{opt}$ are the optimized intercept and slope, respectively. For the removal of the g-mode region, the g-mode initial time is calculated from the fitted line at 200 Hz as $t_{ini}^{g}=\frac{(200-c^{GW}_{opt})}{m^{GW}_{opt}}$. When the slope estimation and/or the slope optimization steps are bypassed, we only select the pixels that are $50$ ms after the earliest pixel in the t-f map of the whole event to be included in the SASI t-f region (or simply, the SASI region). Else, as the g-mode initial time $t_{ini}^{g}$ would be available, only the pixels $50$ ms after it are included in the SASI region. In this way, g-mode removal from the pixel map is performed. We then have a t-f map that only includes the SASI components. This map is further used for SASI parameter estimation described in \ref{sec:GW}.
The results of the slope optimization at different distances are listed in table \ref{tab:sasiIcePfa10} and shown in fig \ref{fig:slope_1}.

Using the pixels remaining in the SASI region, after the removal of the g-mode components, we estimate the SASI parameters: central frequency and duration, using statistical approaches described in section \ref{sec:GW}. We found that with the decrease of the bpp value we used (0.05) with respect to the standard cWB configuration (bpp = 0.1), an extra threshold was needed to remove some noise related pixels in the SASI time-frequency region, as lower bpp (to increase the number of triggers) would also allow more noise related pixels in the triggers. We choose a threshold on the likelihood values of the pixels in the SASI region, as a pixel selection criteria for parameter estimation, which is set at 50\% of the maximum likelihood value among all the pixels in the given event (i.e, an event-dependent threshold).

\bibstyle{aps}
\bibliography{b}

\begin{thebibliography}{48}
\expandafter\ifx\csname natexlab\endcsname\relax\def\natexlab#1{#1}\fi
\expandafter\ifx\csname bibnamefont\endcsname\relax
  \def\bibnamefont#1{#1}\fi
\expandafter\ifx\csname bibfnamefont\endcsname\relax
  \def\bibfnamefont#1{#1}\fi
\expandafter\ifx\csname citenamefont\endcsname\relax
  \def\citenamefont#1{#1}\fi
\expandafter\ifx\csname url\endcsname\relax
  \def\url#1{\texttt{#1}}\fi
\expandafter\ifx\csname urlprefix\endcsname\relax\def\urlprefix{URL }\fi
\providecommand{\bibinfo}[2]{#2}
\providecommand{\eprint}[2][]{\url{#2}}

\bibitem[{\citenamefont{Abbott et~al.}(2016)}]{LIGOScientific:2016aoc}
\bibinfo{author}{\bibfnamefont{B.~P.} \bibnamefont{Abbott}}
  \bibnamefont{et~al.} (\bibinfo{collaboration}{LIGO Scientific, Virgo}),
  \bibinfo{journal}{Phys. Rev. Lett.} \textbf{\bibinfo{volume}{116}},
  \bibinfo{pages}{061102} (\bibinfo{year}{2016}), \eprint{1602.03837}.

\bibitem[{\citenamefont{Abbott et~al.}(2017)}]{LIGOScientific:2017vwq}
\bibinfo{author}{\bibfnamefont{B.~P.} \bibnamefont{Abbott}}
  \bibnamefont{et~al.} (\bibinfo{collaboration}{LIGO Scientific, Virgo}),
  \bibinfo{journal}{Phys. Rev. Lett.} \textbf{\bibinfo{volume}{119}},
  \bibinfo{pages}{161101} (\bibinfo{year}{2017}), \eprint{1710.05832}.

\bibitem[{\citenamefont{Abbott et~al.}(2020)}]{LIGOScientific:2019ryq}
\bibinfo{author}{\bibfnamefont{B.~P.} \bibnamefont{Abbott}}
  \bibnamefont{et~al.} (\bibinfo{collaboration}{LIGO Scientific, Virgo}),
  \bibinfo{journal}{Phys. Rev. D} \textbf{\bibinfo{volume}{101}},
  \bibinfo{pages}{084002} (\bibinfo{year}{2020}), \eprint{1908.03584}.

\bibitem[{\citenamefont{Szczepanczyk}(2018)}]{marek}
\bibinfo{author}{\bibfnamefont{M.~J.} \bibnamefont{Szczepanczyk}}, Ph.D.
  thesis, \bibinfo{school}{Embry-Riddle Aeronautical University, Daytona Beach,
  FL} (\bibinfo{year}{2018}).

\bibitem[{\citenamefont{Lin et~al.}(2020)\citenamefont{Lin, Lunardini, Zanolin,
  Kotake, and Richardson}}]{Lin:2019wwm}
\bibinfo{author}{\bibfnamefont{Z.}~\bibnamefont{Lin}},
  \bibinfo{author}{\bibfnamefont{C.}~\bibnamefont{Lunardini}},
  \bibinfo{author}{\bibfnamefont{M.}~\bibnamefont{Zanolin}},
  \bibinfo{author}{\bibfnamefont{K.}~\bibnamefont{Kotake}}, \bibnamefont{and}
  \bibinfo{author}{\bibfnamefont{C.}~\bibnamefont{Richardson}},
  \bibinfo{journal}{Phys. Rev. D} \textbf{\bibinfo{volume}{101}},
  \bibinfo{pages}{123028} (\bibinfo{year}{2020}), \eprint{1911.10656}.

\bibitem[{\citenamefont{Szczepanczyk et~al.}(2021)}]{Szczepanczyk:2021bka}
\bibinfo{author}{\bibfnamefont{M.}~\bibnamefont{Szczepanczyk}}
  \bibnamefont{et~al.}, \bibinfo{journal}{Phys. Rev. D}
  \textbf{\bibinfo{volume}{104}}, \bibinfo{pages}{102002}
  (\bibinfo{year}{2021}), \eprint{2104.06462}.

\bibitem[{\citenamefont{Halim et~al.}(2021{\natexlab{a}})\citenamefont{Halim,
  Casentini, Drago, Fafone, Scholberg, Vigorito, and
  Pagliaroli}}]{Halim:2021qll}
\bibinfo{author}{\bibfnamefont{O.}~\bibnamefont{Halim}},
  \bibinfo{author}{\bibfnamefont{C.}~\bibnamefont{Casentini}},
  \bibinfo{author}{\bibfnamefont{M.}~\bibnamefont{Drago}},
  \bibinfo{author}{\bibfnamefont{V.}~\bibnamefont{Fafone}},
  \bibinfo{author}{\bibfnamefont{K.}~\bibnamefont{Scholberg}},
  \bibinfo{author}{\bibfnamefont{C.~F.} \bibnamefont{Vigorito}},
  \bibnamefont{and}
  \bibinfo{author}{\bibfnamefont{G.}~\bibnamefont{Pagliaroli}}, in
  \emph{\bibinfo{booktitle}{{16th Marcel Grossmann Meeting on~Recent
  Developments in Theoretical and Experimental General Relativity, Astrophysics
  and Relativistic Field Theories}}} (\bibinfo{publisher}{World Scientific
  Publishing Company, NJ}, \bibinfo{year}{2021}{\natexlab{a}}),
  \eprint{2110.15620}.

\bibitem[{\citenamefont{Halim et~al.}(2021{\natexlab{b}})\citenamefont{Halim,
  Casentini, Drago, Fafone, Scholberg, Vigorito, and
  Pagliaroli}}]{Halim:2021yqa}
\bibinfo{author}{\bibfnamefont{O.}~\bibnamefont{Halim}},
  \bibinfo{author}{\bibfnamefont{C.}~\bibnamefont{Casentini}},
  \bibinfo{author}{\bibfnamefont{M.}~\bibnamefont{Drago}},
  \bibinfo{author}{\bibfnamefont{V.}~\bibnamefont{Fafone}},
  \bibinfo{author}{\bibfnamefont{K.}~\bibnamefont{Scholberg}},
  \bibinfo{author}{\bibfnamefont{C.~F.} \bibnamefont{Vigorito}},
  \bibnamefont{and}
  \bibinfo{author}{\bibfnamefont{G.}~\bibnamefont{Pagliaroli}},
  \bibinfo{journal}{JCAP} \textbf{\bibinfo{volume}{11}}, \bibinfo{pages}{021}
  (\bibinfo{year}{2021}{\natexlab{b}}), \eprint{2107.02050}.

\bibitem[{\citenamefont{Astone et~al.}(2018)\citenamefont{Astone,
  Cerd\'a-Dur\'an, Di~Palma, Drago, Muciaccia, Palomba, and
  Ricci}}]{Astone:2018uge}
\bibinfo{author}{\bibfnamefont{P.}~\bibnamefont{Astone}},
  \bibinfo{author}{\bibfnamefont{P.}~\bibnamefont{Cerd\'a-Dur\'an}},
  \bibinfo{author}{\bibfnamefont{I.}~\bibnamefont{Di~Palma}},
  \bibinfo{author}{\bibfnamefont{M.}~\bibnamefont{Drago}},
  \bibinfo{author}{\bibfnamefont{F.}~\bibnamefont{Muciaccia}},
  \bibinfo{author}{\bibfnamefont{C.}~\bibnamefont{Palomba}}, \bibnamefont{and}
  \bibinfo{author}{\bibfnamefont{F.}~\bibnamefont{Ricci}},
  \bibinfo{journal}{Phys. Rev. D} \textbf{\bibinfo{volume}{98}},
  \bibinfo{pages}{122002} (\bibinfo{year}{2018}), \eprint{1812.05363}.

\bibitem[{\citenamefont{Kotake}(2013)}]{Kotake:2011yv}
\bibinfo{author}{\bibfnamefont{K.}~\bibnamefont{Kotake}},
  \bibinfo{journal}{Comptes Rendus Physique} \textbf{\bibinfo{volume}{14}},
  \bibinfo{pages}{318} (\bibinfo{year}{2013}), \eprint{1110.5107}.

\bibitem[{\citenamefont{Torres-Forn\'e
  et~al.}(2019{\natexlab{a}})\citenamefont{Torres-Forn\'e, Cerd\'a-Dur\'an,
  Obergaulinger, M\"uller, and Font}}]{Torres-Forne:2019zwz}
\bibinfo{author}{\bibfnamefont{A.}~\bibnamefont{Torres-Forn\'e}},
  \bibinfo{author}{\bibfnamefont{P.}~\bibnamefont{Cerd\'a-Dur\'an}},
  \bibinfo{author}{\bibfnamefont{M.}~\bibnamefont{Obergaulinger}},
  \bibinfo{author}{\bibfnamefont{B.}~\bibnamefont{M\"uller}}, \bibnamefont{and}
  \bibinfo{author}{\bibfnamefont{J.~A.} \bibnamefont{Font}},
  \bibinfo{journal}{Phys. Rev. Lett.} \textbf{\bibinfo{volume}{123}},
  \bibinfo{pages}{051102} (\bibinfo{year}{2019}{\natexlab{a}}),
  \bibinfo{note}{[Erratum: Phys.Rev.Lett. 127, 239901 (2021)]},
  \eprint{1902.10048}.

\bibitem[{\citenamefont{Blondin and Mezzacappa}(2006)}]{Blondin:2005wz}
\bibinfo{author}{\bibfnamefont{J.~M.} \bibnamefont{Blondin}} \bibnamefont{and}
  \bibinfo{author}{\bibfnamefont{A.}~\bibnamefont{Mezzacappa}},
  \bibinfo{journal}{Astrophys. J.} \textbf{\bibinfo{volume}{642}},
  \bibinfo{pages}{401} (\bibinfo{year}{2006}), \eprint{astro-ph/0507181}.

\bibitem[{\citenamefont{Mezzacappa et~al.}(2020)}]{Mezzacappa:2020lsn}
\bibinfo{author}{\bibfnamefont{A.}~\bibnamefont{Mezzacappa}}
  \bibnamefont{et~al.}, \bibinfo{journal}{Phys. Rev. D}
  \textbf{\bibinfo{volume}{102}}, \bibinfo{pages}{023027}
  (\bibinfo{year}{2020}), \eprint{2007.15099}.

\bibitem[{\citenamefont{Kuroda et~al.}(2016)\citenamefont{Kuroda, Kotake, and
  Takiwaki}}]{Kuroda:2016bjd}
\bibinfo{author}{\bibfnamefont{T.}~\bibnamefont{Kuroda}},
  \bibinfo{author}{\bibfnamefont{K.}~\bibnamefont{Kotake}}, \bibnamefont{and}
  \bibinfo{author}{\bibfnamefont{T.}~\bibnamefont{Takiwaki}},
  \bibinfo{journal}{Astrophys. J. Lett.} \textbf{\bibinfo{volume}{829}},
  \bibinfo{pages}{L14} (\bibinfo{year}{2016}), \eprint{1605.09215}.

\bibitem[{\citenamefont{Kotake et~al.}(2007)\citenamefont{Kotake, Ohnishi, and
  Yamada}}]{Kotake:2006aq}
\bibinfo{author}{\bibfnamefont{K.}~\bibnamefont{Kotake}},
  \bibinfo{author}{\bibfnamefont{N.}~\bibnamefont{Ohnishi}}, \bibnamefont{and}
  \bibinfo{author}{\bibfnamefont{S.}~\bibnamefont{Yamada}},
  \bibinfo{journal}{Astrophys. J.} \textbf{\bibinfo{volume}{655}},
  \bibinfo{pages}{406} (\bibinfo{year}{2007}), \eprint{astro-ph/0607224}.

\bibitem[{\citenamefont{Kotake et~al.}(2009)\citenamefont{Kotake, Iwakami,
  Ohnishi, and Yamada}}]{Kotake_2009}
\bibinfo{author}{\bibfnamefont{K.}~\bibnamefont{Kotake}},
  \bibinfo{author}{\bibfnamefont{W.}~\bibnamefont{Iwakami}},
  \bibinfo{author}{\bibfnamefont{N.}~\bibnamefont{Ohnishi}}, \bibnamefont{and}
  \bibinfo{author}{\bibfnamefont{S.}~\bibnamefont{Yamada}},
  \bibinfo{journal}{The Astrophysical Journal} \textbf{\bibinfo{volume}{697}},
  \bibinfo{pages}{L133} (\bibinfo{year}{2009}),
  \urlprefix\url{https://doi.org/10.1088%2F0004-637x%2F697%2F2%2Fl133}.

\bibitem[{\citenamefont{Ohnishi et~al.}(2006)\citenamefont{Ohnishi, Kotake, and
  Yamada}}]{Ohnishi:2005cv}
\bibinfo{author}{\bibfnamefont{N.}~\bibnamefont{Ohnishi}},
  \bibinfo{author}{\bibfnamefont{K.}~\bibnamefont{Kotake}}, \bibnamefont{and}
  \bibinfo{author}{\bibfnamefont{S.}~\bibnamefont{Yamada}},
  \bibinfo{journal}{Astrophys. J.} \textbf{\bibinfo{volume}{641}},
  \bibinfo{pages}{1018} (\bibinfo{year}{2006}), \eprint{astro-ph/0509765}.

\bibitem[{\citenamefont{Ohnishi et~al.}(2008)\citenamefont{Ohnishi, Iwakami,
  Kotake, Yamada, Fujioka, and Takabe}}]{Ohnishi_2008}
\bibinfo{author}{\bibfnamefont{N.}~\bibnamefont{Ohnishi}},
  \bibinfo{author}{\bibfnamefont{W.}~\bibnamefont{Iwakami}},
  \bibinfo{author}{\bibfnamefont{K.}~\bibnamefont{Kotake}},
  \bibinfo{author}{\bibfnamefont{S.}~\bibnamefont{Yamada}},
  \bibinfo{author}{\bibfnamefont{S.}~\bibnamefont{Fujioka}}, \bibnamefont{and}
  \bibinfo{author}{\bibfnamefont{H.}~\bibnamefont{Takabe}},
  \bibinfo{journal}{Journal of Physics: Conference Series}
  \textbf{\bibinfo{volume}{112}}, \bibinfo{pages}{042018}
  (\bibinfo{year}{2008}),
  \urlprefix\url{https://doi.org/10.1088/1742-6596/112/4/042018}.

\bibitem[{\citenamefont{Marek and Janka}(2009)}]{Marek:2007gr}
\bibinfo{author}{\bibfnamefont{A.}~\bibnamefont{Marek}} \bibnamefont{and}
  \bibinfo{author}{\bibfnamefont{H.~T.} \bibnamefont{Janka}},
  \bibinfo{journal}{Astrophys. J.} \textbf{\bibinfo{volume}{694}},
  \bibinfo{pages}{664} (\bibinfo{year}{2009}), \eprint{0708.3372}.

\bibitem[{\citenamefont{Guilet and
  Foglizzo}(2012)}]{10.1111/j.1365-2966.2012.20333.x}
\bibinfo{author}{\bibfnamefont{J.}~\bibnamefont{Guilet}} \bibnamefont{and}
  \bibinfo{author}{\bibfnamefont{T.}~\bibnamefont{Foglizzo}},
  \bibinfo{journal}{Monthly Notices of the Royal Astronomical Society}
  \textbf{\bibinfo{volume}{421}}, \bibinfo{pages}{546} (\bibinfo{year}{2012}),
  ISSN \bibinfo{issn}{0035-8711},
  \eprint{https://academic.oup.com/mnras/article-pdf/421/1/546/3136869/mnras0421-0546.pdf},
  \urlprefix\url{https://doi.org/10.1111/j.1365-2966.2012.20333.x}.

\bibitem[{\citenamefont{Scheck et~al.}(2008)\citenamefont{Scheck, Janka,
  Foglizzo, and Kifonidis}}]{Scheck:2007gw}
\bibinfo{author}{\bibfnamefont{L.}~\bibnamefont{Scheck}},
  \bibinfo{author}{\bibfnamefont{H.~T.} \bibnamefont{Janka}},
  \bibinfo{author}{\bibfnamefont{T.}~\bibnamefont{Foglizzo}}, \bibnamefont{and}
  \bibinfo{author}{\bibfnamefont{K.}~\bibnamefont{Kifonidis}},
  \bibinfo{journal}{Astron. Astrophys.} \textbf{\bibinfo{volume}{477}},
  \bibinfo{pages}{931} (\bibinfo{year}{2008}), \eprint{0704.3001}.

\bibitem[{\citenamefont{Foglizzo et~al.}(2007)\citenamefont{Foglizzo, Galletti,
  Scheck, and Janka}}]{Foglizzo:2006fu}
\bibinfo{author}{\bibfnamefont{T.}~\bibnamefont{Foglizzo}},
  \bibinfo{author}{\bibfnamefont{P.}~\bibnamefont{Galletti}},
  \bibinfo{author}{\bibfnamefont{L.}~\bibnamefont{Scheck}}, \bibnamefont{and}
  \bibinfo{author}{\bibfnamefont{H.~T.} \bibnamefont{Janka}},
  \bibinfo{journal}{Astrophys. J.} \textbf{\bibinfo{volume}{654}},
  \bibinfo{pages}{1006} (\bibinfo{year}{2007}), \eprint{astro-ph/0606640}.

\bibitem[{\citenamefont{Foglizzo et~al.}(2015)}]{Foglizzo:2015dma}
\bibinfo{author}{\bibfnamefont{T.}~\bibnamefont{Foglizzo}}
  \bibnamefont{et~al.}, \bibinfo{journal}{Publ. Astron. Soc. Austral.}
  \textbf{\bibinfo{volume}{32}}, \bibinfo{pages}{e009} (\bibinfo{year}{2015}),
  \eprint{1501.01334}.

\bibitem[{\citenamefont{Walk et~al.}(2020)\citenamefont{Walk, Tamborra, Janka,
  Summa, and Kresse}}]{Walk:2019miz}
\bibinfo{author}{\bibfnamefont{L.}~\bibnamefont{Walk}},
  \bibinfo{author}{\bibfnamefont{I.}~\bibnamefont{Tamborra}},
  \bibinfo{author}{\bibfnamefont{H.-T.} \bibnamefont{Janka}},
  \bibinfo{author}{\bibfnamefont{A.}~\bibnamefont{Summa}}, \bibnamefont{and}
  \bibinfo{author}{\bibfnamefont{D.}~\bibnamefont{Kresse}},
  \bibinfo{journal}{Phys. Rev. D} \textbf{\bibinfo{volume}{101}},
  \bibinfo{pages}{123013} (\bibinfo{year}{2020}), \eprint{1910.12971}.

\bibitem[{\citenamefont{Mueller et~al.}(2013)\citenamefont{Mueller, Janka, and
  Marek}}]{Mueller:2012sv}
\bibinfo{author}{\bibfnamefont{B.}~\bibnamefont{Mueller}},
  \bibinfo{author}{\bibfnamefont{H.-T.} \bibnamefont{Janka}}, \bibnamefont{and}
  \bibinfo{author}{\bibfnamefont{A.}~\bibnamefont{Marek}},
  \bibinfo{journal}{Astrophys. J.} \textbf{\bibinfo{volume}{766}},
  \bibinfo{pages}{43} (\bibinfo{year}{2013}), \eprint{1210.6984}.

\bibitem[{\citenamefont{Lund et~al.}(2010)\citenamefont{Lund, Marek, Lunardini,
  Janka, and Raffelt}}]{Lund:2010kh}
\bibinfo{author}{\bibfnamefont{T.}~\bibnamefont{Lund}},
  \bibinfo{author}{\bibfnamefont{A.}~\bibnamefont{Marek}},
  \bibinfo{author}{\bibfnamefont{C.}~\bibnamefont{Lunardini}},
  \bibinfo{author}{\bibfnamefont{H.-T.} \bibnamefont{Janka}}, \bibnamefont{and}
  \bibinfo{author}{\bibfnamefont{G.}~\bibnamefont{Raffelt}},
  \bibinfo{journal}{Phys. Rev. D} \textbf{\bibinfo{volume}{82}},
  \bibinfo{pages}{063007} (\bibinfo{year}{2010}), \eprint{1006.1889}.

\bibitem[{\citenamefont{Lund et~al.}(2012)\citenamefont{Lund, Wongwathanarat,
  Janka, Muller, and Raffelt}}]{Lund:2012vm}
\bibinfo{author}{\bibfnamefont{T.}~\bibnamefont{Lund}},
  \bibinfo{author}{\bibfnamefont{A.}~\bibnamefont{Wongwathanarat}},
  \bibinfo{author}{\bibfnamefont{H.-T.} \bibnamefont{Janka}},
  \bibinfo{author}{\bibfnamefont{E.}~\bibnamefont{Muller}}, \bibnamefont{and}
  \bibinfo{author}{\bibfnamefont{G.}~\bibnamefont{Raffelt}},
  \bibinfo{journal}{Phys. Rev. D} \textbf{\bibinfo{volume}{86}},
  \bibinfo{pages}{105031} (\bibinfo{year}{2012}), \eprint{1208.0043}.

\bibitem[{\citenamefont{Takeda et~al.}(2021)}]{Takeda:2021hmf}
\bibinfo{author}{\bibfnamefont{M.}~\bibnamefont{Takeda}} \bibnamefont{et~al.},
  \bibinfo{journal}{Phys. Rev. D} \textbf{\bibinfo{volume}{104}},
  \bibinfo{pages}{084063} (\bibinfo{year}{2021}), \eprint{2107.05213}.

\bibitem[{\citenamefont{Roma et~al.}(2019)\citenamefont{Roma, Powell, Heng, and
  Frey}}]{PhysRevD.99.063018}
\bibinfo{author}{\bibfnamefont{V.}~\bibnamefont{Roma}},
  \bibinfo{author}{\bibfnamefont{J.}~\bibnamefont{Powell}},
  \bibinfo{author}{\bibfnamefont{I.~S.} \bibnamefont{Heng}}, \bibnamefont{and}
  \bibinfo{author}{\bibfnamefont{R.}~\bibnamefont{Frey}},
  \bibinfo{journal}{Phys. Rev. D} \textbf{\bibinfo{volume}{99}},
  \bibinfo{pages}{063018} (\bibinfo{year}{2019}),
  \urlprefix\url{https://link.aps.org/doi/10.1103/PhysRevD.99.063018}.

\bibitem[{\citenamefont{Abbott et~al.}(2018)}]{Abbott_2018}
\bibinfo{author}{\bibfnamefont{B.~P.} \bibnamefont{Abbott}}
  \bibnamefont{et~al.}, \bibinfo{journal}{Living Reviews in Relativity}
  \textbf{\bibinfo{volume}{21}} (\bibinfo{year}{2018}),
  \urlprefix\url{https://doi.org/10.1007%2Fs41114-018-0012-9}.

\bibitem[{\citenamefont{Klimenko et~al.}(2016)}]{Klimenko:2015ypf}
\bibinfo{author}{\bibfnamefont{S.}~\bibnamefont{Klimenko}}
  \bibnamefont{et~al.}, \bibinfo{journal}{Phys. Rev. D}
  \textbf{\bibinfo{volume}{93}}, \bibinfo{pages}{042004}
  (\bibinfo{year}{2016}), \eprint{1511.05999}.

\bibitem[{\citenamefont{Kuroda et~al.}(2017)\citenamefont{Kuroda, Kotake,
  Hayama, and Takiwaki}}]{Kuroda:2017trn}
\bibinfo{author}{\bibfnamefont{T.}~\bibnamefont{Kuroda}},
  \bibinfo{author}{\bibfnamefont{K.}~\bibnamefont{Kotake}},
  \bibinfo{author}{\bibfnamefont{K.}~\bibnamefont{Hayama}}, \bibnamefont{and}
  \bibinfo{author}{\bibfnamefont{T.}~\bibnamefont{Takiwaki}},
  \bibinfo{journal}{Astrophys. J.} \textbf{\bibinfo{volume}{851}},
  \bibinfo{pages}{62} (\bibinfo{year}{2017}), \eprint{1708.05252}.

\bibitem[{\citenamefont{Andresen et~al.}(2019)\citenamefont{Andresen, Müller,
  Janka, Summa, Gill, and Zanolin}}]{10.1093/mnras/stz990}
\bibinfo{author}{\bibfnamefont{H.}~\bibnamefont{Andresen}},
  \bibinfo{author}{\bibfnamefont{E.}~\bibnamefont{Müller}},
  \bibinfo{author}{\bibfnamefont{H.-T.} \bibnamefont{Janka}},
  \bibinfo{author}{\bibfnamefont{A.}~\bibnamefont{Summa}},
  \bibinfo{author}{\bibfnamefont{K.}~\bibnamefont{Gill}}, \bibnamefont{and}
  \bibinfo{author}{\bibfnamefont{M.}~\bibnamefont{Zanolin}},
  \bibinfo{journal}{Monthly Notices of the Royal Astronomical Society}
  \textbf{\bibinfo{volume}{486}}, \bibinfo{pages}{2238} (\bibinfo{year}{2019}),
  ISSN \bibinfo{issn}{0035-8711},
  \eprint{https://academic.oup.com/mnras/article-pdf/486/2/2238/28488247/stz990.pdf},
  \urlprefix\url{https://doi.org/10.1093/mnras/stz990}.

\bibitem[{\citenamefont{Murphy et~al.}(2009)\citenamefont{Murphy, Ott, and
  Burrows}}]{Murphy_2009}
\bibinfo{author}{\bibfnamefont{J.~W.} \bibnamefont{Murphy}},
  \bibinfo{author}{\bibfnamefont{C.~D.} \bibnamefont{Ott}}, \bibnamefont{and}
  \bibinfo{author}{\bibfnamefont{A.}~\bibnamefont{Burrows}},
  \bibinfo{journal}{The Astrophysical Journal} \textbf{\bibinfo{volume}{707}},
  \bibinfo{pages}{1173} (\bibinfo{year}{2009}),
  \urlprefix\url{https://doi.org/10.1088%2F0004-637x%2F707%2F2%2F1173}.

\bibitem[{\citenamefont{Torres-Forné et~al.}(2017)\citenamefont{Torres-Forné,
  Cerdá-Durán, Passamonti, and Font}}]{10.1093/mnras/stx3067}
\bibinfo{author}{\bibfnamefont{A.}~\bibnamefont{Torres-Forné}},
  \bibinfo{author}{\bibfnamefont{P.}~\bibnamefont{Cerdá-Durán}},
  \bibinfo{author}{\bibfnamefont{A.}~\bibnamefont{Passamonti}},
  \bibnamefont{and} \bibinfo{author}{\bibfnamefont{J.~A.} \bibnamefont{Font}},
  \bibinfo{journal}{Monthly Notices of the Royal Astronomical Society}
  \textbf{\bibinfo{volume}{474}}, \bibinfo{pages}{5272} (\bibinfo{year}{2017}),
  ISSN \bibinfo{issn}{0035-8711},
  \eprint{https://academic.oup.com/mnras/article-pdf/474/4/5272/23231251/stx3067.pdf},
  \urlprefix\url{https://doi.org/10.1093/mnras/stx3067}.

\bibitem[{\citenamefont{Morozova et~al.}(2018)\citenamefont{Morozova, Radice,
  Burrows, and Vartanyan}}]{Morozova:2018glm}
\bibinfo{author}{\bibfnamefont{V.}~\bibnamefont{Morozova}},
  \bibinfo{author}{\bibfnamefont{D.}~\bibnamefont{Radice}},
  \bibinfo{author}{\bibfnamefont{A.}~\bibnamefont{Burrows}}, \bibnamefont{and}
  \bibinfo{author}{\bibfnamefont{D.}~\bibnamefont{Vartanyan}},
  \bibinfo{journal}{Astrophys. J.} \textbf{\bibinfo{volume}{861}},
  \bibinfo{pages}{10} (\bibinfo{year}{2018}), \eprint{1801.01914}.

\bibitem[{\citenamefont{Torres-Forn\'e
  et~al.}(2019{\natexlab{b}})\citenamefont{Torres-Forn\'e, Cerd\'a-Dur\'an,
  Passamonti, Obergaulinger, and Font}}]{Torres-Forne:2018nzj}
\bibinfo{author}{\bibfnamefont{A.}~\bibnamefont{Torres-Forn\'e}},
  \bibinfo{author}{\bibfnamefont{P.}~\bibnamefont{Cerd\'a-Dur\'an}},
  \bibinfo{author}{\bibfnamefont{A.}~\bibnamefont{Passamonti}},
  \bibinfo{author}{\bibfnamefont{M.}~\bibnamefont{Obergaulinger}},
  \bibnamefont{and} \bibinfo{author}{\bibfnamefont{J.~A.} \bibnamefont{Font}},
  \bibinfo{journal}{Mon. Not. Roy. Astron. Soc.}
  \textbf{\bibinfo{volume}{482}}, \bibinfo{pages}{3967}
  (\bibinfo{year}{2019}{\natexlab{b}}), \eprint{1806.11366}.

\bibitem[{\citenamefont{Sotani and Takiwaki}(2020)}]{Sotani:2020dnh}
\bibinfo{author}{\bibfnamefont{H.}~\bibnamefont{Sotani}} \bibnamefont{and}
  \bibinfo{author}{\bibfnamefont{T.}~\bibnamefont{Takiwaki}},
  \bibinfo{journal}{Phys. Rev. D} \textbf{\bibinfo{volume}{102}},
  \bibinfo{pages}{023028} (\bibinfo{year}{2020}), \eprint{2004.09871}.

\bibitem[{\citenamefont{Richardson et~al.}(2022)\citenamefont{Richardson,
  Zanolin, Andresen, Szczepa\'nczyk, Gill, and
  Wongwathanarat}}]{Richardson:2021lib}
\bibinfo{author}{\bibfnamefont{C.~J.} \bibnamefont{Richardson}},
  \bibinfo{author}{\bibfnamefont{M.}~\bibnamefont{Zanolin}},
  \bibinfo{author}{\bibfnamefont{H.}~\bibnamefont{Andresen}},
  \bibinfo{author}{\bibfnamefont{M.~J.} \bibnamefont{Szczepa\'nczyk}},
  \bibinfo{author}{\bibfnamefont{K.}~\bibnamefont{Gill}}, \bibnamefont{and}
  \bibinfo{author}{\bibfnamefont{A.}~\bibnamefont{Wongwathanarat}},
  \bibinfo{journal}{Phys. Rev. D} \textbf{\bibinfo{volume}{105}},
  \bibinfo{pages}{103008} (\bibinfo{year}{2022}), \eprint{2109.01582}.

\bibitem[{\citenamefont{Jolien and Warren}(2011)}]{Jolien:2011}
\bibinfo{author}{\bibfnamefont{D.~E.~C.} \bibnamefont{Jolien}}
  \bibnamefont{and} \bibinfo{author}{\bibfnamefont{G.~A.}
  \bibnamefont{Warren}}, \emph{\bibinfo{title}{{Gravitational‐Wave Physics
  and Astronomy: An Introduction to Theory, Experiment and Data Analysis}}}
  (\bibinfo{publisher}{Wiley, NY}, \bibinfo{year}{2011}).

\bibitem[{\citenamefont{Oppenheim and Ronald}(1975)}]{Oppenheim:1975}
\bibinfo{author}{\bibfnamefont{A.~V.} \bibnamefont{Oppenheim}}
  \bibnamefont{and} \bibinfo{author}{\bibfnamefont{W.~S.}
  \bibnamefont{Ronald}}, \emph{\bibinfo{title}{{Digital Signal Processing}}}
  (\bibinfo{publisher}{Pearson, NY}, \bibinfo{year}{1975}).

\bibitem[{\citenamefont{Andresen et~al.}(2017)\citenamefont{Andresen, M\"uller,
  M\"uller, and Janka}}]{Andresen:2016pdt}
\bibinfo{author}{\bibfnamefont{H.}~\bibnamefont{Andresen}},
  \bibinfo{author}{\bibfnamefont{B.}~\bibnamefont{M\"uller}},
  \bibinfo{author}{\bibfnamefont{E.}~\bibnamefont{M\"uller}}, \bibnamefont{and}
  \bibinfo{author}{\bibfnamefont{H.-T.} \bibnamefont{Janka}},
  \bibinfo{journal}{Mon. Not. Roy. Astron. Soc.}
  \textbf{\bibinfo{volume}{468}}, \bibinfo{pages}{2032} (\bibinfo{year}{2017}),
  \eprint{1607.05199}.

\bibitem[{\citenamefont{Szczepa\'nczyk and
  Zanolin}(2022)}]{Szczepanczyk:2022lni}
\bibinfo{author}{\bibfnamefont{M.}~\bibnamefont{Szczepa\'nczyk}}
  \bibnamefont{and} \bibinfo{author}{\bibfnamefont{M.}~\bibnamefont{Zanolin}},
  \bibinfo{journal}{Galaxies} \textbf{\bibinfo{volume}{10}},
  \bibinfo{pages}{70} (\bibinfo{year}{2022}).

\bibitem[{\citenamefont{Abbott et~al.}(2022)}]{LIGOScientific:2022enz}
\bibinfo{author}{\bibfnamefont{R.}~\bibnamefont{Abbott}} \bibnamefont{et~al.}
  (\bibinfo{collaboration}{LIGO Scientific, VIRGO, KAGRA})
  (\bibinfo{year}{2022}), \eprint{2209.02863}.

\bibitem[{\citenamefont{Lin et~al.}(2022)\citenamefont{Lin, Zha, O'Connor, and
  Steiner}}]{Lin:2022lck}
\bibinfo{author}{\bibfnamefont{Z.}~\bibnamefont{Lin}},
  \bibinfo{author}{\bibfnamefont{S.}~\bibnamefont{Zha}},
  \bibinfo{author}{\bibfnamefont{E.~P.} \bibnamefont{O'Connor}},
  \bibnamefont{and} \bibinfo{author}{\bibfnamefont{A.~W.}
  \bibnamefont{Steiner}} (\bibinfo{year}{2022}), \eprint{2203.05141}.

\bibitem[{\citenamefont{Drago}(2008)}]{drago}
\bibinfo{author}{\bibfnamefont{M.}~\bibnamefont{Drago}}, Ph.D. thesis,
  \bibinfo{school}{Universit`a degli Studi di Padova} (\bibinfo{year}{2008}),
  \urlprefix\url{http://hdl.handle.net/11577/3422378}.

\bibitem[{\citenamefont{Necula et~al.}(2012)\citenamefont{Necula, Klimenko, and
  Mitselmakher}}]{Necula:2012zz}
\bibinfo{author}{\bibfnamefont{V.}~\bibnamefont{Necula}},
  \bibinfo{author}{\bibfnamefont{S.}~\bibnamefont{Klimenko}}, \bibnamefont{and}
  \bibinfo{author}{\bibfnamefont{G.}~\bibnamefont{Mitselmakher}},
  \bibinfo{journal}{J. Phys. Conf. Ser.} \textbf{\bibinfo{volume}{363}},
  \bibinfo{pages}{012032} (\bibinfo{year}{2012}).

\bibitem[{\citenamefont{Warren et~al.}(2020)\citenamefont{Warren, Couch,
  O'Connor, and Morozova}}]{Warren_2020}
\bibinfo{author}{\bibfnamefont{M.~L.} \bibnamefont{Warren}},
  \bibinfo{author}{\bibfnamefont{S.~M.} \bibnamefont{Couch}},
  \bibinfo{author}{\bibfnamefont{E.~P.} \bibnamefont{O'Connor}},
  \bibnamefont{and} \bibinfo{author}{\bibfnamefont{V.}~\bibnamefont{Morozova}},
  \bibinfo{journal}{The Astrophysical Journal} \textbf{\bibinfo{volume}{898}},
  \bibinfo{pages}{139} (\bibinfo{year}{2020}),
  \urlprefix\url{https://doi.org/10.3847%2F1538-4357%2Fab97b7}.

\end{thebibliography}

\end{document}